\documentstyle[12pt]{article}
\newcommand{\citenum}{\cite}

\newcommand{\bcn}{\begin{center}}
\newcommand{\beq}{\begin{equation}}
\newcommand{\beqn}{\begin{eqnarray}}
\newcommand{\ecn}{\end{center}}
\newcommand{\eeq}{\end{equation}}
\newcommand{\eeqn}{\end{eqnarray}}

 \def\lsim{\mathrel{\rlap{\lower4pt\hbox{\hskip1pt$\sim$}}
    \raise1pt\hbox{$<$}}}
\def\slash#1{\setbox0=\hbox{$#1$}#1\hskip-\wd0\hbox to\wd0{\hss\sl/\/\hss}}

\begin{document}
\begin{center}
{\Large Quantum Color Transparency and Nuclear Filtering}
\end{center}
\vspace*{.25cm}
\begin{center}
{\bf Pankaj Jain}\\
Department of Physics \\
Indian Institute of Technology$^\dagger$\\
Kanpur, INDIA 208016 \\
and International Center for Theoretical Physics, Trieste 34100 ITALY\\
\vspace*{.25cm}
{\bf Bernard Pire}\\
Centre de Physique Theorique \\
Ecole Polytechnique \\
F91128 Palaisseau FRANCE\\
\vspace*{.25cm}
{\bf John P. Ralston}\\
Department of Physics \\
University of Kansas \\
Lawrence, KS 66045 USA
\end{center}
\vspace*{.25cm}

\begin{center}
{\bf Abstract}
\end{center}
Color transparency is the proposal that under certain circumstances the strong
interactions can be controlled and in some cases reduced in magnitude. We give
a comprehensive review of the physics, which hinges on the interface of
perturbative QCD with non--perturbative strong interactions. Color transparency
is expected to occur in many kinds of quasi--exclusive reactions with either
electron or hadrons beams. We review the interplay of color transparency with
{\it nuclear filtering}, which is the conversion of quark wave functions in
hadrons to small transverse space dimensions by interaction with a nuclear
medium. A complete description of the phenomena as a multi--scale QCD process
is given in terms of light--cone matrix elements. We also review a number of
other approaches, including pictures based on modeling the time evolution of
hadronic wave--packets and the use of hadronic interpolating states. Spin plays
an intrinsic role in hard exclusive reactions and the role of spin in testing
and understanding color transparency and nuclear filtering is discussed at
length. We emphasize the use of data analysis procedures which have minimal
model dependence, while comparing and contrasting various models.
We review existing experimental data and the experimental program planned at
various facilities. We also discuss making use of the subject's strong
scientific complementarity and potential for progress in exploring hadron
physics at current and future facilities, arguing that the combination of
different experiments can lead to a broad new program to study the strong
interactions and hadron structure.

\vspace*{.15cm}
\hbox to1.5truein {\hrulefill}
\vspace*{.15cm}

\noindent $^\dagger$ Permanent address

\baselineskip=18pt
\newpage
\pagenumbering{roman}							
\tableofcontents

\newpage

\pagenumbering{arabic}

We will begin our survey with an overview of the basic idea of color
transparency and how the concepts have evolved. Following an introduction,
we will present the
basic idea of ``survial of the smallest", then ``back up" to add explanation,
and answer some common questions. The Section is closed with a brief discussion
of
a program in strong interaction physics at current and future facilities.

\section{ Introduction}

\setcounter{equation}{0}

	Color transparency is the predicted phenomenon of reduced
strong interactions under certain conditions.  The prediction
hinges upon what is known about hard scattering processes, the
coherent cancellation of perturbative interactions, and a body of
experience in strong interaction physics.  Color transparency
presents many new experimental opportunities as well as many
interesting theoretical paradoxes.  It  challenges our understanding
of the  quark and gluon-based picture of strong interactions as
it confronts an established picture based on nucleon and meson
degrees of freedom. While the new and old pictures are
complementary, it seems that color transparency would never have
been anticipated in conventional strong interaction physics, because
it represents an entirely new approach to how hadrons interact.

	Recall that the proton-nucleon inelastic cross section is about
$30 -35$ mb in the energy region of $2-30\; GeV$.  The mean free path
before having an inelastic collision inside nuclear matter should be
about 2 fm, so that a proton would almost never escape intact from
a nuclear target in the conventional theory. However, QCD says that
under certain circumstances we can prepare a state to make it pass
through many Fermi of nuclear matter.  For all practical purposes,
{\it sometimes the strong interactions can be turned off}. This
causes a drastic conceptual shift in the interpretation of the
physically realized strong interactions.  It is a direct consequence
of hadrons being composite objects made of internal degrees of
freedom, the quarks.

	Color transparency theory is based on the rather well
established methodology of perturbative QCD.  However, the
experimental situation is not settled; there is much discussion at
this time whether color transparency has actually been observed, or
will be observed, at laboratory energies and in realistic
experimental circumstances.  The subject is quite lively and
controversial; no matter which way it is presented it sometimes
seems to contain fantastic and unbelievable ingredients.  If research
in this area will probably test the limits of perturbative QCD, it will
also explore more about the ``real" strong interactions using QCD as
a tool.

	We hope to learn fundamentals of the strong interactions by
having a new universe, namely reactions inside nuclear targets, to
compare with the usual universe of experiments made in ``free
space".  The new methodology, of using the nucleus as a test medium
to compare and contrast with the ``control" reactions of hadrons in
free space, may eventually be one of the most significant advances
ever in learning about strong interaction physics.  At the
present time, color transparency plays a central role in the growing,
broad subject of the interface of perturbative (p)QCD with the non-perturbative
structure and interactions of hadrons. The ultimate
goal of this and other current study at the interface is:

\medskip

\centerline{\it to solve the physics of the strong interactions.}

\medskip

\subsection{The basic idea: survival of the smallest}

\medskip

	Hadrons are made of quarks and gluons.  While hadrons are
asymptotic energy eigenstates, the quarks and gluons are not.  It
follows that hadrons fluctuate between different quark and gluon
configurations when they interact; the process of
measurement involves time dependent dynamics which we wish to
understand.  To observe color transparency one needs to measure a
hadron when it has fluctuated to a configuration which will have
small interactions, by definition.

	The key to having unexpectedly small interactions in a gauge
theory is coherence, in which emissions from different parts of a
composite system can cancel with destructive interference. Because
coherence is ubiquitous in electrodynamics, there are several
instructive precedents for color transparency.   An early theoretical
prediction by King\cite{K50} was studied by Perkins\cite{P55}. He observed
tracks in emulsion  following $e^+e^-$ pair production in cosmic ray
experiments, and found reduced ionization in the spatial region close
to production of the pair, where the opposite electric charges were
close together in space.  This is a case where the continuum
spectrum of the $e^+e^-$ system is important, and the bound state
contribution irrelevant.   Anomalous transmission of electrically
neutral systems such as positronium through thin foils has also been
observed.  In this case properties of the bound state have been used
to make order of magnitude estimates.  Comparing these two
examples, it should be clear that both bound state and continuum
states can be important in coherence-the main thing is that an
important region of the wave functions produce canceling results.

	In elementary scattering theory, coherence also has a major
effect on the calculation of cross sections.  The total cross section
for scattering of an electron by a bare Coulomb field is infinite.
However, the elastic scattering of low energy electrons on a neutral Hydrogen
atom has
a finite, geometrical cross section, given approximately by
\begin{eqnarray}
{d \sigma \over d\Omega} & =  & \mid f(\vec q)\mid^2\ \ ;\ \ f(q) = {e^2\over
\vec q^2}\bigg(1-
{(2/a)^4\over[(2/a)^2  + \vec q^2 ]^2}\bigg); \nonumber \\
\sigma_{tot} & = & 4\pi a^2\; ,\nonumber\\
\nonumber
\end{eqnarray} where $a$ is the Bohr radius. This comes from the Born
approximation to the screened Coulomb potential of a central proton
surrounded by an electron cloud.\cite{PC65}  This illustrates the rule that if
a
neutral system in a gauge theory is spatially small, then its reaction
cross section will be small.  Obviously this is not the only way for
coherence to reduce interactions, so the argument cannot be
reversed.  A small cross section does not necessarily imply a
spatially small object.   We will see the consequences of this later.

	In QCD useful calculations have been done \cite{L75,N75,GS77}
for the analog of neutral ``atom-atom"
scattering. As in QED the cross section is found to be small if
associated dipole moments are small, cutting off the amount of
radiation exchanged.  The scattering cross section $\sigma(b)$ for
the region where two fermions in neutral ``atoms" are separated by a
transverse position separation $b$ scales like $b^2$.  This suggests that
reduced
interactions, and color transparency, should occur for configurations
of quarks which have a transverse separation small on the typical
hadron size scale of about 1 Fm.  As we will see, defining these
regions is somewhat subtle, and there are different renditions in the
literature of what this means.   Observing color
transparency may require tuning of the experiment, so that
particular regions of the quark wave function can make sufficiently
large contributions to an observable process.

	 It follows that the search for color transparency is a search
for short distance regions where simple color coherence will be
found. According to current theoretical work the short distance
regions might  dominate in certain hard exclusive reactions.  (``Hard" means
more than high energy;  it means that every invariant momentum
transfer $Q^2$, the Mandlestam variables $s$, $t$ and $u$ for
$2\rightarrow 2$ reactions is big compared to a few GeV$^2$.
``Exclusive" means that all the momenta of all participating hadrons
are measured.)  The relevance of short distance properties to hard
exclusive reactions is a very important question, controversial for
something like twenty years.  By now, the theoretical statement of
short distance dominance at large $Q^2$ is controversial mainly in
detail: how big does $Q^2$ have to be for an asymptotic argument to
apply?  In what experiments does it work? A second ingredient from
hard scattering pQCD is that at large $Q^2$ the most important
contributions come from the minimal Fock space number projection
of the hadrons: 3 quarks in a proton, 2 in a meson.\cite{BF75,LB80}  Again, how
large
does $Q^2$ have to be to enforce this?  How reliable are the
perturbative estimates?

	The concepts of coherence and the ``dominant region" of a wave
function are intrinsically quantum mechanical, and depend on the
measurement conditions. ``Short distance" has an intuitive meaning,
but also has a precise definition in QCD in the impulse
approximation, as we discuss below.  Because of all these
conditions, the requirements for color transparency are far from
guaranteed in general, and may or may not be obtained under
conditions such as diffractive scattering.  As for semiclassical or
classical arguments, we will find that a correct view of color
transparency cannot be obtained from their consideration.

In the classic hard scattering conditions, the typical quarks
selected are separated by a transverse distance $b$ that gets
smaller as $Q^2$ increases, with the important integration region
being $b^2 \cong 1/Q^2$.  To test whether this is the case, a nuclear
target provides the ideal detector.  Let an experiment arrange for a
hard exclusive reaction right inside a nucleus, which will be
standing there to attenuate the participating particles.   An example
experiment would be knocking a proton out of a nucleus with a fast
incoming proton, the reaction $pA\longrightarrow p' p'' (A-1)$, where in this
case $Q^2\sim\mid t\mid$.   A
signal  of color transparency is a reduced attenuation cross section
in the nuclear target as $Q^2$ increases.   Ideally, we would measure
the cross section $d\sigma/dt$  at fixed cm angle as a function of
$Q^2$, extract the nuclear attenuation rate from the data, and see
the typical ``size" of the quark separations decreasing rapidly with
$Q^2$.  This is ``part one" of the basic idea, as originally expressed
by Brodsky and Mueller\cite{B82,M82,BM88}.

	The nucleus is ``passive" if we think of it as measuring the
event;  it is ``active" if it plays an essential role in giving us the
configurations sought. ``Part two" of the color transparency story
exploits this aspect of quantum mechanics; this is the concept we have
denoted  ``nuclear filtering".\cite{BEA81,RP88,RP89}  If a spatially large
region
would
contribute much, it would also proceed to have large inelastic
interactions, and the experiment will simply reject it with a trigger
on quasi-elastic data.  This is quantum mechanics, which depends
crucially on the experimental conditions.  Nuclear filtering again is
utterly quantum mechanical and cannot be replaced by a classical
picture.

	There is a precedent to nuclear filtering in Sudakov effects,
reviewed in Section 4.   In the past, Sudakov effects have been
considered in ``free space".   But free space is
a physical medium, much like the nucleus because most of the
probability ends up in highly inelastic events.  In the free-space
Sudakov analysis, the rare and special amplitudes which remain
exclusive, (through inability to radiate away gluons), must come
from limited transverse separations between the charged particles
(or small dipole moments in QED language).  This tends to select
short distance even when kinematics might not suffice.  Similarly,
in a nuclear medium, only the amplitudes which have small
interactions will survive to be measured.  The regions involved must
strongly select short distance between quarks, because the nuclear
medium is dense (compared to free space.)   There are therefore
strong parallels between traditional Sudakov effects and the pQCD
treatment of color transparency which we will explore.  An extra
parameter, and an extra experimental handle,  is the nuclear target
length $R_A\sim 1.2 A^{1/3} Fm$, which regulates the transverse
smallness for partons not to have an interaction.

	With large $Q^2$ and large $A$, we have two excellent tools
that tend to select the short distance regions of quark wave
functions, and study them.

\subsection{Back Up!}

In presenting the basic idea, we have skipped over many essential
details, which will be clarified below. We have also presented the
idea from the picture in the quantum mechanical basis of quarks, as
originally conceived and formulated.  Needless to say, theory can
employ different bases, leading to seeming disagreements in the
physical picture.  The basis of hadrons as a complete set has been
extensively used, and deserves a discussion on an equal footing.

        In perturbing the system to measure color transparency, it is
quite sensible to consider the resulting system as a superposition
over asymptotic hadronic states rather than the fundamental Fock
space.  In this basis, pQCD has to be reinterpreted.  The most
common interpretation is that traditional strong interaction physics
applies.  This is a dynamical assumption, implying that under
conditions of color transparency the decreased strong interactions
will be due to destructive interference among the different hadronic
modes in a wave packet.   This is far from trivial: there is no
systematic way to go from predictions in the quark basis to those
made in a basis of asymptotic states.  Thus, models in the hadronic
basis represent new hypotheses to be tested separately from the quark
picture.  Underlying this is a broader conceptual question: to what
extent can asymptotic hadrons actually be
used to imitate the predictions of QCD?  The answer is unknown.

        Current calculations in the hadronic basis are intrinsically
neither more nor less rigorous than those using the quark and gluon
Fock space.  In the hadronic basis one has to rely on a postulate of
completeness, and a host of unknown, uncalculable matrix elements.
Completeness is a pre--QCD postulate from the days of the S-matrix.
However, it has never been quite clear how the collective
coordinates, the hadrons, are to reproduce the internal coordinates,
the quarks, in the system with confinement.  Setting aside this
question for now, the hadronic viewpoint's advantage is that one can
use some measured information about the interactions of hadrons to
model the system.  This is powerful, and many theorists have
exploited it.  A corresponding disadvantage is that any quantum
mechanical superposition being proposed has to be ``tuned" to
reproduce the perturbative results, where they apply, and as much as
they are understood.    It has often been assumed that a spatially
small wave packet made by superposing hadronic states must
reproduce the quark picture.   But as we shall see in Section (2.2) below,
tuning the hadronic wave packet is not straightforward: contrary to
some expectations, the pQCD dominant region of $b^2\leq 1/Q^2$ is
not the same as a wave packet with $<b^2>\leq 1/Q^2$.

\subsection{Common Questions}

{\it Why are exclusive processes emphasized?  Couldn't one observe
color transparency in all kinds of reactions?}

	The answer to this is maybe so, but probably not.  Inclusive
reactions sum over channels and disparate kinds of configurations of
the quarks inside the hadrons (or hadronic final states). In the sums
there are many color flows, and different degrees of inelasticity.
Perhaps some of the configurations are ``transparent", but it is
reasonable that these should be a small part of most inclusive cross
sections.   An active theoretical occupation is to seek ``solid gold"
reactions which are color transparent while relaxing the difficult
experimental conditions of exclusivity.

        On the positive side, exclusive processes are a wonderful
frontier.  Their study is well justified as a field in itself.
Exclusive processes are difficult to measure, but this is because
there are many, many channels, and much work to be done.   At large
momentum transfer, they involve projections of hadrons onto just a
few quarks. These are exactly the projections we are interested in
learning about on the interface of non-perturbative QCD physics.

{\it Even if the perturbative treatment becomes self consistent,
isn't there still room for doubt in its predictions? }

Absolutely yes.  Many interesting questions - such as what happens
to a hadron's ``cloud" of soft pions and/or soft gluons, the way it
sits in the chirally broken vacuum, and the way a hadron reforms
after being struck hard - are all questions beyond perturbation
theory.  It is conceivable that perturbation theory will fail.  But if
color transparency does not work, then we will have observed where
perturbation theory fails in a region where it might have worked.
This is a way to learn the fundamentals of the strong interactions
better.

{\it Why are small transverse separations between quarks involved
in hard scattering reactions?}

This is a straightforward but detailed technical question.  The true
answer is: the association of large momentum transfer with short
distance is not always true. Understanding when short distance
between quarks is involved or not is a current research problem.  One
must check for it on a reaction-by reaction basis.  By studying it, we
hope to learn more about the way the quarks themselves are
arranged.   It is important to note, however, that we should not, and
do not, always insist on having short distance.  If different
distances are involved in different reactions,  we can take a
snapshot of the quark wave functions to see how the hadrons are
made.

\subsection{Exclusive Reactions: a Program for Solving Strong Interaction
Physics}

QCD is now more than 20 years old, and maturely established as the fundamental
origin of the strong interactions. The time is ripe to begin using that theory,
and understanding it fully, to solve the many outstanding puzzles of hadron
structure and nuclear interactions.  Central to these puzzles is confinement,
not understood theoretically, a topic in which experiment must play a central
role. Physics needs a bridge between the non-relativistic picture of nucleons
making up a nucleus on the one hand, and the ultra-relativistic perturbative
approach used to establish QCD on the other.  Color transparency provides a
means to make this transition.  The subject interpolates between quarks
interacting weakly at short distances to the ``real" strong interactions of
hadrons at long distances.

To study the interface of perturbative and non-perturbative QCD, and move
toward
the goal of solving the physics of the strong interactions, a re-focusing of
experimental emphasis will be needed.  The key element in understanding hadron
structure will be exclusive reactions.  By systematically studying exclusives,
important questions of how quarks are confined in hadrons will become
answerable.  Many questions are extraordinarily interesting: how ``big" in
space
are quark and gluon wave functions spread out, hadron by hadron, Fock-space
channel by channel?  How good is a few-quark approximation, and at what value
of
$Q^2$?  How do gluons dress quarks in detail, and how does their color flow as
mesons are exchanged?  What is the role of quark orbital angular momentum, in
making up the spin of hadrons and in exchanges between them?  A singular thing
about these questions is that they cannot possibly be answered in the bulk,
inclusive-averaged experimental framework.  Instead, theory and experiment must
study channel by channel, spin by spin, in as much quantum mechanical detail as
possible.

	 The framework of hard exclusive reactions, along with nuclear filtering,
promises to make such a study systematic and theoretically well defined
(Sections 3 and 4). A bonus from the ``parton revolution" of the past 20 years
is a strong theoretical structure in defining what matrix elements are
measured,
how, and when. Entirely new matrix elements are measured with exclusive
interactions, and can be used constructively to compare and make cross-links
between diverse reactions.  This is future science we hope to see.   But at the
moment color transparency is still an emerging field, with active discussions
and widely different interpretations.  A calculation now using meson exchange
might be better than perturbative QCD for some reactions, or maybe worse.
Diversity of ideas and interpretations is part of the fun: the subject is open
and interdisciplinary, without any picture yet commanding the field.  In this
report we have done our best to unite the valid quantum mechanics used in
nuclear physics approaches to the much different Green function methods of
perturbative QCD.  From these interdisciplinary roots a new discipline of
strong
interaction physics is ready to spring.

	Investigating the strong interactions in these new ways will require several
experimental technologies.  What is needed now for a program in quark-hadron
structure is development of experimental facilities.  Every different beam and
target has its advantages.  Hadron-initiated reactions are of extraordinary
interest in identifying the color, flavor and angular momentum flow between
particles.  For fixed-angle reactions, a beam energy in the 5-30 GeV region is
ideal, with high flux, high acceptance, and state-of-the-art polarization and
spectrometers the most important requirements.
As will become clear (Section 3), the
comparatively large data taking rates of hadron initited reactions, and the
fact
that several particles interact with the nuclear medium, also gives hadron
initiated reactions a special attractiveness.
The Indiana LISS concept
\cite{LISS95} includes many of these features and seems ideal.  We refer the
reader to a very thorough study preliminary to a formal proposal\cite{LISS95}.

At the same time, facilities at BNL are expected to continue providing
pioneering measurements, which so far are the foundation of our belief that
color transparency has been observed \cite {CEA88}.
A second round of hadron beam experiments has been approved and
a new detector, named EVA, with much higher acceptance has been taking
data with beam energies in the 6-20 GeV range for about one year\cite{BNL850}.
Due to the revolutionary detector design and increased beamtime, this is hoped
to expand the amount of data taken by a factor of perhaps 200, allowing a
larger
energy range and an analysis at different scattering angles. It is also planned
to study meson-nucleus scattering. In particular, the reaction
$\pi A\rightarrow\rho (A-1) p$ at 90$^\circ$ is very interesting as a test of
short
distance physics expected from nuclear filtering (see Section 3.5).

Many interesting results can come from the diffractive regime.  Already there
is
valuable  data \cite{AEA95,F94} from Fermilab experiment E665. This also
provides evidence for observation of color transparency, and will be reviewed
in
Section 3.3.  The Hermes detector \cite{HERMES} at HERA is also beginning
operation. It is a nuclear gas target installed on the $25 - 30 $ GeV electron
(or positron) beam.  The use of light, possibly polarized nuclei is foreseen in
the next few years. The experiment may enable extension of the E665 data (see
Section 3.3) on $\rho$ meson diffractive production at moderate $Q^2$ values to
smaller values of energies $10 \leq \nu \leq 22$ Gev. Hermes' small luminosity
will only allow integrated measurements. Due to limitations of energy
resolution, the experiment  cannot assure that
diffractive events are not mixed with inelastic events, much like FNAL. In the
near future one does not expect detection of the recoiling proton, but
nevertheless there will be much to learn from the results.

Electron-initiated reactions also have extraordinary
appeal: the electromagnetic probe is the cleanest tool to
investigate the structure of hadrons at short distance. Accelerator
technology has progressed dramatically.  Despite the relatively small signal -
due to the size of $\alpha_{em}$ - it is now practical to conduct
precise experimental measurements of exclusive reactions with electron beams.
The breakthrough is the availability of high duty factor and high luminosity
electron machines at Mainz, Bonn and now CEBAF.
CEBAF has accepted several proposals to study color transparency and hopefully
will continue to add more in the future.  There have been many ingenious ideas;
elsewhere Section 3) we will point out that the high precision facilities of
CEBAF have special advantages which have not been exploited.  It is simply not
established whether color transparency will be observed or not at CEBAF
energies.  For this reason, and also due to the intrinsic interest of the
subject, it would seem logical to answer experimental questions answerable at
CEBAF by experiments.

	The ideal choice of the energy range for an electron accelerator suitable for
color transparency studies is fixed by three constraints. (1) One must have
sufficiently high energy and momentum transfer to describe the reaction in
terms
of electron-quark scattering.  High energy creates a very fast process where
the
struck quark can propagate quasi-free.  High momentum transfers are necessary
to
probe short distances. (2) The energy of the incident electron beam should be
tunable to match a characteristic interaction time $\tau$ to the diameter of
the
nucleus.  Starting from the rest frame time $\tau_o\sim 1$ Fm/$c$ and taking
into account a  Lorentz dilation factor $\gamma = E/M$ this means a time $\tau$
of several fm/$c$'s in the laboratory.  If the energy transfer is too large,
the
building-up of hadrons may occur outside the nucleus which can then no longer
be
used as a microscopic detector.  Theory cannot reliably determine the
interaction time yet, which remains an experimental question to be
investigated.
(3) Charm production, if desired, requires a minimum electron beam energy of 15
GeV to have reasonable counting rates.

To get a large enough rate and avoid a prohibitively large number of accidental
coincident events a high duty cycle is imperative.  Finally, good energy
resolution is necessary to identify specific reaction channels.  A typical
experiment at 15 GeV (quasielastic scattering for instance) needs a beam energy
resolution of about 5 MeV.  At 30 GeV the proposed experiments only require one
to be able to resolve pion emission.  These requirements are satisfied by the
ELFE proposal, an exciting and ambitious prospect to which we refer the reader
for more detail. \cite {AP95,AS93}

There have been a large number of studies of accelerator and detector
requirements for a future program in hadron physics. In general, these studies
point at measurement capabilities in quality of tracking, particle
identification, and high acceptance which have previously been used in high
energy physics experiments, and are comparatively new to the nuclear/particle
physics interface.  We refer the reader to several recent specialized meetings.
\cite {MEET}  Sufficiently high momentum  resolution
($5  \times  10^{-4}$)  and high luminosity ($10^{38}$~nucleons/cm$^2$/s) can
be
achieved by magnetic focusing spectrometers.  For semi-exclusive or exclusive
experiments with more than two particles in the final state, the largest
possible angular acceptance ($\sim 4\pi$) is highly desirable.  Quality and
reliability of large acceptance detectors have improved substantially in the
last two decades.  Research and development studies, however, are still needed,
as the literature will show.\cite{MEET}.

We now turn to our main topics, beginning with a deeper look at the basic
quantum mechanics in Section 2.  Section 3 is concerned with defining the
process experimentally, as well as summarizing existing experimental results,
and different ways to extract a signal.  Many suggestions for future
experiments
also appear here.  Section 4 is a small ``handbook" for the beginner to
understand better the methods of pertubative QCD.  There are also some new
results that may interest the experts.  Hopefully the reader will see the link
between this physics and the (mostly non-relativistic) quantum mechanics of
Section 2.  Section 5 is concerned with reviewing diverse models of
propagation.

\newpage

\section{Quantum Mechanical Basics}
\setcounter{equation}{0}
\medskip
Color transparency is an intrinsically quantum mechanical phenomenon with some
semi--classical interpretations. The correct picture has been
controversial.\cite{R94} In
this section we will examine these issues, mostly using elementary quantum
mechanical arguments. Important points include the motivation for short
distance in hard exclusive processes, the different roles of elastic and
inelastic channels, the expansion time scale problem, and use of the hadronic
basis as a substitute for the quark basis. To make this section more
accessible, we have separated detailed discussion of QCD into Section 4, and
used non--relativistic concepts when practical. As we
will see, some basic discrepancies between different theoretical approaches can
be traced to different underlying quantum mechanical assumptions. The way in
which these assumptions are implemented in specific calculations is also a
separate topic, to appear in Section 5 on Models of Propagation.

It turns out that there are plenty of basic quantum mechanical issues to
resolve before turning to detailed calculations. We hope that this review of
elementary ideas can be understood by the beginner, while refreshing the
understanding of the experts.
\medskip

\subsection{What is the right picture?}

\medskip
        It is often asserted that large momentum transfer in a reaction
will select short distance by the uncertainty principle, $\Delta Q
\Delta X \geq 1$.   A widespread notion also exists that simply
absorbing a virtual photon with large $Q$ guarantees some kind of
short distance.  But certainly one must specify what is going to be
measured!  There is no dispute that a {\it hard inclusive process} in
QCD such as deeply inelastic scattering probes short distance
aspects of hadron physics. However,  it worth reviewing the role of
short distance in {\it exclusive}  reactions, and in some cases of
current interest in QCD, the bland assertion of generic short
distance can be utterly {\it wrong}.

\medskip

\noindent {\bf Exclusive versus inclusive processes}

\medskip

        Consider a typical exclusive process such as $\gamma^*+ p
\longrightarrow p'$.  The matrix element involved can be represented
by two form factors:
\beq
\langle p+Q\mid  J^\mu_{em}  \mid p\rangle  =  \bar
u(p+Q,s')[F_1(Q^2)\gamma^\mu+{i\over
2m}F_2(Q^2)\sigma^{\mu\nu}Q_\nu ]u(p,s)
\eeq
The process is illustrated by Fig. (2.1).

Suppose we do not interpret this process yet in terms of ``quarks" or
QCD.  On its own, objectively, there is no kinematic information in
the matrix element to indicate short distance.  Without a model for
the substructure, the matrix element simply stands alone (as it did
in the 1960's). The standard assertions of short distance are
motivated by fairly detailed properties of the proton as a composite
object.  These are specific to QCD and the proton as made of {\it
quarks}.

	Depending on the kinematics, the proton can also absorb a
virtual photon and ``explode" into a multitude of final states in a
hard inclusive process, namely deeply inelastic scattering.  The
matrix element measured in that case does not consist of a single
amplitude, but rather a sum over many amplitudes to produce many
final states:
\beqn
W^{\mu \nu} & = & \sum\limits_N \int d^4 ye^{iQ\cdot y}
< p,s \mid J^\mu(y) \mid N><N\mid J^\nu(0) \mid p,s  > \nonumber \\
&  = & \int d^4 ye^{iQ\cdot y}<p,s\mid
J^\mu(y)J^\nu(0)\mid p,s> \\
\nonumber
\eeqn

The appearance of a Fourier transform with respect to the large
momentum transfer $Q$ is enough to begin motivating a small
spatial separation as $Q \longrightarrow \infty$.  However, care is
still required;  in the parton model, the matrix element above is
reduced to a correlation function of quark fields. This occurs through
the use of the impulse approximation and the ``hand-bag" diagrams
(Fig. 2.2)

The expression then for  $W^{\mu \nu}$ is
$$
W^{\mu \nu}  = \int d^4k {\rm Tr}[\gamma^\mu(\slash k+\slash
Q)\gamma^\nu\Phi(k,Q)]\delta (k^+-x p^+)\; \;
$$
where $\Phi$ is a correlation function to measure quark fields $\psi(y)$, given
by\cite{RS79}
\beq
\Phi_{\alpha\beta}(k,Q)= \int {d^4y\over
(2\pi)^4}<p,s\mid\psi_\alpha(0)\bar\psi_\beta
(y)\mid p,s>e^{ik\cdot y}\; \; ;
\eeq

$$\int d^4 k\delta (k^+-xp^+)\Phi_{\alpha\beta} (k,p)=  \int {dy^-\over 2\pi}
<p,s\mid\psi_\alpha(0)\bar\psi_\beta (y_\bot =y^+=0,y^-) e^{iy^-
(xp^+)}\mid p,s>
$$
``Short distance" means that the quark corellations are measured
practically at zero transverse and null plane time separation, up to
scaling violations.  Nevertheless, not all the distances involved
decrease as $Q\longrightarrow \infty$: the remaining $y^-$
coordinate is finite even at large $Q$. This shows
that ``short distance" is never automatic, and can be misunderstood. Even for
$Q^2>>\; GeV^2$, short distance  can
fail in some kinematic regions, such as Feynman momentum fraction $x\rightarrow
0$.

\medskip

\noindent {\bf Color Transparency Involves Both Exclusives and
Inclusives}

\medskip

        In color transparency we will be concerned with both
microscopically exclusive and inclusive processes.  Consider quasi-
exclusive scattering in a nuclear target.  The process will consist
of a ``hard" collision, and subsequent propagation through the nuclear
medium. There are two distinct channels:

1) the purely exclusive process of knocking a proton into a proton
which exits the nucleus with only elastic interactions.

2) the contribution of microscopic inelastic scatterings, which by
further interaction, can feed back into the exclusive process.

 These two are separate theoretical subprocesses.  While relating
them separately to observables is very intricate, a proton
propagating through the medium and a proton making two transitions
are distinctly different.  Many of the disputes in the interpretation
of color transparency originate because of basic disagreements on
the importance of these different channels. Needless to say, we will
not settle those disputes here, but instead we hope to illustrate the
merits of different viewpoints in a growing field.

	Let us discuss the separate channels separately, first
illustrating the physics of elastic collisions in free space, and then
bring in the complexity added by the attenuating
nuclear medium.

\subsection{Scattering in Free Space}

        We again want to see step by step which assumptions are
behind the assertions of short distance.  We will present a fresh
analogy, temporarily removing the ``virtual photon" bias entirely
from the discussion.  Consider instead a pedagogical example of a
non-relativistic atom scattering off a plane surface we can call a
``partially silvered mirror".   The setup is shown in Fig. (2.3):  an
atom enters carrying momentum $P$, and leaves carrying momentum
$P' = P+Q$.

The reason for introducing the mirror is that it is precisely {\it not}
a short-distance, microscopic probe, but instead a crude
macroscopic object. When the atom reflects from the mirror its
momentum component parallel to the mirror is unchanged.  The
atom's perpendicular component of momentum $P_T$ is reversed, so
we have $Q = 2P_T$.  Let the mirror be arbitrarily large, in fact
translationally invariant in the direction along its surface.  Let the
mirror's reflectance be very low, so that at most one of the
constituents is reflected: the probability of more than one
scattering from the mirror as an ``external field" will be considered
negligible.  The mirror interacts so weakly with the
constituents that it can be treated perturbatively, and its distortion
of bound state wave functions (except for the scattering itself) is
also totally negligible.
Then the picture of scattering for the atom's constituents
looks like Fig. (2.4).

Observing the initial and final momentum of the system, one
can not tell where the collision with the mirror occurred, which
could have been anywhere on its surface...an obvious point due to
translational invariance.  Since we do not know the location of
scattering, we cannot sensibly compare the relative position of the
reflected constituent with the unscattered one.  The mirror
therefore shows us that we could have an arbitrarily  large
momentum transfer and not necessarily probe the short distance
separation of anything inside the atom.

        In fact, the set-up simply gives us a concrete physical picture
of the interaction of a virtual photon carrying space-like momentum
in a $Q^0 =0$ ``Breit Frame" interacting with one of the atom's
internal constituents or ``quarks".  Although the photon interacts
only at a single point, the photon itself is not well localized; like
the mirror, it has translational symmetry properties and serves only
to deliver a definite momentum $Q$.  The photon is just like a mirror
for this reason.  (One proviso should be made: in an ultrarelativistic
system, only two transverse ``invariant" coordinates behave non-
relativistically, as in this treatment.  The third, and the time-like
``longitudinal" coordinates have a relativistic dependence that
scales with the overall momentum of the particle and which should
be treated separately.)

         Let us analyze the mirror scattering in more detail. The
momentum flow of the internal atomic constituents is shown; a
constituent's momentum is half the center of mass
momentum $(P/2)$ plus or minus  a relative momentum (called
$k/2$). With at most one constituent striking the mirror, the other
constituent must fit itself into the outgoing atom through the
fluctuations in relative momentum in the wave function.  Let the
incoming atomic wave function be $\phi_P( P/2+k/2,\ P/2-k/2)$, as
shown. Before the atom strikes the mirror, this is just the ground
state wave function.  Because the collision occurs very fast, the
wave function has no time to evolve to a different state.  This is the
impulse approximation, which is a very good approximation in QCD.
(It is so standard in QCD that it is often formalized into ``hard
scattering kernels" and ``distribution amplitudes" , a notation
carrying concepts equivalent to the ones being used here, and which
will be introduced in Section 4.

        After striking the mirror, the scattered outgoing constituent
has  momentum $P/2+k/2+ Q$.  The full amplitude $M(P, Q)$ for the
two particles to look like an outgoing atom is given by the overlap
back onto the ground state wave function, given by the Feynman
rules of perturbation theory as a loop integral:
\beqn
M(P, Q) & = &  (2\pi)^3\delta( P'-P-Q) \int d^3k
\phi^*_{P'}(P/2+k/2+Q,P/2-k/2)\nonumber \\
& \times & \phi_P( P/2+k/2, P/2-k/2)  \\
& = & \int d^3k
d^3k'd^3k^{\prime\prime}\delta^3(P'-k-k^{\prime\prime})\phi^*_{P'}
(k,k^{\prime\prime})\delta^3(P-k-k')\nonumber \\
& \times & \phi_P(k,k')\delta^3(k^{\prime\prime}-k'-Q)\nonumber \\
\nonumber
\eeqn
In the second line we have simply inserted some delta functions for
later convenience. We have included the overall momentum
conserving delta function, often suppressed in the set-up of
amplitudes, in order to see everything in the calculation.

	The ``wave functions" are actually certain Green functions.  By
translational invariance, for the atom carrying total momentum $P$,
then the wave function of the quarks at positions $x$ and $x'$ should
be given by $\exp(i P\cdot R) \tilde\phi(r)$ where $r = x-x', R= (x+ x')/2$ are
relative and center of mass coordinates.   This wave function is
related to the momentum space wave functions above by:
$$\delta^3(P-k-k')\phi_P(k,k')=\int {d^3xd^3x'\over (2\pi)^6}e^{-ik\cdot x-
ik'\cdot x'} e^{iP\cdot (x+x')/2}\tilde\varphi({x-x'\over 2})$$ The
relativistic case is exactly analogous, replacing the 3--dimensional variables
by 4--dimensional ones. To see the spatial regions of the wave function that
are
involved we
insert these coordinate space representations into Eq. (2.4):
\beqn
M(P,Q) & = &\int {d^3xd^3x'\over (2\pi)^6} {d^3yd^3y'\over (2\pi)^6} \int
d^3kd^3k'd^3k^{''}e^{-ik\cdot x-ik'\cdot
x'+ik\cdot y+ik^{''}\cdot y'} e^{iP\cdot(x+x')/2-iP'\cdot (y+y')/2}\nonumber \\
& \times & \tilde\phi(x-x')\tilde\phi(y-y')\delta^3(k^{''}-k'-Q)\nonumber \\
\nonumber
\eeqn
Doing the momentum integrals we have
$$k^{''}=k'+Q;\hskip .35in \int d^3ke^{ik\cdot(x-y)}=(2\pi)^3\delta^3(x-y)\;
;\hskip .35in \int d^3k'e^{ik'\cdot(x'-y')}=(2\pi)^3\delta^3(x'-y')$$
This leaves some coordinate space integrals to do, which can be
written
$$\int {d^3xd^3x'\over (2\pi)^3} \tilde\phi^*(x-x')\tilde\phi(x-x')e^{iQ\cdot
x'}e^{iP\cdot
R-iP'\cdot R{''}}=\int d^3r{d^3R\over (2\pi)^3}\; e^{i(P'-P-Q)\cdot
R}\tilde\phi^*({r\over 2})\tilde\phi({r\over 2})e^{iQ\cdot r/2}\; .
$$
Finally we obtain an expression we can call a ``mirror form
factor" $F(Q)$:
\beqn
M(P,Q) & = & 8\; \delta^3(P'-P-Q)F(Q)\; ;\nonumber \\
F(Q) & = & \int d^3r \ \ e^{iQ\cdot r} \tilde\phi^*(r)\tilde\phi(r)\\
\nonumber
\eeqn

The form factor which tells us how the external, observable process
of rebounding from the mirror depends on the internal wave function
of the atom.  This quantity would be fairly informative if the atom's
constituents were hidden from us, as is the case with quarks in
hadrons.   The reader can recognize the mathematical expression for
the mirror form factor as a familiar result, the same as the
electromagnetic form factor for absorbing a virtual photon, because
the fundamental kinematic ingredients are the same.

\medskip

\noindent {\it Short Distance}

\medskip

        Let us can examine the expression of Eq. (2.5) for the dominance of
``short
distance".  Take the momentum transfer $Q$ to be very large.  From
what integration region in $r$ does the form factor get its main
contribution?  It is not quite safe to say, ``$Q\rightarrow
\infty$, then $r\rightarrow 0$, by the uncertainty principle."   This
is incorrect because there are many mathematical functions whose
most rapidly varying sections are not at the origin!  Consider, for
example, the Fourier transform of a Gaussian with a node, as one
finds for the first excited state of a harmonic oscillator  (See
Figure 2.5). In a saddle point approximation, the dominant contribution as
$Q\rightarrow \infty$ to the Fourier transform for the form factor
can be shown to come from a region near the ``bumps".
\beq
\int\limits^\infty_{-\infty}dx\; x^2e^{-x^2/2a^2+iQx}\sim
\sqrt{2\pi}Q^2a^5e^{-Q^2a^2/2} \\
\eeq
\noindent In this case the dominant integration region for the large $Q$
behavior is  $r\cong a$ and not the $r\longrightarrow 0$ region.
This is an elementary result, but good to keep in mind. Note that the large $Q$
dependence is exponentially damped because the wave function is ``smooth" in
this
region.

        To make a serious argument about short distance in QCD, one
evidently must supply information about the wave function to find
the quarks.  We can look for inspiration in a more familiar gauge
theory, ordinary electrodynamics.   In atoms the wave function  has a
``kink" near the origin which dominates the large $Q$ Fourier
transform.  The hydrogen atom, for example has a ground state wave
function going like $\exp(-r/a)$. The corresponding form factor using such a
wave function is given by
\beqn
\int d^3r \exp(-i Q\cdot r)  exp(-r/a)  & = &  8\pi a/(Q^2 + 1/a^2)\nonumber \\
& \sim&  8\pi a/Q^2\; \; ,\; \;  Q^2>>1/a^2\; .\\
\nonumber
\eeqn

It is an elementary exercise to verify that here the dominant region
actually is $r\longrightarrow 0$: the power-law decrease with
$Q^2$ can be traced directly to the ``kink" at the origin.
The same occurs in QCD but in that case only a perturbative model of
the wave function (valid at short distance) is known.  The result in transverse
momentum space $k_T$ is
\beq
\phi_N(k_T)\sim {1\over k_T^2}log^{\gamma_N}(k_T^2/\Lambda^2_{QCD})\; ,
\eeq
\noindent where $N$ is an $x$--space projection and $\gamma_N$ is an anomalous
dimension.
(See Section 4). It is no coincidence that Eq. (2.7) and (2.8) show the same
power--law
decrease, because both QED and QCD have dimensionless couplings and vector
exchange; perturbation theory reconstructs the short distance part of the bound
state wave function in both cases.

So here is the ``short
distance"; bin by bin in integrating over the relative coordinates $r$,
the important  region is $r\longrightarrow 0$, due to both large $Q$
and the special short distance properties of the wave functions in
the gauge theory.

\medskip
\noindent
{\it Remarks}
\medskip

The mirror-photon example illustrates several points:

*       In free space it is not necessary for the system to ``compress to
a small size" to undergo a hard scattering.  While self-compression
is an attractive and often quoted semi-classical picture, to hold to
such an interpretation would be a basic misunderstanding of the
quantum mechanical impulse approximation. \footnote{The impulse
approximation is sometimes mistakenly called the ``frozen
approximation". As we have seen, the impulse approximation  applies
when the time scale of the scattering event is much faster than the
time scale of the internal constituents.  The ``frozen approximation"
is a separate concept applied to the relativistic diffractive limit --
an issue to be discussed below.}

        In the example, the atomic wave functions are full sized, and
ignorant of the fact that they are going to smack into the mirror.
They do not time evolve significantly during the short time of the
collision.  Nor is it necessary to model the interaction of the system
with the ``mirror" in terms of complicated initial state
superpositions.  One state, the ground state, is all that is relevant in
the scattering of a ground state, in the free space case.  The same
conclusion follows from inserting a complete set in matrix
elements:

$$<P+Q| J^\mu | P > = \sum\limits_N  <P+Q| N><N| J^\mu| P >= <P+Q| J^\mu|P>$$

This may seem to be a trivial point, but it is worth mentioning
before considering the interactions with the medium.

*       The uncertainty principle is inadequate to predict much here,
because it does not have enough detail.  If we used the uncertainty
principle to estimate a dominant region, it would generally fail. The
precise statement is that{\it an integration region as small} as the
one indicated by the uncertainty principle can be a dominant region -
in the Hydrogen atom case, the region $\Delta r \approx 1/Q$ turns
out to dominate, as much due to dimensional analysis as anything
else.

*       The observations made so far apply to more processes than
just the scattering of a single photon.  The scattering of two atom
on each other can go through ``two mirrors". In the center of mass
frame, the picture looks like Fig. 2.6. Needless to say, the naive
inference of ``short distance" in this case is utterly wrong, because
it ignores the need for information on the relative real space
positions of the ``mirrors" or scattering centers.  It is obvious that
one can have large momentum transfer and not have short distance in
this case.  We will discuss this interesting process of ``independent
scattering" in hadronic reactions in Section 4.

*       Finally, the form factor discussion is simply an example of
``classic" color transparency, in which (as we will show) all of the
ingredients of conventional QCD come together.  This does not
exclude other possibilities, and other motivations for small color
singlet systems even if the short distance motivations from the
impulse approximation are not evident.  Certainly the validity of any
proposed ``small" hadronic system, a point like configuration, and so
on, depends on the model and the context.  It may be eminently
reasonable and give correct predictions under the proper
circumstances.

\medskip

\noindent {\it Inelastics, Again}

\medskip

        Now we return to discuss the inelastic processes, which form
the bulk of all amplitudes possible.  What usually happens when an
atom moving fast strikes the mirror?  The atom explodes (Fig. 2.7).

We seek the amplitudes to form various final states $\mid N>$ , $N \neq$
the proton.  Suppose we propagate into the future the set of
amplitudes $\mid s(0) >$ that are defined  to strike the mirror at time
$t=0$.  The future amplitudes are given by:

$$\mid s(t)>=e^{-i\hat H t} \mid N><N(0)  \mid s(0) >$$ where $\hat H$ is the
Hamiltonian operator.   The time evolution of each energy eigenstate is given
by   $\mid N(t)> = \mid N(0)>exp (-iEt)$.   The outgoing wave packet has a
complicated
time
evolution.  It expands as time progresses.  We can calculate the
``size" of the outgoing wave packet in the following way:
\begin{eqnarray}
<r^2(t)> & = & <s(t)\mid r^2\mid s(t)>=\sum\limits_{N,N'}<N(0)\mid r^2\mid
N'(0)>e^{-i(E_{N'}-E_N) t}\nonumber \\
&\times &<N'(0)\mid s(0)> <s(0)\mid N(0)>\nonumber \\
\nonumber
\end{eqnarray}
If, for example, we truncate the sum over states to two states for
simplicity, then we find a time dependence given by
\beq
\langle r^2(t)\rangle  = A  \cos (E'_N-E_N)t+B ,
\eeq
\noindent where $A$ and $B$ are certain matrix elements. This indicates a
characteristic
time for the packet's expansion is set
by energy differences.  The uncertainty principle would give an acceptable
estimate of this ``coarse grained" flow of probability.

        However interesting, this line of reasoning is unfortunately not
relevant to the elastic scattering already calculated.  The size of
the inelastically scattered wave packet has almost nothing to do
with the elastic process.  This can be illustrated in Figs (2.8, 2.9),
showing the wave packet (in transverse separation coordinates)
``before" and ``after" the scattering from the mirror (or virtual
photon).  The figure shows that the final state wave packet is quite
broad in transverse size: it is not particularly small!  What has been
changed is relative phases among the real space components, so that
the wave packet is broad but has become wildly oscillating.
Because of these oscillations, what matters is the {\it value
of
the overlap from the region near the origin}. This region
is an entirely different thing from the size of the full wave packet
or its configuration, which is destined for inelasticity.   If, for
example, we mixed a symmetric ground state with a typical excited
state, the excited state would be quite important for the bulk
expansion of the inelastic wave packet. Yet, the node at the origin of
the excited states would contribute almost no overlap to the dominant
region at the origin for the elastic process. Finally, neither of these
is the same concept as a point-like configuration in which the
system is quantum mechanically prepared to exist entirely within a
small region!

        We have reviewed an elementary yet significant point: the value
of the ground state wave function at the origin is not at all the same
thing as the average size of the inelastically scattered wave packet
size.  A time scale for the expansion of the inelastically scattered
wave packet simply has little to do with the elastic scattering in
free space.

\medskip
\noindent
{\it More Remarks}
\medskip

	There are a few additional remarks:

*  In many experiments the direction of the momentum transfer ${\bf Q}$ is well
measured.  Suppose this is the $\hat x$ direction, so ${\bf Q}=\hat x Q_x$. The
``x" coordinate
space variable conjugate to $Q_x$ is the naive region where ``small"
separations
are
measured in $\exp(-i Q_x x)$.  Nevertheless, all transverse regions that
generically
contribute have ``small" dimensions.  To see this, project  $\exp(-i Q_x
x)$
into
SO(2) spherical harmonics and find the coefficients in terms of $b_T$,  the
magnitude of the transverse space radial coordinate:
$$\int d \varphi e^{-im\varphi}e^{-b_T\mid Q_T\mid \cos\varphi}=2\pi
i^{-n}J_n(b_T\mid Q_T\mid)$$
Oscillations in the Bessel functions in $\mid Q_T\mid b_T$  generally wipe out
long distance contributions in the magnitude $b_T$.
As we will see in Section 4, the same conclusion is reached in
the pQCD hard scattering formalism, in that case because the gluon propagator
is
isotropic, and selects participants from regions nearby in space regardless of
the direction of ${\bf Q}$.

*  Relativity allows one to repeat this type of analysis in a large set of
reference frames.  Each frame has its own interpretation. For example, for an
electromagnetic form factor it is convenient to use null plane coordinates
(reviewed in Section 4) for the fast proton with momentum
$P^\mu= ( P^+,  P_T, P^- ) =(P^+, 0, m^2/ 2 P^+)$. In this frame, one can set
$Q^\mu= ( 0,  Q_T =\sqrt{(-Q2)} , Q^- = Q^2/2 P^+)$.  The point of this
reference frame is that  $Q^\mu$ is practically ``all
transverse" in the limit of $P^+\rightarrow$ infinity. This focuses attention
on the
transverse space coordinates, $b_T$. The transverse coordinates
also have the advantage of being invariant with respect to boosts along the
``z"
axis.  The kinematic analysis of the previous results then applies.  As a good
estimate, the region of transverse separation $b_T< 1/Q$ is the important one
to
estimate overlaps of wave functions in hard scattering calculations.

	Because of the scaling of the ``+" coordinates under a boost, it is
also useful
to express momentum fraction of quarks using the Feynman $x_F$ variables, which
are
also invariant. In Fig (2.9) we have plotted the wave function of a typical
``pion"
in a mixed (transverse space, $x_F$) representation.  This coordinate system
turns
out to be very useful and will be discussed further in Section 4.  As one sees,
the result of absorbing a virtual photon is creation of oscillations in the
transverse direction without any decrease in the average size of a
state. Compressed states postulated in semi-classical argument are not the
result of kinematics nor calculations in QCD, but instead are dynamical
proposals made on a case-by case basis.

          Next, one might ask, what happens in an attenuating medium?

\subsection{Attenuating Medium: Diffractive Mixing of
Inelastic Production}

        The above discussion might give the impression that inelastic
processes are irrelevant.  This is far from obvious;  in fact, many
current models of color transparency quite reasonably emphasize the
possible importance of inelastic contributions.  The ``probe" of color
transparency, which is the nucleus, also interacts significantly with
the wave packet in the process of making a measurement.  In so
doing, the intermediate channels which are irrelevant in a free space
exclusive experiment may feed into the
elastic process finally measured. A further complication is that at
high energies many states of quite different masses can contribute
through diffractive excitation. This realization has led to the
application of a pre-existing body of theoretical expertise,
developed on both sides of the iron curtain before QCD.  Since the
theory of diffractive processes has not evolved significantly in the
last twenty years, the potential for conceptual clash between the
``pre-QCD" diffractive picture and the ``post QCD" hard scattering
picture is considerable.  Neither picture is a substitute for the
other;  each is useful when appropriate.

        The kinematics of the diffractive limit is one where the
momentum transfer to a spectator is small, or at least fixed, as the
energy of a beam becomes large. Contributions from each
momentum transfer channel become focused kinematically into the
forward small angle region, creating a highly coherent quantum
mechanical problem.  It is worth noting that the diffractive problem
has nothing to say about the ``hard scattering" part of color
transparency experiments, if such is present.  Consequently the
impulse approximation and the concept of ``short distance", which
have become part of QCD, are not contained in the diffractive
discussion.  Diffractive scattering is relevant to the ``propagation"
part of interaction with the nuclear medium before or after a hard
scattering occurs.

        In a classic paper in 1956, Feinburg and Pomeranchuk\cite{FP56}
discussed
the quantum mechanics underlying high energy perturbative
calculations in electrodynamics. This paper is particularly useful
for understanding the concept of ``formation time" in inelastic
phenomena.  Some time later, Good and Walker\cite{GW60} discussed the physics
of strongly interacting diffractive production with deep quantum
mechanical understanding.

\medskip

\noindent {\it Good and Walker's Polarizer}

\medskip

        Feinburg and Pomeranchuk (F\&P) and Good and Walker (G\&W)
observed that the relevant quantity for time evolution of single
particle states with mass $m$ is not the energy $E$, but the
differences in energies $E$ for states with the same momentum
$P_z$.  The relation for single particles is $E= \sqrt{( p^2 + m^2)}$.
In the ultrarelativistic limit, the energy differences go like the
inverse of the big momentum:

\beq
E_1- E_2 = \Delta E = (m^2_1 - m^2_2)/ 2P_z
\eeq

This strange, literally ``asymptotic" result of relativity can be
easily visualized by an energy-momentum diagram (Fig (2.10)).
(Note that this energy difference is proportional to that obtained
from the null plane Hamiltonian $P=(m^2+ p_T^2)/ 2 P^+$.)  It is
slightly more precise to think in terms of the null plane formalism.
For a given momentum transfer $q^-$ to the beam, we can make an
arbitrarily  large invariant mass-squared $m^2 \cong 2 q^-
P^+$ simply by having large enough $P^+$. From this we can construct
a time scale, the so-called formation time, from the inverse energy.

	This concept generalizes to the case of the nuclear medium,
but the free space formula  (2.9) for it certainly does not.  The
problem is that the dispersion relation for hadrons propagating in
the nucleus is unknown.  The problem has the potential to be serious
in the high energy limit, because one is asking for a small difference
between two big, relatively uncertain numbers. We will postpone
discussion of this question and possibly important dependence on
transverse momentum, set to zero for the discussion, to Section 5.

        The next step in the argument is to observe that, due to
relatively small effective energy differences, many different
particle states become effectively degenerate and can be considered
to strongly mix under an interaction.  Due the relativistic effects,
the effects of strong interactions become even {\it stronger}.

        The quantum mechanics of such a strongly mixed system is
quite interesting.  G\&W make an instructive analogy with
diffraction of light from a circular disk of polarizer. (Fig. 2.11)  The
degenerate states which occur for light are the two
orthogonal polarization states.  Suppose first that an unpolarized
electromagnetic wave is moving along the $z$-axis and impinges on
the polarizer.  Then, one component of the electric field is
transmitted with no change, while the other component is absorbed,
leading to a diffraction pattern with the corresponding polarization.

	Suppose, next, that the incoming wave is already polarized in
the $\hat x$ direction. If the polarizer is set in the $\hat y$
direction, then it is a black disk for the incoming wave and leads to
a diffraction pattern.  This also is quite simple.

        Imagine finally the case where the polarizer is set at
45$^\circ$ to the $x$ axis.  Then the outgoing wave with this
projection of polarization is undisturbed.  The outgoing wave in the
perpendicular polarization direction shows a black disk absorption
pattern. It is elementary, but somewhat disturbing, to express the
outgoing amplitudes again in the parallel and perpendicular
coordinates of the original wave.  In the final state, both $\hat x$
and $\hat y$ polarizations occur.  Yet there was no $\hat y$
component in the initial state; the polarizer has produced it, by
filtering away waves from the incoming $\hat x$ polarized ones!

        All this is in accord with elementary quantum mechanics.  If
the initial state, polarized at $0^\circ$ to the $x$ axis, is called
$\mid E_0 > $, and a state polarized at 45$^\circ$ to the $x$ axis is
called $\mid E_{45} >$,then the polarizer corresponds to the
operator $\mid E_{45} ><E_{45}\mid$ that covers a certain spatial
region. The polarizer is diagonal in this basis.  It could also be
written as

$$\mid E_{45} ><E_{45}\mid = 1/2 ( \mid E_0 > + \mid E_{90} >) (
<E_0\mid +< E_{90}\mid )$$

The transition from 0 to 90 degrees polarization is clear as the off-
diagonal cross terms $\mid E_0 ><E_{90}\mid+\mid E_{90}>< E_0\mid$
in the operator.  Thus, conclude G\&W, the
interaction with the polarizer has, through the process of
measurement, produced a component of the polarization that did not
exist before.  In the same way, they reason, a number of particles
which are practically degenerate with the initial state are quantum
mechanically resolved and appear to be produced by the process of
scattering on a strongly interacting target.

        The G\&W argument reminds us that quantum mechanics is
needed to make a description of amplitudes.  The strange results of
quantum mechanics cannot be described by semi-classical or
probabilistic arguments.  A ``classical filter", for example, could
exponentially attenuate the {\it power} carried by a wave.  Yet a classical
filter could not cause the creation of the new amplitudes as
described above.  A ``quantum filter", by resolving different
components in different ways at the amplitude level, will appear to
create observable phenomena merely through the process of
measurement.

\medskip

\noindent {\it The Quantum Filter in Color Transparency}

\medskip

         We have seen that a picture invoking bulk attenuation of hadron
numbers crossing a nucleus would be inadequate to describe the
relative propagation of different amplitudes.  Instead, color
transparency involves a quantum filter in the form of the nuclear
medium.  To visualize this, let us continue with the analogy of the
atom scattering from a mirror. This time we immerse the ``mirror
experiment" in an atom-absorbing medium (Fig. 2.12).

	In the quantum medium, we will specify the dependence of the
absorption on the internal configurations of the atom's constituents
at the amplitude level.  The theoretical object carrying this
information will be called the survival propagator  $G_A$. $G_A$ is the
amplitude for two (or three or more) particles
entering the medium at $x$ and $x'$ to be found again at positions
$y$ and $y'$ (Fig. 2.13).  This 4 (or 2 N) point Green function
consists of

$$G_A( x, x'...y,y') = <A\mid T\lbrack\psi^\dagger(y')\psi^\dagger(y)\ldots
\psi(x') \psi(x)\rbrack\mid
A>$$
where $\mid A>$ is the target state and $(\psi^\dagger)\; \psi$ are the
field operators to create (annihilate) the quarks. $G_A$ is defined to be
2PI (2 particle irreducible) or 3PI
depending on the number of quarks being propagated. We will be
concerned with local color singlet parts of the 2N-point function suitable for
hadron propagation.
Flavor indices are suppressed,
and we can assume we are examining the propagation of
two (or more) unlike quarks, such as a $u$ and $\bar d$, which
propagate together to finally form a hadron such as a $\pi^+$ meson.  As
remarked
earlier, the transverse spatial separation coordinates $(\vec b_T=\vec x_T-\vec
x'_T,\; \vec b_T'=\vec y_T-\vec y'_T)$ are
the important ones in a relativistic treatment, while the
longitudinal coordinates are considered separately.

	The survival propagator $G_A$ is the object that contains information about
interactions with the medium.  Let $G_A( k, k';\ell,\ell'; w )$ be the Fourier
transform with respect to the quark momenta, with $w$ the total amount of
momentum
transferred from the nucleus (Fig. 2. 13).  The effects of interaction can be
expressed in two steps.  First, the outgoing wave function is converted to an
interacting object we call $\phi_A(k, k'; w)$:
\beq
\phi_A(k, k'; w) = \int d l \; G_A(k,k'\; ;\; l,l',w)\phi(l,l')
\eeq

The integration over the variable l is indicated by the loop in the diagrams
and
in a relativistic treatment (Section 4) is 4-dimensional.  The object $\phi_A$,
which can be called a ``filtered wave function,"  acts inside the momentum
convolutions just as the original wave function does. In terms of its Fourier
transform in the relative coordinates $k-k'$ conjugate to quark separations
$r$, we
again have for the full amplitude
\beq
M(P,Q;w)=\delta (P'-P-Q-w)\int dr\; \tilde\phi_A^*(r)\tilde\phi(r)e^{-iQ\cdot
r}
\eeq

Short distance in the separations of $\phi_A$ are again selected by the
oscillations
in $\exp(-i Q\cdot r)$.

\noindent {\it The Survival Propagator in QCD}

Obviously, the relation between $\phi_A$ and the original wave functions $\phi$
depends on
the model for $G_A$.   Current ideas based on QCD hold that $G_A$ can be
predicted
perturbatively in the region where all pairwise transverse separations between
the quarks are small compared to a Fermi.  The diffractive approach, and
hadronic basis calculations (Sections 4, 5) are different models with the same
goal. For example, if one proposes that many intermediate states are created by
the
fast momentum $P+Q$ plus the nuclear momentum transfers totalling $w$, then the
model
for $G_A$ and  $\phi_A$ will reflect this.  The outputs of these models can
usually be
re-expressed as models for $\phi_A$, so the decomposition above is rather
general.

	Due to the integrations over the internal coordinates in Eq. (2.5), one
sees
that the hard scattering is tied into the filtered wave function in a
non-trivial way.  In particular, it would be an approximation of unknown
validity to propose that the hard scattering simply multiplied the nuclear
effects.

	Color transparency is obtained using the perturbative picture for
$G_A$.  As a
first remark, the survival amplitude is ``unity" minus the scattering
amplitude.
For the scattering amplitude, we practically have an almost entirely imaginary
(absorptive) object, closely related by the optical theorem to previous
calculations of the total cross section between ``atoms". We can adapt the
eikonal method of Ref. \citenum{GS77} to calculate the scattering amplitude in
pQCD.

	Consider, then,  the scattering of two quarks (system 1) on a spectator
atom
in the medium (system 2) made of two quarks.  Consider the 2-gluon exchange
diagrams shown in Fig. (2.14).  Flavor is ignored; since helicity is conserved
by a vector interaction, spin indices can also be suppressed.  Each gluon
attachment is accompanied by a color matrix $(\lambda_a)_{ij}$.  Taking the
singlet projection
indicated, one finds conveniently that all matrix orderings in the same system
are the same, and simply produce an overall factor $Tr[\lambda_a \lambda_b] =
C_F \delta_{ab}$.

	 The amplitude itself can be set up directly in coordinate space. Let
$B$ be the
impact parameter between the two systems, with net momentum transfer $w$. Let
the
quarks be located at positions $x_i, x_i'$, given by
\beqn
x_1=B/2-r\hskip .5in x_2& = & -B/2-s\nonumber \\
x_1'=B/2+r\hskip .5in x_2'& = & -B/2+s\nonumber \\
\nonumber
\eeqn
In the relativistic case these are transverse vectors.
Let the coordinate space kernels of the gluon exchange between points $x_i$
and $x_j$
be denoted $V(x_i-x_j)$. There are 16 possible combinations of the pairwise
interactions, of the form $V(x_1-x_2)V(x'_1-x'_2)$, etc.  These combinations,
denoted
$K(x_i, x_j)$, will be listed momentarily.  Then the amplitude is
\beq
G_A(x,x';\; y,y';\; w)=\delta(x-y)\delta(x'-y')\bigg[1-\int dBe^{-iwB}\int
ds\mid\varphi(s)\mid^2 K(x_i,x_j')\bigg]
\eeq
Now we consider the kernel $K$.  Each exchange between opposite color charges
comes with a relative ``minus" compared to the exchanges between like charges.
As the reader can check, the total of all the combinations of the form
\beqn
K(x_i,x'_j) & =&  V(x_1-x_2)[V(x_1-x_2)-V(x_1-x'_2)+V(x_1'-x_2')-V
(x_1'-x_2)]\nonumber \\
& -& V(x_1'-x_2)[V(x_1-x_2)-V(x_1-x'_2)+V(x_1'-x_2')-V(x_1'-x_2)] \nonumber \\
& + & \ldots \\
\nonumber
\eeqn
can also be written
\beq
K(x_i,x_j')=[V(x_1-x_2)-V(x_1-x_2')+V(x_1'-x_2')- V(x_1'-x_2)]^2
\eeq

One sees the results of coherence in the ``minus" signs, which represent
destructive interference between emissions from opposite color charges. We
obtain the perturbative expression for $G_A$ scattering from one atom
spectator:
\beqn
G_A(x,x',y,y',w) &=&  \delta(x-y)\delta(x'-y')\bigg[ 1-\int dBe^{-iwB}\int d\;
s\mid\varphi(s)\mid^2\nonumber \\
&\times & [V(x_1-x_2)-V(x_1-x_2')+V(x_1'-x_2')V(x_1'-x_2)]^2\bigg]\nonumber  \\
\eeqn

We are interested in this in the short distance region selected by $\exp(-iQ
r)$.
$A$ power series approximation to the interaction part of $G_A$ in this
variable,
with $r=\mid x_1-x_1'\mid$, yields
\beq
K(x_i,x_j')=[r\cdot\nabla(V(x_1-x_2)-V(x_1-x_2'))]^2
\eeq
This goes like $r^2$,  showing the characteristic ``geometrical" behavior of
dipole-dipole scattering (and the associated dipole moments in the expansion of
$V$).

	An exactly similar result follows in the relativistic treatment, if one
expresses space variables using $B_s$, the transverse separation between the
center
of $P^+$ of the right-moving quarks, and center of $P^-$ of left moving
spectator, in
the center of momentum frame.  One recovers in that case the results of
Ref.\citenum{RP90}.  Although the perturbative kernels seem to suffer from
infrared
divergences, the calculation is finite.  For example, the transverse Fourier
transform of a massive gluon
is given by
\beqn
V(x-x') & = & \int d^2k_T{e^{ik_T(x_T-x_T)}\over\vec k^2+m^2}\nonumber \\
& = & {1\over 2} \mid x-x'\mid m\; K_1(\mid x-x'\mid m)\; , \\
\nonumber
\eeqn
where $K_1$ is a Bessel function of imaginary argument, and where $m$ is a
gluon mass.  This is badly behaved as $m\rightarrow 0$.  Yet,
the region of transverse position sensitive to this is erased
by coherence, and replaced by scales characteristic of the bound state, so the
massless gluon limit is ``safe".\cite{L75,N75}

	The main result is that the effects of interaction in the perturbative
treatment are a factor of the spatial variable ``$r^2$", at least as
$r\rightarrow 0$.  Because
this is the region selected, and it also vanishes in the region of large
$Q$, we
have color transparency as a rather general result in this limit:
\beq
\lim_{Q\rightarrow \infty}\int dre^{-iQr} \varphi_A^*(r)\varphi(r)=\int dr\;
e^{-iQr}\varphi^*(r)\varphi(r)
\eeq
Color transparency in pQCD is thus a feature of kinematics, coherence, and the
overlaps of wave functions, rather that of superposition over states; a ``one
hadron" universe is clearly sufficient for it to occur.  The fact that a
completely perturbative treatment of color transparency is self consistent has
not been widely appreciated.  Because the interaction probability is small,
even
the presence of many targets (a complicated nucleus) does not disturb the
result asymptotically. Detailed models of the nucleus as a target could, in
principle, lead to straightforwardly calculable pertubative results.

\noindent {\it A Model Independent Framework}

	The framework here can accomodate many models; we have devoted the
entire
Section  5 to ``models of propagation". Perhaps the most controversial item at
finite $Q$ is
the effects of the nuclear momentum transfer $w$.  In the hadronic basis
treatment, many physical states are used to create a superposition which moves
through the medium, created by the large $P^+$ and moderate $w$--values from
the
nuclear target.  This can be expressed as a superposition of states model for
$G_A$.  In addition, integration over the energy-like variable of $w$ is
thought to
create a time-scale (or many time scales) which might give complicated
dependence on
the target length.  These ideas will be discussed in detail in Section 5, while
general remarks on the time scale issue will be given soon. There seems to be
no
disagreement that the formalism of Ref.\citenum{RP90} is the correct
description at very
large values of $Q$.  There is, of course, plenty of controversy whether this
can
apply at current laboratory values.\cite{FFS91,RP912}

        Before closing this subsection we point out an amusing point.
Among all bound states, the value of the wave function at the origin
is the {\it largest} in the lowest lying states, and generally decreases
monotonically as the states number increases, at least for atoms.

Since the quantum filter of $G_A$ has caused the region near the
origin to propagate most favorably, it is bound to dominate the
measurements.  This again is called ``nuclear filtering".  If, for
example, the value of $Q$ were too small to make the Fourier motivated
approximation
of short distance very good, then the region near the origin might
anyway be selected by sheer ``survival of the smallest".\cite{RP90}
Transitions
to excited states involving less wave function at the origin and more
wave function in the peripheral regions getting ``filtered away"
would be suppressed. Thus, large $A$ and large $Q$ provide two complementary
but independent motivations for short distance.

\medskip

\noindent {\it The Time Scale Problem}

\medskip

	So far we have not discussed whether the path taken in the
attenuating medium is ``long" or ``short".  If the attenuating medium
is shallow, then it hardly matters what the attenuating interaction
is, and color transparency may require cleverness to measure. (As
we will see, electroproduction has to contend with this problem.) The path
length is this context is long or short compared to some
characteristic scale. What scale? Are we to think of the size of the nucleus
in
units of some hard scattering, short distance scale, an interaction
scale, a mass scale, a scale modified by a Lorentz boost, or what?

	Closely associated with this problem is the ``time scale"
problem, addressing with non-relativistic concepts the dynamics of
the time evolution of a hadronic wave packet.  In most models, if the
wave packet expands rapidly, it is assumed to lose the capacity for
exhibiting color transparency.  In the same models, the observable
energy dependence of the processes, which of course includes the
effects of the Lorentz boost, is often  interpreted rather directly as
a measure of the time evolving dynamics.

	The details of the initial state become crucial in this
discussion.  In many treatments, the initial state is set up as a boundary
condition.
There are two extremes: a geometrically small wave
packet might be formed at time $t=0$. This proposal is ad-hoc, but
worth considering.  On the other hand, in the impulse approximation,
a hadron-sized wave packet is disturbed by the addition of large
momentum transfer oscillations at time $t = 0$.  This is what
happens in the theory of  hard scattering QCD.

	Because color transparency may exist in many circumstances,
and the clear--cut kinematic conditions of hard scattering theory may often not
be
fulfilled experimentally, we will consider each of the above extreme
cases here.  For example, in charmed quark pair production the
initial state is unknown and a small wave packet can be invoked as
an idea to be explored.  To complicate things further, the time
evolution in an attenuating nuclear medium is actually unknown,
because the energy eigenstates in interaction with the medium are
unknown.  Thus, many disparate pictures are possible.

	We now turn to more specific remarks. Many authors
have proposed the initial
state is ``compressed" or quantum mechanically prepared in some
way to exist in the region $r< 1/Q$.  From the uncertainty principle,
the dominant momentum space region of such a wave packet
has $k\cong Q$, leading to a rapid expansion in size.  On the other
hand, in the impulse approximation we have emphasized here, the final state
wave function is
$exp(i(P+Q) )\phi(k+Q)$.  The relative coordinate momentum space
wavefunction is centered at $P+Q$, a ``big" number. Because this is just a
translation, the
fluctuation in momentum is still set by the hadronic size. To see this, let
$\Delta k^2=<k^2>-<k>^2$. Then
\beqn
\Delta k^2 & = & \int dk \varphi^*(k+Q)k^2\varphi (k+Q)- \bigg[\int
dk\varphi^*(k+Q)k\varphi(k+Q)\bigg]^2 \nonumber \\
& = &\int dk\varphi^*(k)k^2\varphi(k) \\
\nonumber
\eeqn

It follows that the case of the impulse approximation has a different time
evolution,  because the fluctuations are not large.
(Fig. 2.15). We will present a simple calculation to verify this.

	Of interest are the short time dynamics, for which we can use the power
series
solution of the operators in the Heisenburg picture as a function of the time
$t$:
\beq
\hat r(t)=\hat r(0)+it[\hat H,\hat r(0)]+(it)^2[\hat H,[\hat H,\hat
r(0)]]/2!+\ldots
\eeq
The Hamiltonian in such a non-relativistic framework (which is adequate to
illustrate our point) is written as $\hat H = \hat p^2/2m+ V(\hat r)$. A short
calculation shows
that the time dependence of the relative coordinate-squared $r^2$ is given by:
\beq
\hat r^2(t)=\hat r^2(0)+(t/m)[\hat p(0)\hat r(0)+\hat r(0)\hat p(0)]+t^2[\hat
p^2(0)-(1/2m)\partial V/\partial r]+\ldots
\eeq

Now consider the average $r^2$  as the struck wave function $\phi_s(r)$
time-evolves
towards the final state wave function $\phi_f(r)$, measured by
$<r_{sf}^2(t)>$, where
\beqn
<r^2_{sf}(t)> & = &  {1\over 2}\int d^3r\phi^*_f(r)e^{+i\hat Ht}\hat
r^2e^{-i\hat
Ht}\phi_s(r)/F+c.c.\; \; \nonumber \\
& = & {1\over 2}\int d^3r\phi^*_f(r)\hat r^2(t)
\phi_s(r)/F+c.c.\; \; ; \\
F & = & \int d^3r\phi^*_f(r)\phi_s(r)\; , \nonumber \\
\nonumber
\eeqn
where $c.c.$ stands for complex conjugate. (We keep the real part; the reader
may
verify that the imaginary part works the same way.) The reason for using
$<r_{sf}^2(t)>$ rather than a diagonal expectation value in the struck state is
to
probe the regions involved in the ``particular probability" to go from the
struck
state and reach the final state, and not the ``coarse grained probability" for
the whole wave packet to expand to any old inelastic state whatsoever. Of
course,  $<r_{sf}^2(t)>$ is also an average $r^2$ measured in the form factor,
which we
have already seen is the relevant thing for interaction with the nucleus via
the
survival propagator $G_A$.

	Because we want representative short distance behavior, the final state wave
function will be been chosen to be Coulomb-like, as opposed to other
possibilities such as a Gaussian:
\beq
\phi_f(r)=e^{-r/a}
\eeq
where $a$ is a bound state distance scale (the Bohr radius for an atom). The
two
pictures of the wave function $\phi_s$ just after being struck are:
\beq
\phi_s(r)=e^{-r\mid Q\mid} \hskip 1.75in {\rm compressed\; case}\; ;
\eeq

\beq
\phi_s(r)=e^{-r/a -iQ\cdot r}  \hskip 1.5in {\rm impulse\; case}\; ;
\eeq
We have used the fact that in the pQCD kinematic analysis the wave function
$\phi\rightarrow  \phi \exp(-iQ\cdot r)$ after being struck, while in the
compressed case it is assumed
to reside in a region $r<1/Q$.

  	The details of the calculation depend on the choice of the Hamiltonian,
but
the main features are easy to anticipate.  Suppose we use a harmonic oscillator
Hamiltonian, with $V(r) = m w_o^2 r^2/2$.  Upon doing some integrals, we find
for
the two cases:
\beq
<r_{sf}^2(t)>=  {12\over (\mid Q\mid +1/a)^2}+t^2({Q\over
am^2}-{6w_0^2\over (\mid Q\mid +1/a)^2})+\ldots \hskip .25in {\rm compressed\;
case}
\eeq

\beq
<r_{sf}^2(t)>= {8Q^2-24/a^2\over ( Q^2 +4/a^2)^2}+t^2\left({1\over
m^2a^2}-{4Q^2w_0^2-1 2w_0^2/a^2\over (Q^2+4/a^2)^2}\right)+\ldots \hskip .25in
{\rm impulse\; case}
\eeq

In both cases short distance is being selected, in one (impulse case) via the
Fourier oscillations, and the other (compressed case) by simply imposing it.
However,
the relevant rates of expansion  of the ``compressed" wave packet and the
``hard struck" wave packet are strikingly different.  The expansion of
the compressed wave packet is set by the initial size and goes like
$Q$.  The constant expansion rate of the wave packet in the impulse case is
due to the hadronic time scale.\footnote{The reader may wish to repeat a
similar calculation for the time evolution of $\hat r^2(t)$ in the diagonal
case $\int d^3r\varphi^*_sr^2\varphi_s$, which measures the coarse--grained
probability.}

	 Relativity makes the situation even more interesting.  In  a
typical high energy hard scattering observation, the momentum
transfer $Q^2$ goes like the beam energy $E$  (for example,  $Q^2 =$
const $s$ = const $2mE$, in fixed angle hadron scattering; $Q^2
=2mE$ where $E$ = the photon energy $\nu$ in electroproduction where $m$ is a
mass scale).
The Lorentz boost to the lab frame expands every time scale in the
struck hadron rest frame by a factor of $E/m$.  In the lab frame we
will observe the above formulas with $Q^2\rightarrow 2mE$, $t\rightarrow tm/E$:
\beq
<r^2_{sf}(t)>_{LAB} = 6/(mE)+\sqrt{2m}\sqrt{E}t^2/(E^2a) +...\hskip .25in {\rm
compressed\; case}
\eeq

\beq
<r^2_{sf}(t) >_{LAB} = 4/(mE)+  t^2/(a^2E^2)  +... \hskip
.5in
{\rm impulse\ \ case}
\eeq
In the compressed case one does not gain as much by going to higher energy,
because the system simply expands more rapidly to compensate.  This has caused
many difficulties for this proposal, and led some authors to conclude that
large $Q^2$ leads to ``rapid expansion." But, in
the hard scattering case from the impulse approximation kinematics, there is an
advantage to high energy,
because every effective time scale -- even including the increased
hardness of the scattering -- increases in the relativistic high
energy limit.

	Perhaps more effort has gone into studying the time scale
issue in color transparency than any other aspect. The example
shows not only that  no unique ``time scale" exists;  even the
functional dependence on energy depends on the assumptions.  The
resolution is subtle and depends on the experiment.  It is reasonable
to suppose that the true situation might be intermediate between the two
pictures
with
quantum mechanical superposition of the two pictures playing a role.
Nevertheless, the rapid expansion dynamics  of a ``compressed state"
force it to be the most difficult in which to observe color
transparency.   By the same token, the dynamics favors selection of
hard scattering impulse amplitudes because these expand at a
slower rate.  This is another case where the quantum mechanics of
nuclear filtering are important, because even if hard scattering
conditions seem to be fulfilled only marginally, there is every reason to
believe that
selection by
triggering on elastic survivors should make this contribution
relatively more important.

\medskip

\noindent {\it Color Transparency in the Hadronic Basis }

	Much of the discussion of color transparency in the hadronic
basis focuses on the role of inelastic intermediate states.  Since
there are no ``quark relative coordinates" in this basis, it is not
clear how to interpret the perturbative prediction for $G_A$.  Most
workers have tried to understand how color transparency might
happen by creating various assumptions or parameterizations of
$G_A$.

	The role of the hadronic basis and inelastic channels can be
explored\cite{HBB91} by inserting a complete set of states in the $s$-channel
cut
of $G_A$.  Note that a complete set means all physical processes,
namely all numbers of particles of all types, including resonances,
continuum, all possible momenta, the works.  One can write an $s$-
channel dispersion relation (Fig. 2.16):

$$G_A(s)={1\over 2\pi i}\int ds' {disc\; G_A(s')\over s-s'+
i\varepsilon}$$
where $disc$ is the $s$--channel discontinuity (aka ``imaginary
part").
Such expressions should be model independent and exact. State by state in $s'$,
we can
imagine doing the integrals over the internal coordinates of the
constituents for each on--shell final state.  The resulting expression
makes no references to quark coordinates, and should permit one to
survey the contributions of the various hadronic state channels.

	Recall the discussion of the wave function after the hard
scattering (Eq. (2.5)).  The wave function of the struck state is
just the original one multiplied by $\exp(-iQ\cdot r)$. The
contribution of this state onto each final state $<N|$  is given by the
overlap  $<N|exp(-iQ\cdot r) |P>$.  This, of course, is the transition
form factor to the particular state:

$$<P+Q| J^\mu |P>   =  \sum\limits_N <P+Q|  N><N| J^\mu |P>$$ where the field
theory operator $J^\mu$ was just replaced by $e^{-iQ\cdot r}$ in
non--relativistic
one photon exchange.
The set of channels which can be inserted is potentially very large:
for a typical momentum transfer of 300 MeV, and with a beam
momentum of 10 GeV, one can involve all states with the proper
quantum numbers up to an invariant mass of 6 GeV$^2$!   From deeply
inelastic scattering we know that the continuum states make a large
contribution. (Conversely, saturation by a few resonances is a poor
approximation.)
With so many states the possibility of difractive
rescattering back into the original state may be considerable, and
the hoped-for understanding of color transparency in terms of
hadrons might occur.

	 Unfortunately, there is no universally agreed proceedure for
making the calculation. Successful models exist, but almost all of
them make the ``compressed state" boundary condition as an initial assumption,
leading to a rapid
(non-pQCD-like) expansion rate.  Because this is an initial condition in the
hadronic basis treatment, subsequent calculations do not determine whether it
is ``right" or ``wrong".
An exception to the compressed initial state  assumption is the recent work of
Markovoz and Miller\cite{MM94}.  We will discuss various models in more
detail in the section on ``models of propagation".

\medskip

\noindent {\it Few Particle Unitarity}

\medskip

	We finish the section with an observation.  Ordering the set of
intermediate states by increasing energy, the first state to
contribute is the proton itself.  In the quark picture this ``diagonal"
contribution is sufficient for predicting a significant part of the
color transparency.  No other states are needed, because the wave
function for small transverse separation propagates almost freely,
and has a comparatively strong overlap with the final state.  	It is a very
detailed question whether the single state would be a ``good"
approximation. Nevertheless, the quarks say that {\it there is no
strict need for mixing among hadronic states to have color
transparency}. We can drop all the other states, truncating the
universe to the diagonal contribution, and still  see the effect! We
need only to refer back to  the impulse approximation in the quark
basis Eq. (2.5) to verify this.

This is a very peculiar thing, because if we had independently
measured the attenuation of Hydrogen atoms in a medium without a
hard scattering mirror experiment, we might not anticipate it.  But
it is actually simple.  The total attenuation cross section of an
atom not undoing hard scattering would be set by its overall size, of order one
angstrom.  For the total cross section experiment the sum over all
regions of tranverse separation is needed; the region of the largest size in
the wave function is the most important.  The total cross section for
attenuation is a square Angstrom.
This is a measurement
based on one, ``center of mass" coordinate as determining the
dynamics.  Compare this to the hard scattering experiment followed by
attenuation. This depends sensitively on the transverse region with the
{\it smallest} spatial size, which
is entirely different, and which probes the internal coordinates.
Should we attribute this to the role of inelastic processes?   No.
The relation is simply unitarity in the single atom sector. (Fig. 2.17).

Because an atom is made of only two constituents (a very
good approximation, for atoms), then we will get the total cross
section by integrating over all possible 2 particle final states of the
constituents.  This relation is

$$\sigma_{tot, atom} = \int dr'\; disc\; G_A^{(1)}(r=r')$$
where $G^1_A$ is the one--state contribution.  Since $disc\; G_A^1(r=r')$ is a
positive definite quantity, the integral over it is certainly larger
than the contribution from the short distance region of $r'<1/Q$.  Thus it is
rigorous that a process selecting short distance will make a
comparatively small  contribution to $\sigma_{tot}$. The
one particle contribution will kinematically give color
transparency (Fig. 2.17).

	The same conclusion is very difficult to obtain by simply
considering the ``atomic center of mass" (hadronic) basis alone.  This is
because
{\it the matrix
elements needed are new diagonal contributions not readily
accessible from bulk ``center of mass" hadron cross section
measurements}.  Moreover, one can easily see that the mixing of the
different channels might be entirely irrelevant. Imagine a hypothetical
case where the spacing between excited states of an atom (the
different ``hadrons") were made arbitrarily large in energy.  For the
Hydrogen aton this can be done by taking the Rydberg energy $me^4/ 2
\hbar^2\longrightarrow \infty$.  With an arbitrarily large separation of
levels,
mixing of levels will {\it not} occur.  At the same time, we can arrange
{\it not} to have the wave function at the origin change at all.  We simply
take the limit in a way so as to keep fixed the Bohr radius $a_0=
\hbar^2/me^2$.  The joint limit is arranged by taking $e^2 \longrightarrow
0$
and $m \longrightarrow \infty$ with $m^2e^2$ fixed.  According to our earlier
calculation, the
scattering of the atom via the elastic ``ground state" channel not change in
this limit, while the
mixing of off-diagonal channels would decrease substantially.  We can
even take the opposite limit: let the separation between levels be
reduced as much as desired, while still keeping the wave function at
the origin fixed.  In this case a host of levels become available for
mixing.  They do not matter; same story, same result, for the
diagonal contribution. Yet, by rescattering, these might add probability
separately in this case.

	Some arguments on color transparency miss this point
completely.   The ``quantum filter" of the nuclear medium can operate
this way.  It occurs because the different theories based on quarks
and hadrons are not the same;  quarks act as if they have more degrees of
freedom than hadrons.   But {\it as a separate issue}, the mixing of states
might nevertheless be important when we do the
experiment in an attenuating medium.  The diagonal contribution does not
exhaust all possible contributions to the ``$s$-channel cut" of $G_A$ by any
means -- there are many other channels. It might be possible to create an
observable which can
separate diagonal from off-diagonal transitions. In the absence of any clear
separation, the
relative strengths of the different processes constitute a research
problem. In Section 5, we review the various models.
\newpage

\subsection{Figure Captions}

\begin{itemize}

\item[2.1]  The elastic form factor.  Assumptions are required to show that
short
distance is probed by large $Q$.
\item[2.2]  The handbag diagram for deeply inelastic scattering.  The sum over
the
final state (dashed line) runs over all number of all kinds of constituents,
including both the struck quark and undetected final states of all kinds.
\item[2.3]  An atom scattering off a weakly interacting mirror (left) receives
momentum in the same way it absorbs a virtual photon with spacelike momentum
transfer (right).
\item[2.4]  Kinematics for scattering with momentum transfer $Q$ from a mirror,
or a
virtual photon. The wave functions $\phi$ are indicated by blobs.
\item[2.5]  A wave function's large $Q$ Fourier transform may receive its
dominant
contributions (arrows) from regions not representing small separation (left) or
the region of small separations (right) depending on where the most rapidly
varying region occurs.
\item[2.6]  Independent scattering from two ``mirrors".  Assuming large
momentum
transfer, short distance is nevertheless not guaranteed; in particular, the
separation of the two mirrors in coordinate space can be large.  The same
phenomenon can occur when quarks exchange gluons.
\item[2.7]  An inelastic event in the quark basis (left) and in the hadronic
basis
(right).  Sums over all final states in the quark basis are done using
unitarity.  In the hadronic basis, one can attempt to model all the terms
individually.
\item[2.8]  (a) The relative coordinate wave function of a full-sized Hydrogen
atom, an exponential.  (b) The real part of the same function multiplied by
$\exp(-i Q x)$, which is the result of absobing a virtual photon.  Plot is to
scale
with $Q=10$.  The overlap of this wave function onto the ground state gives the
form factor; the result is dominated by $\Delta x = 1/Q$. (c) A hypothetical
wave
function (arbitrary normalization) described by $\exp(-10 x^2)$  from the
``compressed intial state" hypothesis, in which a wave function is quantum
mechanically prepared to exist in a region $\Delta x = 1/Q$.
\item[2.9]  Plots of a typical wave function for quarks in a pion in a mixed
representation, as a function of the longitudinal momentum fraction $x_F$ and
the
transverse space coordinate $b_T$.
(a) The wave function is $\exp(-b_T) x (1-x)$, where $b_T$ is the transverse
coordinate
in units of Fermis. (b) Real part of same wave function after absorbing a
photon
in the $b_T$ direction with momentum $Q = 1\; GeV$.   The width of the wave
function in
$b_T$ is unchanged, but the photon produces oscillations which select the
region $b_T
< 1/5$ Fm from the region near the origin.
\item[2.10]  Even large mass differences become small energy differences for
states
moving close to the speed of light.  To predict the small energy differences
used in the hadronic basis approach, one must know the dispersion relation with
great accuracy.
\item[2.11] An incoming electromagnetic wave polarized in the $\hat {\bf x}$
direction
hits a
polarizer oriented at 45 degrees.  The outgoing wave is measured to have both
$\hat {\bf x}$
and $\hat {\bf y}$ polarizations, illustrating the creation of new degrees of
freedom
by a
quantum filter.
\vskip .25in
\item[2.12] The mirror experiment is placed in an absorbing medium, which
creates
a quantum filter.  Regions of the wave function with small transverse
separation
propagate well, while large separation regions are more attenuated. The
survival
propagator $G_A$ represents this interaction.   Momentum $Q$, received from the
hard
interaction, is distinguished from momentum $w$ transferred from the medium.
\item[2.13] (a) the survival propagator $G_A$  for a quark anti-quark. (b) the
survival propagator $G_A$ for 3 quarks.
\item[2.14]  Diagramatic expansion of $G_A$. The spectator atoms (system 2)
have wave
functions indicated by blobs; a propagating color singlet (system 1) is to be
attached to a hard scattering on one side, and a wave function for the final
state on the other.  Other spectators are not shown.  Space coordinates used in
the calculation are indicated; the separations between quarks in system 1 and 2
are {\bf r} and {\bf s}, respectively.
\item[2.15]  Momentum space models for the struck wave function.  (a) A
full-sized
wave function $\phi(k)$ before being struck. It is centered at $k=0$, with
typical
hadronic fluctuations of relative momentum of order 300 MeV. (b) The same as
(a)
after receiving large momentum $Q$, which causes a translation.  The relative
momentum is order $Q$, but fluctuations remain of order 300 MeV. (c) Momentum
space picture of a ``compressed" state, which is centered at large $Q$ and also
has
a large fluctuation of order $Q$.  Large fluctuations cause rapid expansion not
occuring in the impulse approximation.
\item[2.16]  The survival propagator $G_A$ can be written using a sum over on
shell
intermediate states $N$, an $s$-channel dispersion relation, for each fixed
momentum
transfer $w$ to the nucleus. If the intermediate states are very massive, a few
resonances will not be a good approximation.
\item[2.17]  Unitarity and color transparency. (a) A single hadron contribution
to
the imaginary part ($s$-channel discontinuity) of the survival propagator
$(G_A^{(1)})$ for a quark-antiquark. The momentum transfers and relative
separation
coordinates $r$ of the  quark antiquark are held fixed; the region
$r\rightarrow 0$ is of
interest.  (b) Unstitching diagram (a) and reassembling it backwards, the
integral over all $r$ will give the scattering amplitude $\sigma_{tot}$ in this
channel.
Since the integrand is positive definite, the integral over any finite region
of
$r$ of $disc\; (G_A^{(1)}(r)  )$ is always less than  $\sigma_{tot}$,  final
state hadron by hadron.

\end{itemize}

\newpage

\section{Experimental Definitions and Criteria}
\noindent
\subsection{ Overview}
\setcounter{equation}{0}
\medskip
In this section we review some of the ideas which have been put forward
for finding color transparency and review them critically. This subject is
bubbling with ideas, and it is
important at present to have a liberal view of what might be an
interesting experiment.

A basic requirement for color transparency is that the color
flow should be controlled. For example, in exclusive processes we believe we
can account for the flow of color charge by studying Feynman diagrams
of the microscopic processes. When a color singlet flows coherently
through a reaction, as occurs when we trace the quark lines in an
elastic scattering, then one can argue that all that is needed is short
distance dominance and the reduced attenuation signal of color transparency
should be observed. Insisting on large $Q^2$ may not be necessary because
attenuation
in the `quantum filter' may be able to create short distance
(or small color separation) conditions on its own.

In the earliest work, the conditions of color transparency were fairly
tight: to enforce exclusivity, one desired sufficiently restrictive cuts so
that no pions would be produced beyond those discussed in the reaction.
As the field has evolved, many experiments (e.g. at FNAL energies)
are being examined in which this restriction cannot be enforced.
This seems to be healthy development to the
extent that something can be discovered in kinematic regions where theory
may not be well developed. The search for color transparency is primarily
an experimental task; like gold, it is going to be `where you find it'.
We will begin by organizing the discussion under common questions,
and then turn to reviewing suggestions for experiments.

\medskip
\noindent
{\it What cuts are required to define an acceptable exclusive reaction?}

\medskip
  In the first BNL
experiment on color transparency in pp scattering \cite{CEA88},
 only processes which satisfied
the nominal conditions of elastic scattering within
an energy resolution of about 20 MeV were
included. This cut, of course, implied that there would be no pions in the
final state on a target at rest. Since the targets in the
nucleus are not at rest due to Fermi motion, the experiment also
included a veto
detector designed to reject cases where pions were emitted by the final state
nucleus. Although such restrictive definition were necessary for a
pioneering experiment, we argue that the experimental cuts can be made less
restrictive and still represent coherent color flow. At Fermilab energies,
for example, such restrictive cuts would require unrealistically precise
determination of the energies of all the particles. A small error in
the measurement of energy $\Delta p$ of the incident particle would lead
to a large uncertainty in the squared invariant mass, $\Delta M^2\approx
\Delta p\sqrt{s}$, where $s$ is the center of mass energy squared. Therefore
at Fermilab energies it would be very difficult to require that no
additional particles are produced unless specific detectors are added.

The upgraded BNL experiment will measure all the momenta
of incoming and outgoing particles, allowing for an overdetermined
kinematic situation. This is a significant experimental accomplishment, and may
also make possible many unusual experimental cuts (such as direct observation
of
the system ``center of mass").
The interpretation of the data can be somewhat
ambiguous, however, because the system being measured interacts with
the nucleus. Typical momentum transfers in a strong interaction scattering of
a few hundred MeV are comparable to typical  Fermi momenta, so some
modeling of data which `measures' Fermi momenta directly is
necessarily involved. Depending on the observable- we will discuss
several in the following sections- a combination of modeling ordinary
nuclear effects and clever choices of what to measure certainly
extends the interesting allowable energy resolution into the 100's
of MeV range. The opposite extreme, of requiring very high resolution
so that individual nuclear shell structure can be picked out, may
be interesting but seems at the present time to be ``overkill".
Finally, if there is some information available from
production of pions, then such processes need not necessarily
be dismissed; theory may get to the level of understanding the color
flows.

In high energy diffractive $\rho$ production at Fermilab
\cite{F93,F94,AEA95}
there is an
indication that color transparency might be observed. The kinematics
of this case are such that final states with huge invariant masses
may likely sneak through the cuts imposed on the detection process.
This again is not necessarily catastrophic, because it may be perfectly OK
for a nucleus to ``explode" after the event has occurred.
One should be cautious, however.
In some cases, such as hard scattering, the criteria
of no pion production seems feasible and straightforward. In other
cases, and until various theories reach a high enough level of
accountability, it is rather difficult to claim that one is making
a definitive test.

There are two characteristic limits: hard scattering and diffractive.
Generally, in hard scattering all scales are large; in the diffractive limit,
$s>>$ other scales. Perturbative QCD generally applies in the hard scattering
limit, its application in the diffractive limit is a research problem.

\noindent
\subsection{ Hard Scattering Experiments}

\medskip
 There currently exists two different hard--scattering experiments that may
show
the phenomena of color transparency. The first type of
experiments involve wide
angle hard hadron-nucleus scattering, the second lepton-nucleus scattering. The
BNL pA
experiment and the SLAC eA experiment fall under these categories.
The phenomenon of color
transparency  apparently
has been already observed in at least one of these types of experiments but
requires further confirmation.

The hard scattering experiments require
$s>>1$ GeV$^2$ and $Q^2>>1$ GeV$^2$ with $s/Q^2$ of order 1, where $\sqrt{s}$
is
the center of mass energy and $Q^2$ is the momentum transfer. The impulse
approximation applies to these experiments and therefore it should
be possible to factorize the hard and soft scattering kernels for such
processes.

\bigskip
\noindent
{\it BNL: a brief review}

\bigskip
The pioneering color transparency experiment involving hard scattering
was done at BNL by the group of Carroll et al. \cite{CEA88}.
A proton beam was used to study
$pA\rightarrow p'p''(A-1)$,
(Fig. (3.1)). Data was taken simultaneously on 6 targets,
namely $^1H$(in the form of polyethylene),
 $^7$Li, $^{12}$C, $^{27}$Al, $^{65}$Cu
and $^{198}$Pb. The beam energies were 6, 10, and 12 GeV, although the data
was reported at 12 GeV only for Al and Cu. The data was measured at a
quasi-elastic
point of 90$^o$ cm scattering. (At this point the data
depend only on one scale $Q^2 = -t = -u \sim s/2$.) The spectrometers
were set at the correct angle and energy for an elastic collision, within
the resolution of the instruments, and a veto was placed on the nuclear
targets to exclude events where a large amount of inelastic energy was
deposited. In practice any events in which pions might have been produced
were substantially reduced. Nevertheless one could not determine
what excited nuclear state was left behind. The four--momentum of
one of the final state protons
and the track direction of the second proton were measured.
The Fermi momentum of the target proton was then deduced from the kinematic
information available by assuming that the
event was a two body elastic collsion. Fig. (3.2) shows the projection of the
nucleon momentum distribution along the direction perpendicular to the
two body nucleon-nucleon scattering plane \cite{CEA88,H90}.
 The incident momentum is 6 GeV for
Li, C, Al, Cu and Pb and 10 GeV for Al.

Fermi motion of the target nucleon inside
the nucleus makes the experimental extraction of transparency
rather
complicated.
It  introduces a separate kinematic variable which has to be deduced from the
available experimental information and assumptions involving elastic
kinematics.
To give
an idea of the magnitude of the
effect we compute the shift in cm energy-squared $s$ due to Fermi motion
of the target nucleon in BNL kinematics. Let $p'$ be the four momentum of a
proton at rest.
Including Fermi momentum $k_F$, the initial momentum of the struck proton is
$p'+k_F$. The cm energy-squared for a collision with a beam proton carrying
momentum $p_b$ is given by,
$$ s=(p_b+p'+k_F)^2 = 2m^2+2 E_bm - 2p_bk_F\ .$$
where $m$ is the mass of the proton. Taking $p_b$ equal to 10 GeV and $k_F$
equal to 0.3 GeV the change in $s$ from its value if the target nucleus were
at rest is about 3 GeV$^2$.
A small change in the value of $s$ can make a big
difference
in the cross section. An experimental cure for this problem is to
measure enough variables to determine the Fermi momentum of each struck
proton, which was done in the BNL experiment. Thus Figs. (3.3) and (3.5)
show
data for values of the effective beam energy which range around the values
actually used, as the Fermi motion effects were incorporated by the
experimentalists. A weakness in the procedure is that one must know the Fermi
momentum distribution: the BNL experiment used its own measured out-of-plane
momentum distributions as well as standard models for this.
Needless to say, the role of the Fermi motion may be more subtle than this
and it has received much theoretical attention.

 The experiment introduced and
reported a ratio $T(Q^2,A)$ called the ``transparency ratio", defined by
\beq
T = {(d\sigma/dt)(pA\rightarrow p'p^{''}(A-1)\over Z(d\sigma/dt)(pp\rightarrow
p'p^{''})}
\eeq
The experimentally measured values of $T$ as a function of incident momentum
are shown in fig. (3.3).
The reason for reporting the transparency ratio is that certain systematic
experimental uncertainties in the data cancel out.

It is often loosely
said that the free space $pp\rightarrow pp$ scattering in the denominator
scales like $s^{-10}$. The transparency ratio, then, conveniently takes out a
very
rapidly varying function of the energy. The factor of $1/Z$ represents the
expectation that the hard scattering is incoherent: the only interference that
should occur at large momentum transfer comes from one proton being struck
at a time. The transparency ratio is an observable quantity which takes
into account a basic idea of factorization, which roughly means that the
hard scattering effects multiply the soft nuclear interaction effects. However,
the transparency ratio is tricky to interpret and other observables can be
constructed.
This is discussed in more detail later.

\bigskip
\noindent
{\it SLAC NE18: a brief review}

\bigskip
The second experiment involving hard scattering was performed at SLAC and
involved quasi-elastic electron scattering on nuclear targets as
shown in fig. (3.4)\cite{MEA94,OEA94}.
In this case five
nuclear targets $^1H$, $^2D$, $^{12}C$, $^{56}Fe$ and $^{197}Au$ were used.
The momentum
transfers reported were $Q^2 = 1,3,5$ and 6.8 (GeV/c)$^2$.
The experimentalists reported a transparency ratio which is defined slightly
differently
than in Eq (3.1). We denote this transparency ratio as
$T'$. It was defined as
\beq
T' = {\sum_{\cal R} N_{data}\over \sum_{\cal R} N_{PWIA}}
\eeq
where $N$ refers to the number of events in the kinematic region $\cal R$
and PWIA refers to the theoretical calculation using the plane wave impulse
approximation.
The region $\cal R$ for the SLAC experiment was chosen to
be ${\cal R}: [-25 <E_m <100 {\rm MeV}; 0 <p_m <250 {\rm MeV/c}$], where $E_m$
and
$p_m$ are the missing energy and momentum due to the Fermi motion of
the target nucleon inside the nucleus. In the plane wave impulse approximation
the proton is ejected from the nucleus without final state interactions by
exchange of a virtual photon with the incident electron. It therefore
corresponds to treating the target protons as being free.
The experimental results for the transparency ratio $T$
are shown in figs. (3.5) and (3.6). Fig. (3.6) shows the comparison of the
results for Carbon nuclei with different theoretical
predictions.
The transparency ratio was found to be roughly flat with a very slow
increase with energy for
all the nuclei studied by NE18. The Fermi momentum distribution extracted
by the NE18 group is shown in fig. (3.7).

\bigskip
\noindent
{\it  phenomenological remarks}

There have been many theoretical studies to predict or explain the
results of color transparency experiments involving hard scattering.
The rise and fall of the transparency ratio with increasing energy, seen
by the BNL experiment, apparently
contradicted the expectations of color transparency
as well as of the Glauber model \cite{CEA88,FLF88,LM92}.
After some inital confusion it became clear
that some of the observed structure was caused by oscillations with energy in
the denominater of Eq. (3.1).
This behavior of the transparency ratio is
understandable \cite{BT88,RP88}
if there are long distance contributions to the exclusive
wide angle $pp$ elastic scattering in free space. In the Brodsky-de
Teramond \cite{BT88} (BdT) model the long distance contributions are postulated
to
arise due to several charmed dibaryon resonances. These resonant contributions
are filtered by the nuclear medium leading to the observed
bump in the transparency ratio.
Ralston and Pire \cite{RP88} instead
 argued that exclusive processes generically
get considerable contributions from long distance components, an example
being given by the independent scattering process \cite{L74}.
 Interference of these long
distance amplitudes with short distance contributions can explain
the
oscillations seen in the free space exclusive $pp$ wide angle elastic
scattering cross section. The long distance processes will be absent in the
nuclear medium
due to
 ``nuclear filtering".
 With oscillations absent in the nuclear process, the theory predicts
oscillations in the transparency ratio $180^o$ out of phase with
the free space oscillations \cite{RP88}. Future experiments
should be able to clearly distinguish between these two explanations of the
BNL data.

Farrar et al. \cite{FLF88} have argued that at the hard scattering
a wave packet of very small spatial
dimensions is formed. The origin of this wave packet is ascribed to the hard
scattering.
Such a wavepacket is not an eigenstate of the strong
interaction Hamiltonian and should expand during its propagation
through the nucleus. The authors use two semiclassical models of this expansion
namely i) the naive model in which the transverse separation increases
linearly with time and ii) a model in which the expansion goes like the square
root of the time.
The results of Farrar et al are shown in Fig. (3.8) for the BNL
experiment. In this paper and
subsequent work no mention was made of the oscillations in the denominator of
the transparency ratio. Nevertheless, data for the transparency ratio was fit
to a model of the nuclear effects in the numerator of the ratio, thereby
leading
to a monotonically increasing transparency ratio.
The agreement with experimental results shown in Fig. (3.3), in which the
theory predicts
a dramatic decrease beyond E$_{\rm beam}=12$ GeV,  is clearly very poor. The
same is also true for
the predictions for the SLAC experiment as shown in Fig. (3.6). Other
consequences of the idea is presented in several
reviews\cite{FMS92,FMS93,FMS94} of this approach.

Exploring the effects of the denominator in the transparency ratio,
 Hepplemann \cite{H90}
presented a plot of the numerator, that is the cross section for
$pA\rightarrow p'p^{''}(A-1)$ (up to a normalization), a result shown in Fig.
(3.9).
Due to the experimental
procedure this was obtained by multiplying $s^{10}T$ by the denominator in Eq.
(3.1). The results are a reasonably flat $s$--dependence, supporting the idea
that whatever causes the oscillations is filtered away in the nuclear target.

Jennings and Miller \cite{JM90,JM91,JM921,JM922,JM93,JM94}
 have given detailed treatments of modelling the
expansion of the wavepacket, assuming it begins from a compressed initial
state.
They use non-relativistic quantum mechanical
models and also attempt a less model dependent treatment in which their
point--like wavepacket is expressed in terms of experimentally
observed hadronic states. They calculate the transparency ratio using
several assumptions such as the  BdT\cite{BT88} and the RP
\cite{RP88} mechanism. They find that their calculations do not
agree with the results of BNL experiment with \cite{JM921,JM922}
and without \cite{JM91} the inclusion
of BdT  or RP mechanism. In reference \citenum{JM93},
however, by inclusion
of a more careful treatment of Fermi motion and of the longitudinal component
of the momentum of the detected proton, the authors found agreement
with the BNL data.

Similar conclusion was also obtained by
Anisovich et al \cite{ADN92} who used the RP \cite{RP88} mechanism
within a quark-diquark model of proton to calculate color transparency. They
find that the BNL results are consistent with color transparency.

Kopeliovich, Nikolaev, Zakharov and collaborators
\cite{KZ911,KZ912,K92,NNZ94}
have also used their quantum mechanical model
to calculate color transparency, with and without the inclusion of the
interference of long distance components. Using theory technology adapted from
the diffractive limit, Kopeliovich and Zakharov \cite{KZ912}
argue that their
calculations are unable to reproduce the BNL results.
In a
recent review of their approach, Nikolaev \cite{N94} favors the
RP explanation but argues that the experiment is kinematically
flawed in coupling $Q^2$ and energy at fixed--angle kinematics. (From the hard
scattering point of view, this flaw is a virtue.)
Moreover, considerable latitude in varying these independently exists in the
new
BNL
experiment due to more complete kinematic measurements.
Nikolaev et al \cite{NSS941}
 also calculate the transparency ratio for the SLAC NE18 electroproduction
experiment within their quantum mechanical model. They find that the onset
of color transparency is very slow in their model in agreement with the
NE18 results. The authors argue that this slow onset of color transparency
is obtained because of the large number of excited states needed to produce
wave packets of small transverse size.

There appears to be a general agreement among the theorists
\cite{BT88,RP88,ADN92,JM93,KYS95}  that
a proper understanding of the BNL results involves some interfering long
wavelength contributions to the free space process, which
are filtered in the nuclear medium. If
this is the case then the hard scattering in nuclear medium cannot safely be
assumed
to be the same as the hard scattering in free space. It is
reasonable to
introduce a new method for analysing color transparency experiments in which
the hard scattering rate can be  treated as an adjustable parameter along
with the attenuation cross section \cite{JR931}.
This procedure has been applied to the BNL
experimental data; details will be given in Section 3.4 below. The best fit
requires that the hard
scattering rate be significantly different from the corresponding rate in
free space and also that the attenuation cross section is considerably smaller
than 40 mb expected due to a Glauber model \cite{JR931,JR932}. The
nuclear medium hard scattering rate extracted by this analysis is in much
better agreement with the rate expected from short distance pQCD predictions of
Brodsky and Lepage \cite{LB80},
compared to the corresponding rate in  free space scattering.
The attenuation cross section decreases as $1/Q^2$
again in agreement with QCD predictions. Based on this analysis of the BNL
data, the transparency ratio for electroproduction in the SLAC experiment was
also studied. In this case use of the pQCD short distance model for the hard
scattering in the nuclear target is again in agreement with the experiment.

There exist several studies which attempt to better understand the role of
Fermi motion in color transparency experiments.
It has been claimed by Jennings and Kopeliovich \cite{JK93}
and by Bianconi, Boffi
and Kharzeev \cite{BBK932} that in electroproduction at the elastic point,
the phenomenon of color transparency is possible only
due to Fermi
motion. However Nikolaev et al \cite{NSS93} do not agree with this
and show that the effects of Fermi motion are much weaker.
 We find no theoretical support for stating that color transparency requires
Fermi motion, and will come back to this point in Section 5.

Since the Fermi motion effects are controversial,
it is clearly useful to reduce the dependence of the observed transparency
signal on the Fermi motion. Such an attempt has been pursued by
   Frankel, Frati and Walet \cite{FFW94} who suggest an experiment
in which color transparency effects are claimed to be separated from nuclear
structure effects. The authors give a definition of
the transparency ratio $T$ which is argued to be independent of the spectral
function.

\bigskip
\noindent
\subsection{ Color Transparency in Diffractive Scattering}

We next turn to another
very interesting set of color transparency
experiments which
involve diffractive scattering. Within this class there are both low $Q^2$
experiments and those with  transverse momenta of final state hadrons
larger than a GeV.  These experiments
are expected to yield important information about
the interface of perturbative and non-perturbative QCD.
A basic process of interest involves the production of quark antiquark pair by
a high energy virtual photon.
The $q\bar q$ pair eventually converts into
a vector meson which could be either $\rho$, $\rho'$, $\psi$, $\psi'$ etc.
The kinematics of this process is shown in Fig. (3.10). The incident
muon with four momentum $k$ emits a virtual photon with
momentum $q$ which eventually forms a vector meson with momentum $r$ with
exchange of a momentum $t$ with the nucleus. The momentum of the scattered
muon is denoted by $k'$. The relevant kinematic variables for this process
are the invariant mass squared of the photon $-Q^2=(k-k')^2$,
the energy lost by the muon in the laboratory frame
$\nu=p\cdot q/M$, the four momentum transfer squared between the
vector meson and the target proton $t=(q-r)^2$ and the fraction of the
energy lost by the incident muon that is carried by the vector meson
 $z=E_\rho/\nu$. Here $E_\rho$ is the energy of the vector meson and $M$
is the proton mass.
The experimentalists also define \cite{AEA95}
 $t'=t-t_{\rm min}$ where $|t_{\rm min}|$
is the minimum value of $|t|$ allowed by the kinematics, which corresponds
to the limit of zero angle between the momentum of the virtual photon $q$
and the vector meson $r$.

Suppose $Q^2>$ a few GeV$^2$. The initial state is domnated by a  $q\bar q$
pair  produced with a transverse separation of about $1/Q$, which is much
smaller
than
a typical hadronic scale. The resulting meson is expected to exhibit color
transparency in the nuclear medium \cite{BM88,BFG94}.
An important consideration in such experiment is the production time
for the $q\bar q$ pair and the time of
formation of the vector meson.
In light cone coordinates the production time $\tau_p$ is given
by $x^+\approx 1/Q^-\approx E_\rho/(Q^2+m_V^2)$, where
 $m_V$ is the mass of the vector meson.
We note that at the kinematic conditions of E665 experiment where
$E_\rho$ was of the order of 100 GeV and $Q^2$ of the order of 1 GeV$^2$,
$\tau_p$ is extremely large. For a wide range of kinematic variables
it is vastly larger than the nuclear diameter.
The time of formation $\tau_F$ of the vector meson
  also turns out to be much larger in comparison to  nuclear dimensions.
  It is roughly equal to the time that the $q\bar q$ pair takes
to reach a transverse separation equal to radius of the hadron $r_H$. This
implies that $\tau_F \approx r_H/v_T$, where $v_T$ is the transverse
velocity of the quark. The transverse velocity is of the order of
$k_T/E_\rho$, where $k_T$ is the transverse momentum and $E$ is the energy
of the hadron. We then find that the formation time is of the order of
$r_HE_\rho/k_T$, which for large $E_\rho$
is considerably larger than even the largest nuclei. Due to these long time
scales, the impulse approximation does not apply. However, the ``frozen
approximation" is reasonably invoked because time evolution is slowed by the
Lorentz boost.
Because of the large formation time the
$q\bar q$ pair is expected to escape
the nuclear medium well before it forms the hadron. This implies that
the state that traverses the nucleus is best thought of as a well localized
$q\bar q$
pair and not an ordinary hadron.

\medskip
\noindent
{\it FNAL E665: A Brief Review}
\medskip

Depending on the value of $t'$
the $q\bar q$ pair can be produced coherently or incoherently from a
nuclear target. In fig. (3.11), we show the
graph from the E665 experiment, Ref. \citenum{AEA95}, which shows the $t'$
distribution of events. The coherent
and incoherent production regions are clearly separable in the curve for
Calcium.
For the events that correspond to coherent scattering the transparency
ratio is expected to go like $A^{1/3}$ in the limit of infinite $Q^2$
when the nucleus is transparent to a point like color singlet object,
since the scattering is now taking place along the entire length of the
nucleus. Again, since the scattering is coherent over the entire
nucleus the impulse approximation is not applicable and the process is not
factorizable in terms of hard and soft scattering kernels. The interpretation
of these experiments is considerably different from the hard scattering
case. For the case of incoherent
scattering the transparency ratio is expected to
go to 1 in the high energy limit.

The results of the  E665 experiment for the case of $\rho$ mesons are
shown in Figs. (3.12-3.14).
The typical values of $Q^2$ reported by the E665 group are of the order of
0.1 GeV$^2$ to about 10 GeV$^2$. The conditions for
color transparency are generally expected to hold only for $Q^2$
greater than about 1 GeV$^2$ but some authors argue otherwise.
 The beam energy is extremely large, 470 GeV. The momentum transfer to the
nucleus, $-t$, ranges from about 0.1 GeV$^2$ to about 1 GeV$^2$.
The experiment observed a rise in the transparency ratio with $Q^2$ as shown in
fig. (3.13),
in the case of coherent scattering and in fig. (3.12) for the
case of incoherent scattering, providing evidence for reduced attenuation
with
increasing $Q^2$.

The $Q^2$ dependence of diffractive $\rho $ production was predicted in
advance of the experiment by
Kopeliovich et al \cite{KNN93}. The authors
  calculate the transparency ratio using the wave function $|\gamma^*>$
of the $q\bar q$ fluctuation of virtual photons. This wave function
was derived in Ref. \citenum{NZ92} in the mixed $( b , z)$ representation,
where $ b $ is the transverse size of the hadron and $z$ is a fraction of the
photon's light-cone momentum carried by the quark, $0<z<1$. The authors
argue that the important quantity that enters the calculation of
transparency ratio is the product $\sigma( b )|\gamma^*>$ which
peaks roughly at $2 b _Q$, where $ b _Q\approx 2/\sqrt{Q^2+m_v^2}$
is the size of the relevant
$q\bar q$ fluctuation and $\sigma( b )$ is its interaction cross section.
This implies that the process measures the wave function at $2 b _Q$
and therefore can be used to scan the wave function.
In the calculation final state diffractive scattering leads
to considerable off diagonal contributions.

The authors give
predictions for the transparency ratio for $\rho$, $\rho'$, $J/\psi$,
$\psi'$, $\Upsilon$ and $\Upsilon'$ states. In contrast
to their predictions for $\rho'$, no significant difference in the transparency
ratio was observed between $\rho$ and $\rho'$
in results presented so far \cite{F94}.

In diffractive scattering one probes an end point region with
$Q^2$ of the order of 1 GeV$^2$ and $s>>Q^2$,
at the interface of perturbative and
non-perturbative regimes. For a wide range of kinematic variables,
factorization
of hard and soft scattering amplitude breaks down, and perturbative QCD is
not applicable in a straightforward way.
Experimental cuts were set to suppress
events in which the photon produced other fast particles in the final state
besides the $\rho$ meson.
 Moreover the experiment did not exclude the possibility of peripheral pion
production. Strictly speaking, there is no guarantee of the exclusive
kinematics
typically necessary to observe color transparency. A further complication is
that in this
end point region the process is not necessarily dominated by any short distance
part of the hadron wave function. Nevertheless,
the process showed a  very interesting rise with $Q^2$ of the transparency
ratio which can be interpreted as evidence for color
transparency. This may be an indication that the extremely
tight kinematic cuts sometimes felt necessary for observing color
transparency are overly  restrictive.

\noindent
\subsection{ Experimental Analysis Procedures}
\bigskip
\noindent
{\bf Ratio method}
\bigskip

 There exists currently two methods to test for color
transparency. The ratio method, which is still widely used, was the
original method introduced in the BNL experiment. The signal for color
transparency
was expected to be a
monotonic rise in T with energy  such that it
asymptotically approaches 1.
This expectation is based on the assumption that in free space the hard
scattering is dominated by the short distance components of the hadron
wave function. If this is true, and if the hard scattering
in the nuclear medium is
the same as the free space scattering and further if the nucleus is indeed
transparent
to incoming and outgoing protons, the nuclear scattering should be
identical to free space scattering. The data from the BNL experiment, shown in
fig. (3.3),
disagree with this expectation and showed a rise and fall of the
transparency ratio with increasing energy rather than showing a monotonic
increase.

This anomalous behavior of the transparency ratio at first caused a great stir.
By now it is clear that the structure is due to the process of making
a division by the free-space $pp\rightarrow pp$ data. On a log--log plot, that
data falls roughly
like $s^{-10}$, Fig. (3.15),  but also are modulated by oscillations with the
logarithm of
the energy (Fig. (3.16a)). The oscillations are rather clear evidence for
quantum
mechanical interference of amplitudes. Apparently the numerator of the
transparency ratio - which is the nuclear scattering - has little, if any,
oscillations, so that the ratio must oscillate. There is good evidence for
this:
consulting Fig. (3.16b), the rise in the transparency ratio is exactly
correlated
with a falling oscillations of the free space data, indicating that the
nuclear target is showing reduced oscillations. Making a plot of the nuclear
differential cross section by itself \cite{H90}, with no division by the free
space
denominator but taking out a factor of $s^{-10}$, one sees that the
oscillations, if present at all, are much reduced in the nuclear target
Fig. (3.9).

\medskip
\noindent {\bf Evidence for Filtering}

\medskip
We have strong reasons for believing the oscillations in the free
space data are understood as part of a QCD phenomenon studied for more than 10
years.  But temporarily setting aside the explanations proposed for the
oscillations, {\it if} there is interference {\it then} the free-space
$pp\to pp$ amplitude has
not settled into its asymptotic form.   It must be different from the processes
occurring inside the nucleus, which do not show the same oscillations. Model
independently,
something interfering in the free space $pp\to pp$ process has apparently been
killed in the nuclear target.  There are two explanations proposed for this:

Brodsky and de Teramond \cite{BT88} propose that there are several charmed
dibaryon
resonances in the region of interest, which are eliminated in the nuclear
case.
The states have not been observed elsewhere, but the value of the threshold
kinematics do coincide with calculations based on charmed quark masses.  By
adjusting several
parameters the shape of the differential cross section and the spin analyzing
power $A$ were reproduced.  Presumably the charmed states could be observed to
test the idea.  It could also be applied to other reactions.

Ralston and Pire \cite{RP88}
 proposed that the free space $pp\to pp$ scattering
consists of
roughly two perturbative QCD regions, which have an energy dependent
``chromo-coulomb" phase difference \cite{PR82}.
One of the regions corresponds to
``large'' quark configurations which would be filtered away in the nuclear
medium,
leading to disappearance of the oscillations in the nucleus.  By proposing that
the oscillations are reduced for A$>>$1 the energy dependence of the BNL
transparency ratio is reproduced with no free parameters.  The fact that such
energy dependent phases occur in QCD has been confirmed by Sen \cite{S83} and
Botts and
Sterman \cite{BS89,B91}.  Spin observables in this model of the free-space
scattering have
been fit by Ramsay and Sivers \cite{RS92}, and Carlson, Myrher and
Chashkuhnashvilli \cite{CCM92}.  However, the fundamental parameters needed to
reproduce the data have not yet been obtained from the (quite complicated)
pQCD calculations. For example, the calculation of Ref. \citenum{BS89}
obtained an oscillation with too slow an energy dependence to fit the data.
This is not too surprising, since the calculation was not set up to search for
more rapidly varying phases, leaving open the possibility that the oscillations
might still be calculable from first principles.

The fact that there are ``large" configurations in the hard scattering
description
of pQCD was not widely appreciated until recently. Such processes, which
involve independent scattering of quarks at separated scattering centers,
simply do not obey the assumptions of the ``basic idea" presented earlier.
Nevertheless, independent scattering occurs in the perturbation theory and must
be dealt with to understand color transparency.   Both of the explanations
above agree on the proposal that the structure in the transparency ratio occurs
due to dividing a smoothly varying, somehow filtered  numerator by an
oscillating denominator.

\smallskip
\noindent {\bf A Crossroad}

\medskip
Conceptually the appearance of an oscillating transparency
ratio brings us to a crossroad. One learned that the experimentally attractive
transparency ratio was not a reliable observable to measure color transparency
on its own. What else might be used?

We know that the experiment on the nuclear target measures a
combination of the hard scattering rate and the nuclear attenuation.  If one
honestly admits that the nuclear hard scattering rates are unknown,  how can
one
extract a signal of color transparency from the nuclear attenuation?  Suppose,
for example, that one overestimated the nuclear hard scattering rate- could
this simply be compensated by a corresponding underestimate of the nuclear
transparency? This ambiguity is not just a problem with the
proton initiated processes but also, although less well appreciated,  a
difficulty with electroproduction experiments.

\medskip

\bigskip
\noindent
{\bf Attenuation Method}
\bigskip

  This method \cite{JR931,JR932} is designed to test color transparency
without making the assumption that the hard scattering in the nuclear medium
is the same as the corresponding free space scattering. Instead, the nuclear
hard
scattering rate
is extracted directly from data.

The attenuation method is based on the observation
that the slope of the curve of transparency ratio T vs. nuclear A contains
information about the nuclear attenuation and can be used to extract an
effective attenuation cross section.
To understand how this is done, consider measuring experimentally an effective
absorption cross section $\sigma_{eff}$ in some material. If the test beam flux
is known, this is easy and can be done by ratios. But if the test beam flux is
unknown -- the case of undetermined hard scattering rates in the nuclear
medium -- then a ratio is not definitive. One can, however, study the
absorption as thicker and thicker targets are used, and fit $\sigma_{eff}$ from
the target thickness dependence. This is the basic idea of the attenuation
method.

Since the target thickness is set by the nuclear number $A$ in color
transparency, one should extract $\sigma_{eff}$ from the $A$ dependence at
fixed kinematic conditions ($Q^2$ and energy). This can be done rather model
independently. In the fit there is an overall normalization, proportional to
the hard scattering rate. There is a separate variation in the shape of the
curve with $A$ (e.g. the slope on a log--log plot). The shape of the $A$
dependence is fit by varying $\sigma_{eff}$, while the hard
scattering rate determines the overall normalization. These two independent
observables are easily separated as a function of two ``probe" variables $Q^2$
and $A$.

To simplify the analysis, we make a few limited assumptions, which can be
relaxed if needed. There must be an assumption that the observed process can be
factored in to hard and soft interactions. For $A >>1$ any variation of the
hard
scattering rate with $A$ after filtering should be negligible compared to the
rapid $A$ dependence of the soft attenuation process. Finally, Fermi motion
effects are modeled as necessary.

Let us illustrate
this in more detail.  The impulse approximation can be used to
separate mathematically the hard scattering and the process of propagation
through the nuclear medium:
\beq
measured\ rate (Q^2,A) = [scattering\ rate\ (Q^2)][survival\
probability\ (\sigma_{eff},A)]
\eeq
Here by ``measured rate'' we mean data for either a cross section or a
transparency ratio. If one studies the transparency ratio, then ``scattering
rate'' is actually the ratio of the hard scattering rates in the nuclear
target to the analogous free-space scattering. On the other hand, if (3.3) is
applied to a
cross section on a nucleus then ``scattering rate'' actually equals the
hard scattering rate in the nuclear target.

We next observe that the $A$
dependence
can be experimentally isolated by taking the logarithm of Eq. (3.3)
at fixed $Q^2= Q_0^2$:
\beq
 log[measured\ rate (Q^2,A)] =  log[survival\
probability]+const (Q^2_0)\ .
\eeq
Here {\it const}$(Q_0^2)$ is a term independent of $A$: it comes from the hard
scattering rate.
Whatever the $const$, the attenuation cross section $\sigma_{\rm eff}$ is the
parameter which
determines the shape of the survival probability
as $A$ is varied. Thus to determine $\sigma_{\rm eff}$ we want a fit to the
shape of the $A$ dependence of the right-hand side, but a fit only up to
an additive constant. In practise this is quite easy. For example,
plot the measured rate as a function of $A$ on log-scale paper: adding an
overall constant is just a rigid translation of the fit up or down on the
paper.

Relating the shape of the survival probability to $\sigma_{\rm eff}$ requires
a theoretical model. If one uses a simple exponential attenuation model this
dependence is quite explicit and straightforward. If this fits the data set
well
then one effective cross section $\sigma_{\rm eff}$ at each $Q_0^2$ seems
to be all that is observable with the data set. Other models are possible:
for a huge class of models of the survival probability the attenuation
procedure applies
just the same.

After determining the survival probability, the $const\; (Q^2$) values contain
the
information about the hard scattering rate. The interpretation depends on
whether (3.3) is applied to the transparency ratio or a cross section on the
nuclear target. If ``measured data'' is a cross section then one finds the
scattering rate in the nuclear target directly. No normalization needs to be
set using isolated hadron events. If there is nuclear filtering we expect the
empirically determined hard
scattering rate generally will disagree with the scattering rate of free
space data.

This method was applied to the BNL data. The results provide evidence for
observation of color transparency and nuclear
filtering.\cite{JR931,JR932} The $A$ dependence of the data at different fixed
$Q^2$ is shown in
Fig. (3.17). The
data was analysed by extracting the cross section in the nuclear target by
multiplying the transparency ratio by the known free space cross section.
This yields the nuclear cross section divided by the number of protons in the
nucleus Z. By assuming some factorization, this quantity can be written
as
\beq
s^{10}{d\sigma_A/dt \over Z} = N(Q^2) P_{ppp}(\sigma_{\rm eff}(Q^2),A)\ .
\eeq
The factor $s^{10}$ and $1/Z$ are inserted to take out the typical order of
magnitude
of the cross section; their usage introduces no bias. The factor
$P_{ppp}$, which is the survival probability, contains information
about the attenuation of the three protons that cross the nucleus in
$pA\rightarrow p'p^{''} (A-1)$. The factor $N(Q^2)$
contains information about the hard scattering. In equation (3.3) we have made
the reasonable assumption that once $A>>1$ then $N(Q^2)$ is independent of
$A$. In other words, the hard scattering rate saturates once $A$ is large
enough. We return to this point below.

To proceed further one must model the survival probability. The simplest model
assumes that protons are exponentially attenuated in their passage
through the nucleus. Thus the survival probability of a proton after
propagating a distance $z$ through nucleus is equal to ${\rm
exp}(-\rho\sigma_{\rm eff}z)$.
The survival probability for the three protons
can then be easily modelled by a Monte Carlo event generator by treating the
$\sigma_{\rm eff}$ as a free parameter. Using the
Monte Carlo calculation the data were fit to determine $\sigma_{\rm eff}$
and $N(Q^2)$ for each value of $Q^2$ at which the data is available.
The resulting fits for the BNL data were found to be:

\begin{center}
\begin{tabular}{ccll}
E &  $\sigma_{eff}(E)$ &\hskip .25in $N(E)$ &\ \ \ $\chi^2$ \\
 & & & \\
6 GeV &  17$\pm$2 mb &  $(5.4\pm 0.4)\zeta$ &  0.28 \\
10 GeV &  12$\pm$2 mb & $(3.3\pm0.4)\zeta$ & 0.53 \\
\end{tabular}
\end{center}

\noindent
where $\chi^2 = \Sigma[(y_i - d_i)/\Delta d_i]^2/N$, $d_i$ are the $N$ data
points,
$\Delta d_i$ is the error in $d_i$  and $y_i$ are the theoretical values
calculated using the Monte-Carlo,
 and $\zeta = 5.2\times10^7$ mb GeV$^{18}$ is a constant
containing the  overall normalization of the  cross section.
The $Q^2$ values for the two beam energies 6 and 10 GeV are 4.8 GeV$^2$ and
8.5 GeV$^2$ respectively.
The fit to the data (Fig. (3.18a,b)) clearly shows that $\sigma_{eff}$
decreases with
energy and is considerably lower than a typical inelastic
cross section of $36\ mb$.

Let us return to the issue of saturation of the hard scattering rate for
$A>>1$. It should be clear that the method is very general, and can easily be
modified to model the $A$--dependence of hard scattering due to filtering.
We will not do that here, but instead look to the data to estimate small--$A$
effects.
 As remarked in Ref. \citenum{JR931}, the effects on
the fit of discarding the Li data point are negligible. We regard the issue of
whether the data point for C might be discarded as an empirical question. It
should be noted that even with the choice to discard $A<26$, the last 3 data
points in Figs. (3.18a,b) are sufficient to eliminate a 30--36 mb absorption
cross section as a contender. The Attenuation method is thus considerably more
definitive than the Ratio method, and a standard ``Glauber Model" has been
ruled
out by the $A$ dependence at each fixed $Q^2$.

Having fit the data's A dependence, the fit can be examined at fixed A to find
the survival factor as a  function of $Q^2$.  The BNL data was taken
only at two energies for a good range of A, so  $N(Q^2)$ and
$\sigma_{eff}(Q^2$) were found at
two points. However, for the Aluminum target the data at intermediate
energies was also given.
One can make a function for $N(Q^2)$ which interpolates with $Q^2$
smoothly between the two endpoints where $Q^2$ was reported. It is given by:
\beq
N(Q^2) = {5.4 \zeta\over  (Q^2/4.8 GeV^2)^{0.86}}\ \ .
\eeq

Then inverting (3.5) one can determine the survival factor $P_{ppp}$ as
well as $\sigma_{eff}$ as a function of $Q^2$ by using the data for
Aluminum at intermediate energies. The
resulting best consistent fit for $P_{ppp}$ is shown in Fig. (3.19).
  The corresponding $\sigma_{eff}$($Q^2$) is shown  in Fig. (3.20). The
survival factor would be flat with $Q^2$ if one used  a traditional Glauber
model.
The fact that the survival probability rises with $Q^2$ is clear evidence for
color transparency.

A final consistency check involves looking at the the normalization factor.
After taking out the  nuclear attenuation effects, according to the
perturbative treatment the $Q^2$ dependence of $N(Q^2)$ is  due to the hard
scattering process.  It was found that $N(Q^2)$ decreases relative
to $s^{-10}$,
meaning  that the hard scattering rate in the nuclear target is decreasing
faster than the naive quark counting  model prediction one would use at short
distance after nuclear filtering.  In perturbative QCD,
however, the quark-counting  prediction is modified,  due  to the running
coupling $\alpha_s(Q^2/\Lambda^2_{qcd})$ and scaling of distribution
amplitudes.  The short distance
QCD prediction goes like $\alpha_s^{10}$ because there are five gluons in the
amplitude:
\beq
d\sigma/dt_{pQCD} \cong
(\alpha_s(Q^2/\Lambda_{qcd}^2))^{10}s^{-10}f(t/s)\eeq
It is interesting to compare the form including powers of $\alpha_s$ with
the hard
scattering rate in  the nuclear target. First, recall that the $Q^2$ dependence
of
$\alpha_s^{10}$ causes
serious disagreement with the data for  isolated $pp$ scattering in free space
with $\Lambda_{qcd} \approx$ 100 MeV for the quark--counting model.  We would
like to see whether the nuclear target has filtered the events down to
something
like the shortest distance component.  Although (3.7) may appear naive, it is
 adequate because of  the usual ambiguity in the choice of scale.  One can
generate a range of reasonable theoretical  predictions by
choosing the $Q^2$ scale
of $\alpha_s$ in a typical range $(-t/2)\ <\ Q^2\ <\ (-1.5\ t)$.
  To improve on this
theoretically requires a calculation of next to leading logarithms.

\medskip

For comparison with the nuclear target, the ratio of the global
$s^{-10}N(Q^2)$ to the pQCD  predictions Eq. (3.7) was considered. The results
are
plotted
as solid lines in
Fig. (3.21). The asymptotic prediction \cite{LB80}
is also  illustrated as a dotted line.
It falls within the region of scale ambiguity of the pQCD
predictions.  For comparison we also show s$^{-9.7}$, a dashed line which is
well outside the
range of the perturbative  predictions.  The solid lines of
$s^{-10}N(Q^2)/(d\sigma/dt_{pQCD}(Q^2)$)
are rather flat, showing good  agreement of the BNL nuclear hard scattering
data
with short distance QCD. For the Aluminum target we have  several data points
at several $Q^2$
which also fall fairly well within the band of perturbative  predictions.
These are rather spectacular results, but we emphasize that they should be
viewed with  caution because the BNL experiment was a pioneering one.  If they
are
confirmed by upcoming higher  precision data,  it will lead a great deal of
strength to the idea that QCD is cleaner after filtering in  a nuclear target.

\medskip
\noindent
{\bf A Scaling Law}
\medskip

We next discuss a scaling law \cite{PR911} predicted for the survival
probability. Because the scaling law predates the attenuation method,
it was first applied to the transparency ratio, under the assumption that
hard scattering rates might cancel out in electoproduction.  The scaling law
says that the
survival probability is a function of a dimensionless variable.
The important dimensionless variable that it can depend on is the
effective
number of nucleons encountered by the protons as they propagate
through the nucleus (Fig. (3.22)).
This is proportional to the length of the nucleus $(A^{1/3}Fm)$
times the nuclear density $n$ times the effective nucleon-nucleon
 cross section $\sigma_{eff}$. If the cross section goes
like $1/Q^2$, then the
survival probability is a function of the dimensionless
quantity $n A^{1/3}Fm/Q^2$.

The advantages of a scaling law in testing dynamical assumptions are very
great. Rather than testing values of ratios which can be fit with many
parameters in some model, the scaling law method allows an almost
model--independent check of the most basic assumptions.

To experimentally determine whether this law is satisfied or not,
consider the survival probability to be a function of
$Q^2/A^\alpha$ and determine $\alpha$ from the experimental data.
A basic complication is the
fact that the experimentally measured transparency ratio $T$
involves an unknown function $f(Q^2)$,
where $f(Q^2)$ is the ratio of the hard scattering in the nuclear
medium to the free space scattering. With a good data set over a range of
the $Q^2$ and $A$ plane one can nevertheless extract functional
forms for both $f(Q^2)$ and the survival probability $P$ by fitting
$T=f(Q^2)P(Q^2/A^\alpha)$. This is attractive because of its
model independence: no theory needs to be used to model $P$, which is
simply determined by the best fit.

The data from the BNL experiment is limited, available only at
two energies for several $A$. We therefore take a somewhat
different approach here to check the scaling law with this
data. We assume that the hard scattering inside the nuclear
medium satisfies the short distance pQCD predictions. This is
based on the idea of nuclear filtering; is supported by
the global fit, and by the fact that the cross section in the
nuclear target does not show any oscillations. We therefore set the
average hard
scattering cross section in Eq. (3.5) to be Z times the short distance
perturbative QCD prediction given by Eq. (3.7).
(In Eq. (3.7) we ignored the
anomalous dimension which is found to make a very small difference
in the results.)
We can then calculate
$f(Q^2)$ and take it out of the transparency ratio to yield
the survival probability.
The survival probabilities for three
different values of $\alpha$ are plotted in Fig. (3.22). The data
points are circles $(Q^2=4.8\ GeV^2)$, squares $(Q^2=8.5\ GeV^2)$
and crosses $(Q^2= 10.4\ GeV^2)$. It
is clear from the figure that $\alpha\ =\ 1/3$ is favored over other
values of $\alpha$.

(There remains one subtlety. Although the scaling law procedure is
quite explicit about the functional dependence on the two variables
$Q^2$ and $Q^2/A^\alpha$, the ansatz (1) has a symmetry: the data and
fit are unchanged under $f(Q^2)\rightarrow f(Q^2)/K$, $P(Q^2/A^\alpha)
\rightarrow KP(Q^2/A^\alpha)$, where $K$ is any constant.
The absolute
magnitudes of the survival probability (or the hard scattering)
and the effective attenuation cross section are therefore
not determined by this method.)

Finally we note that the method of looking for scaling relations does not make
any assumptions about the
free space
scattering. It only assumes a factorization in the nuclear medium. Therefore
it is applicable irrespective of the source of long distance contamination
in the free
space scattering.

The method has
been criticized by several authors
on the grounds that due to the time evolution of the hadron wavefunction as
it  propagate through the nuclear medium, the attenuation cross section
changes with time. These authors claim it is not safe to treat it as a constant
as was
done in the original application of this method. We must point
out that this is not a criticism of the method but of the
model used to calculate the nuclear attenuation. Other models are possible, and
might be used to predict other scaling laws to test these assertions. For a
crude example, if $\sigma_{eff}$ increases linearly with time, then one might
get an
additional $A^{1/3}$ dependence. Such models would be tested by scaling in
$Q^2/A^{2/3}$.
One can always
extract an average attenuation cross section $\sigma_{\rm eff}$; if
there is color
transparency
in the data, this information will be directly available by a plot of
$\sigma_{\rm eff}$
vs. $Q^2$. It would seem logical to explore such broad simple regularities in
theory and data before getting mired in the details of particular models.

To summarize, the BNL data yielded an attenuation cross section which was
consistent
with the theoretical expectation that it should decrease like $1/Q^2$. More
interestingly, the data yielded a hard scattering rate which is consistent
with the theoretical expectations from short distance pQCD.

\medskip
\noindent
{\bf Experimental Analysis of Electroproduction}
\medskip

We next discuss the electroproduction experiments on color transparency.
The NE18 group at SLAC has already published results\cite{MEA94,OEA94} of
measurements
of $eA\rightarrow e'p(A-1)$ on several nuclear targets with maximum
momentum transfer $Q^2$ being 6.8 GeV$^2$.
  There also exists a
proposal at CEBAF to perform a similar experiment with better statistics.
These kind of experiments are beautifully complementary to the hadron
beam experiments. The initial state of the large $Q^2$ virtual photon
is very clean and well understood: one is measuring the photon's absorption
and final state interactions in the nuclear target. There is tremendous
interest in this and other experiments which could measure meson
knockout, transition form factors, etc. A clean theoretical matrix
element is measured in the experiment:
$$<p,A-1|J^\mu_{\rm em}|A>$$
However, the experiment is somewhat insensitive to the attenuation cross
section of the system crossing the nucleus, so that comparatively high
precision data may be needed to extract an attenuation cross section.
This is shown in Fig. (3.23),
which shows some predicted survival probabilities
for several attenuation cross sections $\sigma_{\rm eff}$. Even with cross
sections varying by a factor of about three, the curves on the log-plot
are almost parallel. It follows that even for a substantially decreased
$\sigma_{\rm eff}$,
the effects of color transparency (if any) might be masked by a
simple change in the normalization.
We have already seen that such a change in normalization is equivalent to
a change in hard scattering rate, so it is tricky in electroproduction
to separate the ambiguities of hard scattering inside the nucleus from
the extraction of $\sigma_{eff}$.

\medskip
\noindent
{\it Qualitative Features of $A$--Dependence in Electroproduction}
\medskip

To understand this in more detail one can make an analytic calculation.
Assuming a uniform nucleon number density $n=1/6$ Fm$^3$ and nuclear
radius $R_A=1.2\ {\rm Fm}A^{1/3}$, one can find the survival probability
$P_p$ of one proton taken at random over the volume of a nucleus and crossing
on a straight line with exponential attenuation. It is a function of
$y=R_A/l$, where $l=1/(n\sigma_{eff})$ is the mean free path:
\beq
P_p(\sigma_{\rm eff}, A) = P_p(A^{1/3}\sigma_{\rm eff}) = {3\over 4y}
+{3\over 4y^2}e^{-2y} - {3\over 8y^3}(1-e^{-2y})
\eeq
This formula was used to generate Fig. (3.24). Even though there is an
integration over a sphere, this formula depends on only one variable
$y\sim A^{l/3}n\sigma_{eff}$. In the limit $l>>R_A$ we
have 100\% survival. In the limit of $l<<R_A$ the  problem is also quite
simple. Particles survive if they are knocked through no more than  thin skin
of depth $l$ on the ``back'' side of the struck nucleus. The volume containing
such survivors is $\pi R_A^2l$; the area $\pi R^2_A$ occurs because
we want the projected area along the beam direction. The survival
probability, in this limit, should be the ratio of the volume containing
survivors to the total volume:
$$P_p(l<<R_A) = {\pi R_A^2l\over 4\pi R_A^3/3} = {3\over 4n\sigma_{\rm eff}
A^{1/3}}={3\over 4y}\; .$$
Now the ambiguity in examining data in this case is brought out by considering
a transparency ratio
$$T(Q^2,A)\sim {N_A(Q^2)\over N_1(Q^2)}P_p = {3N_A(Q^2)\over 4nN_1(Q^2)}
{1\over A^{1/3}\sigma_{\rm eff}(Q^2)}\ ,$$
where $N_A(Q^2)$ and $N_1(Q^2)$ are nuclear and free-space hard scattering
rates, and we indicate the possible $Q^2$ dependence of $\sigma_{\rm eff}
(Q^2)$ explicitly. The left-hand side of the equation above is measured.
But in the limit chosen there is no model-independent way to extract
$\sigma_{\rm eff}(Q^2)$! That is because there are two unknowns, and
we can let
$$N_A(Q^2)\rightarrow \lambda(Q^2)N_A(Q^2)\ ;\ \sigma_{\rm eff}(Q^2)
\rightarrow \lambda(Q^2)\sigma_{\rm eff}(Q^2)$$
where $\lambda(Q^2)$ is any function, and still fit the data.

The way out is to use the $A$ dependence to vary the target thickness
over a reasonable range of $l/R_A$. With such a procedure the BNL data
was fit using the shape of the A-dependence, resolving the ambiquity.
The same procedure can be done in electroproduction, but it
is more difficult. One can compare Fig. (3.23) to the survival $P_{ppp}$
of {\it three} particles in a nuclear target, all having the same
exponential attenuation law, averaging over the volume, and with the
90$^o$ kinematics of the BNL experiment. This calculation was also done with
realistic nuclear densities, which show considerable edge effects.
To a surprisingly good accuracy,
the survival of three particles is proportional to the cube of the
survival of one, for $A\geq 7$. (The empirical proportionality
constant is given by $P_{ppp}\approx 1.9P_p^3$.) Because of this cube
effect, the curvature in the A-dependence in the (p, 2p) experiment
is easy to pick out.

However, in electroproduction a high probability of
survival occurs because the attenuated ``particle'' only has to cross
a small part of the nucleus once. This means that the experiment
does not necessarily have a high discriminating power in measuring
the attenuation cross section, unless the data is of comparatively
high precision.  The reason for emphasizing this point is simply to balance the
virtues and complementarity of each experimental approach. The virtues of
electroproduction, in precision and understanding of the probe, are so well
known that it might be helpful to mention the  ironic limitations.
Perhaps a CEBAF experiment could resolve the question.

\medskip
\noindent
{\it $Q^2$ Dependence}
\medskip

The results of the SLAC NE-18 data as a function with $Q^2$
show only a mild increase with
$Q^2$ compared to a plane wave impulse approximation. Based on the scaling of
the spectral densities of the NE18 \cite{E92}
 experiment with $Q^2$, a contant transparency ratio was
anticipated in Ref. \citenum{JR92} well before the experimental results on
the ratio were announced. This paper also preceeded several papers based on
preliminary SLAC results.
Plots\cite{OEA94} of the $A$ dependence  at fixed $Q^2$ do seem to rise
systematically with increasing $Q^2$, but it is not a very big
effect. If the transparency ratio used is interpreted to measure
the survival probability, then the experiment shows only a weak
signal for color transparency. However, {\it we have already seen
that the transparency ratio cannot generally be interpreted as a
survival probability}. All that the NE18 data can determine now is some curve
in the $\sigma_{eff}(Q^2)$ versus $N_A(Q^2)$ plane.
If the electroproduction data becomes good enough to
extract an attenuation cross section along with a hard scattering
rate, then we suggest that would be the preferred procedure.
It is possible that CEBAF (or future facilities such as ELFE) experiments will
have sufficiently high statistics to make this possible.

\medskip
\noindent
{\bf Interpretation of the SLAC Nuclear Hard Scattering Rate}
\medskip

Theoretically, the description of the hard quasi-elastic scattering
in electroproduction begins with exclusive form factors: the elastic
form factor, and transition form factors to whatever states end up
converting to protons after travelling through the nucleus. There
are plenty of reasons to believe that the electromagnetic form factors
used to describe the process in a nuclear targets may not be the same ones
as in free space. Moreover, the free space processes at laboratory
values of $Q^2$ are not truly short distance dominated. Isgur and Lewellyn-
Smith (ILLS) \cite{IL84} and Radyushkin \cite{R84} have made the case for this,
claiming that
{\it none of the perturbative arguments apply at all}. But if this
is so, how can one explain the many regularities of power law behavior
which are observed in many experiments? It would be unscientific to
ignore this qualitative success! More recently, Li and Sterman and Li
\cite{LS92}
have shown that the ILLS arguments focus on an integration region
that does not occur. In the new estimates, less than about one-half of the
amplitudes in free space scattering could come from dangerously large
quark separations.\footnote {This may sound like a negative conclusion
but it an improvement on ILLS who claim that not even a tad of the amplitude
is ever short distance.} The new ingredient is the appearance of Sudakov
effects in the hard scattering form factor. Would these offensive regions
be filtered away in a nuclear target? Nobody yet knows, but there are
several indications that it should happen. We would then see a
repeat of the BNL phenomenon: for experiments inside the nuclear
target there would be a more pure, short distance form factor, and
$\alpha_s$ should make its appearance in the naive way for the hard
scattering.

In Ref. \citenum{JR931}, the transparency ratio
for electroproduction has been calculated
 by making the assumption that the elastic
form factor of nucleon scales like $\alpha^2_s/Q^4$. The free space
form factor scales like $1/Q^4$. Taking the attenuation  cross section,
Fig. (3.21),
 extracted from the BNL data and calculating the survival factors $P_p$
using the formula given above, the authors obtained the curve for
transparency ratio which are reproduced in Fig. (3.24) for the Fe target.
The theoretical calculation
is flat in approximate agreement with data.
The transparency ratio calculated in Ref. \citenum{JR931} is defined slightly
differently
from the one reported by the NE18 group, but this should not
effect the predictions significantly. In Ref. \citenum{JR931} only the
momentum dependence was predicted since the over all normalization
of the elastic form factor in the nuclear medium is not known.

A better procedure to interpret the NE18 data may be to use the
Li and Sterman \cite{LS92}
 analysis for elastic form factor in free space and then model the filtering of
the long distance part of the amplitude
in the nuclear medium. It would be interesting
to extract the effective attenuation cross sections
for the short wavelength pieces
which should be the only amplitudes that survive in a large nucleus.



\subsection { Ideas for future experiments}

In this section we review some ideas that have emerged in the
literature for future experiments related to color transparency and
nuclear filtering. There exists an enormous amount of work in this
direction by many authors. We will also
discuss the near--term experimental viability of some of these proposals.

\medskip

\noindent
{\bf 1) Exclusive Channels}
\medskip

 There exist a large number of different exclusive channels where we expect
color transparency to occur. Examples include
 $eA\rightarrow e'p (A-1)$, $eA\rightarrow
e\pi A$, $eA\rightarrow eK \Lambda (A-1)$ in electron accelerators,
$\pi A\rightarrow p(A-1)\pi'$, $\bar p A\rightarrow p(A-1)\bar p'$,
$\bar p A\rightarrow \pi^+\pi^- (A-1)$, $\bar p A\rightarrow K^+K^- (A-1)$ in
hadronic machines.

Furthermore  the corresponding hadron initiated processes where the nucleus
is replaced
by a proton are expected to show oscillations which will be absent in the
nuclear case due to filtering of the soft components. This will lead to
oscillating transparency in all these cases.
A higher energy $pA\rightarrow p'p''(A-1)$ experiment
will also greatly help in distinguishing between the two proposed explanations
\cite{BT88,RP88}
 for these oscillations seen in the current BNL data. If the independent
scattering interference
mechanism is correct then the transparency ratio should continue to show
oscillations at higher energy, $180^o$ out of phase with the oscillations
in the free space data. The charmed dibaryons explanation will require
no further oscillations.

As discussed in detail in Sections 4 and 5,
 the nuclear collision will also yield the short distance wavefunction of
the participating hadrons, thereby giving important dynamical information about
the hadron. It should be possible to use the wavefunctions measured in these
experiments to predict the results of other experiments.

A high statistics electroproduction
color transparency will be performed at CEBAF \cite{MEL91} which
should help in greatly clarifying the relevance and precision of SLAC data.
The higher energy machine ELFE might, however, be necessary to see a
clear signal of color transparency in electron initiated experiments.

\medskip
\noindent
{\bf 2) Helicity Conservation}
\medskip

 Another set of very interesting nuclear medium effects arise because of the
``hadron  helicity conservation" rule in hard exclusive
collisions. In exclusive large momentum transfer hadron hadron
collision $A+B\rightarrow C+D$, a general prediction of short distance
dominance is
that
\beq
\lambda_A + \lambda_B = \lambda_C + \lambda_D
\eeq
where $\lambda_I$ is the helicity of hadron I. However, this rule has been
observed to be violated in several processes involving free space collisions.
That hadron helicity non--conservation is so widespread proves
conclusively that
many processes must contain considerable contributions from long distances
\cite{GPR95}. A generic test of filtering to short distances in the nuclear
medium is restoration of hadron helicity conservation.

Many simple and direct tests of this idea can be done with polarized beams.
These may be become available at BNL or future facilities, perhaps including
the
LISS facility\cite{LISS95} now under discussion.  It is also quite useful to
use mesons and
baryons which are ``self-analyzing" in the final state.
 The $\Delta$ baryon and $\rho$
meson are standard examples.  Consider, for instance, the reaction
$$
\pi A \rightarrow \vec\rho p (A-1)
$$
\noindent
which can be measured without polarized target or beam.  Hadron helicity
violation has been measured here in the free space case \cite{CEA81}
 at energies $ O(20
GeV) $ and large angles. A direct check of nuclear filtering is the restoration
of hadron helicity conservation in large nuclei.  By our counting this effect
is due to an extra factor of $b^2$, and will scale like $A^{1/3}/Q^4$.
The helicity density matrix of
the produced vector meson is analysed from the angular spectrum of its $\pi
\pi$
decay.

	The situation for electron beams is quite unsettled
experimentally.  A recent
SLAC experiment had made the separation of the electric and magnetic proton
form
factors at high momentum transfer.  The data for the proton's helicity flip
form
factor $F_2$ is consistent with a power law decreasing like $1/Q^6$.
Theoretically, the electric form factor $F_2$ (Eq. 2.1) is a higher twist
object difficult
to analyse within QCD.  However, it has been pointed out in
 Ref. \citenum{JPR92}
that the spin structure of this observable makes it possible to
analyze it within pQCD.  It is claimed that $F_2$ is determined by non-zero
quark orbital angular momentum, which involved wave functions with a node at
the
origin. Because these wave functions are zero in the short distance limit,
nuclear filtering should preferentially deplete them, providing a novel way to
measure the wave function.  Many interesting experiments can be done at
moderate, fixed $Q^2$.  Thus one predicts that the analogue of $F_2/F_1$, when
extracted from measurements in a nuclear target, should show a strong
decreasing
function of A.  One also expects that the magnitude of the analogue of $F_2$
when measured in a large nucleus should be reduced compared to the free space
value.  This idea may be tested in the future as part of a proposal to CEBAF to
measure form factors of nucleons in nuclei by polarization transfer
\cite{GRB94}

\medskip
\noindent
{\bf 3) Recoil Polarization}
\noindent
\medskip

It is well known that proton-proton elastic scattering at a few GeV of energy
and small momentum transfers produces a spin polarization normal to the
scattering plane.  Since this violates the hadron helicity conservation rule,
one can ask whether the effect might be reduced by nuclear filtering.  An
approved CEBAF proposal \cite{SEA91}
 to measure final state polarization from a hard-struck
proton traveling through the nucleus will study $eA\rightarrow e'p(A-1)$.  A
scattering plane is produced by comparing the momentum of the virtual photon
and
the final state proton.  To measure the recoil polarization $P_n$ normal to
this
plane, one needs a polarimeter at the focal plane of the hadron spectrometer.
The theoretical analysis of this experiment is tricky, because it is completely
quantum mechanical; Greenberg and Miller \cite{GM94} have a detailed treatment
within the
context of a compressed-state model.  It is not clear whether the energy at the
first CEBAF machine will be high enough for color transparency to be observed
in
this way.   A study at much higher momentum transfer will require a higher
energy machine such as the European ELFE project \cite{AS93,AP95}.
In that case, making a
precise measurement of  $P_n$ will also require substantial progress to be
achieved in proton polarimetry at high energies.  One might also look for
correlations of polarizations of resonances such as the delta with the
scattering plane to exploit the self-analyzing final state.

\medskip
\noindent {\bf 4) Excited States}
\noindent
\medskip

An important element of color transparency experiments is that only the
short distance part of the hadron wave function survives upon crossing the
nuclear medium. If the short distance wave function for a hadron is close to
zero, then the production of such a hadron will be considerably suppressed in a
nuclear target.  This leads to the very interesting prediction that the radial
and orbital excited states of hadrons will be considerably suppressed when
produced in a nuclear medium. This phenomenon can be observed in
electroproduction experiments where the photon propagates e.g. as a $\rho$ or
$\rho'$ meson through the nucleus.  From time-scale considerations, one guesses
that the formation time of the hadron has to be smaller than the longitudinal
dimension of the nucleus to find this suppression.  Otherwise the hadron is
formed outside the nucleus and no significant difference should arise between
different excitations. We believe that at ELFE and/or CEBAF energies this would
be possible.  Fermilab energies, however, turn out to be too large for this
effect.  A better picture there is that a quark pair rather than a physical
meson propagates through the nucleus; the meson is typically formed long after
the interaction  of the quark pair with the nucleus.

\medskip
\noindent {\bf
5) Scanning the Hadronic Wave Function}
\noindent
\medskip

There exist a large number of diffractive resonance production experiments
which are expected to show color transparency. This effect may have  already
been
confirmed by the experiment E665 on diffractive $\rho$ production. Kopeliovich,
   Nemchik, Nikolaev and Zakharov \cite{KNN93}  have
discussed an interesting set of future experiments involving photoproduction of
other resonances and their excited states.  The authors argue that it is
possible to scan the wavefunction from large to small distances by changing the
momentum of the photon. Some of this is the same as nuclear filtering and the
behavior of large $Q^2$ in selecting the short distance parts of the wave
function, in cases where the impulse approximation can be used (Section 4.5).
Other aspects
are unique features of the authors' theory of diffractive physics. Emphasising
off--diagonal contributions and quantum mechanical mixing, the radially excited
states are predicted not be suppressed to the same extent one might expect if
only the short distance contributions were dominant.

	Quasi-exclusive meson production experiments can also be used to gain
information about the distribution functions of the meson, as mentioned by
Carlson and Milana \cite{CM91}.

Nikolaev has made numerous suggestions for CEBAF experiments, which can be
found in extensive reviews \cite{N94}.  A diffractive picture is
assumed and often used quite imaginatively.  A novel recent paper \cite{BZN95}
predicts color transparency for $s\bar s$ mesons produced under conditions of
extremely low momentum transfer.  This experiment seems to be ideally suited
for
CEBAF and it would be interesting to know the results.

\medskip
\noindent {\bf
6) Diffractive Jets, and Explosive Destruction of the Proton}
\noindent
\medskip

It should be interesting to study diffractive scattering of pions on nuclear
targets at very high energies in the process $\pi A \rightarrow {\rm jets} +
X$.
  Frankfurt, Miller and Strikman \cite{FMS93} have suggested
that color transparency might be observed in kinematic situations where two
opposite isolated jets carry most of the beam energy.  The idea is that the
two-quark Fock component of the pion might contribute significantly,
and that
short-distance should occur due to the large momentum separating two observable
jets at Fermilab energies.  A shortcoming of this proposal is that it is quite
vague, not specifying a definite cross section nor kinematic conditions to make
the experiment very definite.  For example, jet calculations in QCD are usually
quite sensitive to the definition of the ``jet". It follows that the
normalization estimated in Ref. \citenum{FMS93}
     for the process, and claims that the $x$
dependence of the distribition amplitude can be measured, cannot be considered
reliable. The authors have suggested eliminating some of these problems by
finding a signal which goes  like $A^2$, indicating coherent scattering with
the
nucleus, controlling the color flow this way.

	We suggest that the experiment should include a veto in the forward direction
to eliminate co-moving higher Fock components.  Such a veto, plus resolution of
the total jet energies, might be sufficient to extract the 2-quark, short
distance component.  There seems to be no compelling reason to insist on the
super-low momentum transfers of the $A^2 $ dependence, because color is
separated in the event: there is only an implicit assumption that the formation
of jets will not disturb the color transparency of the separating pieces.  The
region of diffractive scattering off separate nucleons therefore need not be
excluded.  Finally, there is no a-priori reason to use a pion beam.  Using a
proton beam, one could also look for 3 jets carrying the beam energy (within
resolution), with a veto on other particles in the forward directions. In this
way one might study ``explosive destruction of the proton" into 3 independent
valence quarks, and correlate this with ideas about the 3-quark Fock
representation.

\medskip
\noindent {\bf
7) Intrinsic Charm }
\noindent
\medskip

	The idea of an intrinsic charm component of the proton wave function has
emerged from the phenomenological analysis of charmonium production at large
$x_F$. It has been suggested \cite{BH91} that the higher twist Fock
component $\mid uudc\bar c\rangle$ was responsible for leading charm
production.
When the proton scatters on the nucleus, the coherence of the Fock components
may be broken and charmonium systems may be formed from the fluctuations. Now,
the nuclear dependence of this process is expected to be typical of soft
interactions, {\it i.e.} $\sigma_A \sim A^{2/3} \sigma_N$ and {\it not} linear
in $A$ as in incoherently adding physics.
Thus, the nucleus again should act as a filter against intrinsic charm
component
contribution to charmonium production, in favor of a perturbatively generated
$c\bar c$ pair from gluon emission.

	However, charmonium production and absorption in nuclear medium is a subject
so
full of controversies that it might be difficult to test color transparency
ideas in these processes. A major barrier is the uncontrolled color flows, and
the multitude of different processes, which contribute to a inclusive processes
producing charm. Closely related is the observed suppression of $J/\psi$ and
bottomonium production on nuclear targets. In that case, it is not settled yet
whether the suppression is due to big absorption of the $c\bar c$ pair in the
nucleus, or reduced production due to energy losses in the intitial state.

\medskip
\noindent {\bf
8) Anomalous Color Transparency}
\noindent
\medskip

	Consider the reaction $\gamma A\rightarrow \pi^0 A$ in the region of small
momentum transfer $-t< m_\pi$ \cite{R91}.  Experimentally a region of $t$ can
be isolated
showing that the production of the $\pi^o$ occurs by the ``Primakoff effect",
namely  the $\gamma \gamma \pi^0$ vertex made by the incoming photon, the
nuclear Coulomb field, and the $\pi^o$.  Most production by this mechanism
occurs just before the $\pi^o$ strikes the nucleus.

	This process may show a signal of ``anomalous" color transparency
\cite{R91} when studied
as a function of $A$ and the $\pi^o$ energy. Although all momentum transfers
are
small, the $\gamma \gamma \pi^0$ vertex is known to be controlled by short
distance: it is the original perturbative ``chiral anomaly".  If the $q\bar q$
pair
contributing to $\pi^o$ production  are close together, then the nuclear
attenuation measured just after the pair is formed should show small
attenuation.  Signals of this would occur using either the ratio or attenuation
method; an example of predicted energy and A dependence of the transparency
ration is shown in Fig. (3.25).  The kinematics of the $\pi^o$ must be tuned to
find a
region where Primakoff is operating while a good fraction of the $\pi^o$'s
produced
penetrate the nucleus, but a window exists where this is possible.  Because the
anomaly is an object of great intrinsic interest, a``snapshot" of its short
distance characteristics would be extremely interesting and a rather direct
probe of fundamental microstructure of chiral symmetry.

\medskip
\noindent {\bf
9) Virtual Photon Polarization Dependence}
\noindent
\medskip

It has been argued that transversely and longitudinally polarized photons may
interact with the proton in different ways \cite{B92},
based on the possible
occurence of protons in``compressed" fluctuations.   For instance, if only
transverse photons interact with compressed protons, the color transparency is
claimed to depend on the degree of virtual photon polarization $\epsilon$
through the equation
 $$
R=R_G +[R(Q^2) - R_G] \frac{\tau \mu_p^2} { \tau \mu_p^2+\epsilon}$$
where $ \tau = \frac {Q^2} {4M_p^2} $ , $ M_p$ and $\mu_p$ are the proton mass
and magnetic moment, $R_G$ the Glauber value for transparency and $R(Q^2)$ a
naively calculated transparency ratio. A different formula is found if only
longitudinal photons interact with protons. The net effect of such a picture is
an electron energy dependence of color transparency at fixed $Q^2$.

\medskip
\noindent {\bf
10) Double-scattering Events}
\noindent
\medskip

Studying the $Q^2$ dependence of the number of final state interactions of
recoil protons in quasi-elastic scattering of electrons on light nuclei is a
type of inverse way to look for color transparency. It is claimed
\cite{EFG94} that rescattering events such as
$$
e + A \rightarrow e'+ p'_f +p_T + (A-2)
$$
\noindent
have a cross-section which decreases much for increasing $Q^2$ in light
nuclei. Model calculations for the $^4He$ case show that measurable effects
can be obtained provided a full acceptance detector is used.

\medskip
\noindent {\bf
11) Missing Transverse Momenta}
\noindent
\medskip

A way to access final state interactions (and its decrease through color
transparency effects) is to study the broadening of the momentum distribution
transverse to Q of scattered protons in $(e , e',p)$ reactions. In the case of
light nuclei, the description of this broadening is dominated by single
rescattering of the struck proton. An accurate theoretical description is
however needed \cite{NSS93} for rescattering without color transparency
to separate CT effects from other $Q^2$ dependences.

\medskip
\noindent {\bf
12) Transparency in semi-inclusive processes}
\noindent
\medskip

Rinat and Jennings \cite{RJ94}
 have extended the concept of transparency to also include
semi-inclusive processes. They argue that the transparency measured in these
processes is not the same as the one measured in purely elastic scattering.
Although this work is a commendable first attempt, we mention it also as an
example of an area where much more conceptual work could be done.

\medskip
\noindent {\bf
13) Color Opacity in the diffractive region of $pp$ scattering}
\medskip

The $pp$ elastic cross section at very high energy $(E\approx 1000$ GeV)
with $t$ of the order of 10 GeV$^2$, Ref. \citenum{KEA77,HEA77}, shows a
remarkable agreement \cite{DL79} with the
$t^{-8}$ power law behavior predicted by the independent scattering
mechanism\cite{L74}. In Fig. 3.26 we show this for the reader's inspection.
Based on this agreement, Donnachie and Landshoff \cite{DL79} have argued that
the
independent scattering amplitude  dominates the scattering cross
section in this region. This idea can be further tested by studing this
kinematic region of $pp$ scattering using nuclear targets. We predict strong
attenuation of the
protons participating in this process, and $t^{-10}$ behavior if nuclear
filtering succeeds in selecting the shortest distance components for $A>>1$.

\bigskip

 In conclusion, there exist a large number of novel and interesting effects in
experiments involving hard scattering or diffractive scattering which can test
the ideas of color transparency and nuclear filtering.

\bigskip

\vfill
\eject

\subsection{Figure Captions }

\begin{itemize}

\item[3.1] Kinematics of the BNL $pA\rightarrow p'p''(A-1)$ experiment.
\item[3.2] The out of plane projections of the missing nucleon momentum
distributions N($p_y$) taken from   Ref. \citenum{H90}. Fig. (3.2a) through
(3.2e) correspond to 6 GeV and Fig. (3.1f) corresponds
to 10 GeV incident momentum.
\item[3.3] The momentum dependence of the transparency ratio for the
process $pA\rightarrow p'p''(A-1)$ reported in Ref. \citenum{CEA88}. The
dashed line at the bottom of Fig. (3.3b) corresponds to Glauber calculation.
\item[3.4] Kinematics of an $eA\rightarrow e'p(A-1)$ experiment.
\item[3.5] The momentum dependence of the transparency ratio for the
process $eA\rightarrow e'p(A-1)$ reported in Ref. \citenum{OEA94} (filled data
symbols). Open symbols are from Ref. \citenum{GEA92}.
\item[3.6] Comparison of the SLAC NE18 experimental results (Ref.
\citenum{MEA94})
 for the transparency ratio $T'$
 for $eA\rightarrow e'p(A-1)$ for Carbon with several theoretical
predictions. The dotted curve is the prediction of the naive parton model
of Ref. \citenum{FLF88} and the dashed curve is the correlated
Glauber/quantum-diffusion model of Ref.\citenum{BFF92}. The solid line is the
Glauber prediction.
\item[3.7] The missing energy (a,b) and
missing momentum (c) distributions for the SLAC $eA\rightarrow e'p(A-1)$
experiment taken from Ref. \citenum{MEA94}. Solid curves are Monte Carlo
predicted with conventional nuclear spectral functions. No radiatve corrections
have been done.
\item[3.8] The theoretical predictions of Farrar, Liu, Frankfurt and Strikman,
Ref. \citenum{FLF88} for $pA\rightarrow p'p''(A-1)$. The model predicts a
monotonically
increasing transparency ratio in clear conflict with the data,
especially for the Al target (not shown by the authors; see Fig. 3.3).
\item[3.9] The nuclear cross section $T\cdot d\sigma_{pp}/dt$ multiplied by
by an overall power law factor $s^{10}$. This plot, taken from Ref.
\citenum{H90},
shows lack of oscillations in the nuclear cross section and provides
evidence for nuclear filtering.
\item[3.10] Kinematics of the Fermilab $\gamma^*A\rightarrow VA^*$ experiment,
where $V$ is a vector meson. The final state $A^*$ is not measured.
\item[3.11] The $t'(=t-t_{\rm min})$ distributions for Hydrogen (lower curve)
and Calcium (upper curve), taken from Ref. \citenum{AEA95}. The curve for
Calcium
clearly shows the separation into coherent and incoherent scattering processes.
\item[3.12] Transparency ratio $T$ reported by the E665 experiment \cite{AEA95}
for photoproduction of $\rho$ mesons for the case of incoherent scattering
from nuclear targets.
\item[3.13] Transparency ratio $T$ reported by the E665 experiment
\cite{F93,F94}
for photoproduction of $\rho$ and $\phi$ mesons for the case of coherent
scattering from nuclear targets.
\item[3.14] Transparency ratio $T$ reported by the E665 experiment
\cite{F94}
for photoproduction of $\rho'$ mesons for the case of coherent scattering from
nuclear targets.
\item[3.15] The differential cross section $d\sigma/dt$ for proton-proton
elastic scattering at fixed $\theta_{\rm cm}$. The solid lines are the
quark counting predictions.\cite{BF75}.
\item[3.16] (a) The energy dependence of $R_1(s)\sim
s^{10}d\sigma/dt(pp)|_{90^o}$ for high energy $pp$ elastic scattering at $90^o$
cm angle. (b) Prediction of oscillating transparency $T(s)$ for $A=27$ from
\citenum{RP88}. There are no free parameters in the energy dependence. Upper
and lower curves show possible effects of nuclear phase shifts on the
prediction, which become negligible at large $A$.
\item[3.17] The $A$ dependence of the transparency ratio $T$ as reported in
Ref. \citenum{CEA88}.
\item[3.18] Fits to BNL data obtained in Ref. \citenum{JR931} for the $A$
dependence of transparency ratio for a) $(Q^2=4.8 {\rm GeV}^2)$ and
b) $(Q^2=8.5 {\rm GeV}^2)$ using the attenuation method. The solid  line
shows the best
fit.  For comparison the dashed lines show constrained fits using permutations
of $\sigma_{eff}$ and best normalizations.
\item[3.19] Survival probability $P_{ppp}$ extracted in Ref. \citenum{JR931}
 from BNL data by use of the attenuation method. The increase with $Q^2$ is
evidence for observation of color transparency.
\item[3.20] The attenuation cross section $\sigma_{eff}$ extracted from the BNL
data.
Solid line: best fit  including the intermediate $Q^2$ data points reported for
Aluminum.  Dashed line:  a 36 mb strong  interaction cross section for
comparison.  The two data points are the global $\sigma_{eff}$ values extracted
from Fig. (3.18).
\item[3.21] Comparison of the fit to the hard scattering rate N($Q^2$) to pQCD
prediction, as given in  Eq. (5), including scale ambiguity.  The ratio
s$^{-10}$N($Q^2)/(d\sigma/dt_{pQCD}$) is plotted as solid lines;  a  flat
curve indicates
agreement.  The lower line uses the scale of the running coupling as ($Q^2$ =
0.5(-t)) in calculating $d\sigma/dt_{pQCD}$.  The upper line uses
$\alpha_s(Q^2$ = 1.5(-t)).
Data points are for the  complete data set (box symbol) and the Aluminum data
(no symbol). The long dashed curve  represents s$^{-9.7}/(d\sigma/dt_{pQCD}$)
and the short
dashed curve is the asymptotic prediction \cite{LB80} for $d\sigma/dt_{pQCD}$
using  $\alpha_s(Q^2 = -t)$ and divided by the right hand side of Eq. (3.7).
\item[3.22] The survival probability/constant as a function of $Q^2/A^\alpha$
for the BNL data for three different choices of $\alpha$. The hard scattering
cross section has been set to be $Z$ times the short distance pQCD prediction,
given by Eq. [3.7]. The data points are circles $(Q^2=4.8$ {\rm GeV}$^2)$,
squares $(Q^2=8.5$ {\rm GeV}$^2)$ and crosses $(Q^2=10.4$ {\rm GeV}$^2)$. The
data shows scaling with the value of $\alpha=1/3$.
\item[3.23] Prediction of the survival probability $P_p$ in quasi-exclusive
electroproduction versus A  using $\sigma_{eff}$($Q^2$) extracted from the BNL
data.  Note that a change in normalization can nearly  compensate a change in
$\sigma_{eff}$.
\item[3.24] $Q^2$ dependence of a ``transparency ratio" consisting of
$\alpha^4_s(Q^2) P_p$ for
Fe under the assumption that the nuclear form factor goes like $\alpha_s^2$.
Lower and upper lines show the range of  theory predictions from varying the
choice of scale $Q^2$ of $\alpha_s$ from $(-0.5 Q_o^2) < Q^2 <(-1.5 Q_o^2)$,
where $Q_o^2$ is the
photon momentum transfer.  Vertical scale depends on overall normalization of
hard scattering rate in nuclear medium which has not been specified.
\item[3.25] (a) Energy dependence of anomalous transparency in $\pi^0$
production for $A=200$, from Ref.\citenum{R91}. $T$ is the transparency ratio.
Curves are shown for different values of parameter $\lambda$, which is the
amplitude for the $q\bar q$ pair to be small. The slope of the curves are quite
different even for $E_{\pi^0}$ as low as few GeV. b) $A$ dependence of the
tansparency
ratio.
\item[3.26] The differential cross section $d\sigma/dt$ for $pp$ elastic
scattering at $\sqrt{s}=400$ GeV (hollow circles) and 494 GeV (filled circles)
is shown on the left. On the right, $\sqrt{s}=1482$ GeV. Both figures are fit
by $d\sigma/dt=0.09t^{-8}$. From ref. \citenum{DL79}.
\end{itemize}

\newpage

\section { Elements of pQCD  }
\setcounter{equation}{0}
\medskip

        Color transparency was proposed and came to be studied only
after a long history of theoretical developments in QCD.  Many differences in
approach between pQCD and nuclear physics have caused confusion and
perplexity when introduced along with color transparency.  To save the reader
some of the  suffering of rediscovering known facts, we include here a
brief history of QCD and the context of various approximations.  We
hope to separate the solid foundations from the theoretical ``opinion
of the day" (reviewed in other sections).  Our
discussion will first be qualitative, providing background which might be
helpful to newcomers. We then go into more quantitative
and more detailed subtopics in the application of QCD.

\subsection {A Brief History }

\noindent
In the earliest days of QCD \cite{M89,ZK93}, the theory was applied only in
terms of the operator product expansion ({\it OPE}).  The {\it OPE} is a method
to express matrix elements of composite operators (known to be
probing short distance) as an expansion using as a basis some
fundamental operators of the theory.  For example, the Fourier transform of
the non-local operator product of two currents encountered in deeply inelastic
scattering
\beq
  \int dy e^{iqy}  <p \mid T(j_\mu(y) j_\nu(o))\mid p > = T_1 (-g_{\mu\nu}+
\frac{q_\mu q_\nu} {q^2} )+...
\eeq

\noindent
 can be expanded in terms of local quark fields by writing
\beq
          T_1  \sim \sum\limits_N A_N c_N(Q^2) (\frac {Q^2} {2 M \nu})^{-N}
\eeq

\noindent
where $Q^2 = - q^2$, and $A_N$ is defined by
\beq
 A_N p^{\mu_1}p^{\mu_2}...p^{\mu_N} = \frac 1 4 \sum\limits_s
<p,s \mid O^{\mu_1\mu_2  ...\mu_N}\mid p,s >.
\eeq

\noindent
The $c_N(Q^2)$ are so called Wilson coefficients and $ O_N$ are matrix
elements of known local operators.  The $c_N(Q^2)$ are defined to satisfy
renormalization group equations, so that their $Q^2$ dependence is
calculable using perturbation theory.  Note that the
procedure assumes that there is only one ``large scale", so that the
renormalization group, which formally controls dependence on an ultraviolet
regularization scale $\mu^2$, can be used to deduce the dependence on
the physically measured scale $Q^2$. (Problems with more than one scale are
treated in the parton model, described briefly below.)
The matrix elements $O_N$  are nonperturbative numbers coding the details
 about the particular hadronic state on which the current acts.
In this, and most perturbative QCD approaches, the perturbative
theory does not pretend to model the hadron, and makes no initial
statement about it.   Instead, the matrix elements are to be
extracted from experiments or supplied from some non-perturbative
model.  This is a very important point.

        This separation of what could be calculated perturbatively
from other unknowns made the first systematic tests of QCD
possible.  The simplest series of operators employ the lowest
number possible of quark fields  sandwiching a number of kinematic
operator coefficients, namely (forgetting color matrices and indices)
\beq
O_N  = \bar \psi \gamma_{\mu_1} D^{\mu_2}... D^{\mu_N} \psi + permutations
\eeq

\noindent
where $\psi$ are quark fields and $D^\mu = (i \partial^\mu-gA^\mu)$ is
the gauge-covariant derivative containing the gluon field $A^\mu$.
The series is organized into  a power series in $1/Q^2$.  The leading
terms in this series are called ``leading twist". The word ``twist" =
(engineering dimension - spin) originated in the {\it OPE} notation for
organizing the $O_N$ into like combinations. The terminology ``leading
twist" and ``higher twist" have subsequently been generalized away
from its {\it OPE} origins to simply mean any case where a supposed
leading order calculation (in terms of a $1/Q^2$ expansion) is done,
in contrast to a supposed power suppressed correction.

        The series can be summed in a suitable gauge where all of the
matrix elements of the minus component of $A$ vanish, to give
\beq
q_{\alpha\beta}\biggl(x,Q^2\biggr)=\int dy^-e^{ixp^+y^-}<ps\mid
\psi_\alpha(y^-,0_T,0^+)\bar\psi_\beta (0)\mid ps)>
\eeq

\noindent
which is the same definition as the parton corellation function, Eq (2.3).
In general, hard inclusive experiments at leading order in $1/Q$ can be
related in much the same way to corellation functions of quark or gluon
fields, which are
the parton distributions.  By using translational invariance and
inserting a complete set of states, we can relate the matrix element
above to a sum of probabilities:
\beq
q_{\alpha\beta}(x)=\sum\limits_N \mid <N\mid\psi_\alpha(O)\mid
ps>\mid^2\delta (x p^+-p^+-N^+)
\eeq

\noindent
This shows that we actually sum over all channels of all numbers of
quarks, anti-quarks, and gluons in the target. Subsequent measurements show
hadrons to be immensely
complicated objects in the parton basis.  The (leading twist) quark
and gluon distributions turn out to be universal quantities; different
experiments can be shown to measure the same matrix elements in
combinations that can be unravelled through theory and experimental
deduction.  This is the strongest meaning of the word
``factorization", which more generally means that the theory can
separate the process into certain generalized products of two or more
subprocceses for study.

\medskip

\noindent {\it The Parton Model}
\medskip

        The {\it OPE} is supplemented by the more intuitive and flexible
parton model,\footnote{ what we call here parton model is
often referred to as the pQCD improved parton model in contrast with the
 original or  {\it naive} parton model.} originally due to Feynman and Bjorken
\cite{F72}{}.  In the early
days of the theory, the parton model was thought to be a weak sister
and not up to the rigor offered by the {\it OPE}.  This changed when it was
realized that the parton model was actually more general and
powerful than the{\it OPE}, which was formalizing a limited aspect of
the parton picture.  The key papers in engineering this breakthrough
were the so-called DGLAP articles \cite{D73}{}.  This established the physical
 origin of leading order
scaling violations to be the substructure of a quark or gluon at a given
resolution scale in terms of yet more quarks and gluons.
It is important to realize that a theoretical state containing  ``three
quarks and no gluons" refers to  quarks defined at
some scale of resolution.  At a finer scale of resolution, that is at
larger $Q^2$, the same objects will be seen to contain more and more detailed
 quark and gluon field combinations.  A gauge theory cannot be
beaten and will not converge in the definition of ``what is a quark."
Gauss' law requires that the quarks be ``dressed" with their Coulomb
field at shorter and shorter distances, a procedure extending
{\it ad infinitum} in resolution and to all orders of pertubation theory.

        The parton model is  more flexible than the one-scale {\it OPE}
because problems with several scales and kinematic regimes can be
treated.  For example, the Drell Yan process can be measured as a function of
 the invariant center of mass energy squared $s$, the lepton mass squared
$Q^2$,  the lepton transverse momentum $Q_T$, its rapidity $y$ and an azimuthal
 angle, besides numerous spins.  Using all these variables there may be regions
 of ``easy" perturbative problems and others with almost untractable problems.
In general, processes with more than one large scale can be
considered, but problems involving big ratios of scales (such as the
limit $Q^2/ Q_T^2>>1)$ are often problematic.

         The scientific process of deduction and measurement of the
numerous parton distribution functions during these last twenty
years \cite{ZK93}{} is one of the most magnificent achievements of physics.
Far
more than static observables such as the hadron mass spectrum, these
 matrix elements contain secrets about how Nature makes hadrons in a confining
theory. However, no theoretical model is yet able to reproduce these
distributions from the QCD Lagrangian or a suitable approximation scheme.
Conversely, these data have not been able to guide theorists toward a
physical understanding of confinement mechanisms. The reason might be that
these observables, coming from inclusive measurements, are  quantities
averaged over many unobserved states. The difficult study of exclusive
 processes, to which we now turn, is a way to make future progress
in understanding hadron structure.
\medskip

\noindent {\it Exclusive Processes}
\medskip

The early study of hard scattering and  exclusive processes \cite{BF75,MMT73}{}
  predates QCD. By using a
generic  renormalizable vector theory with dimensionless coupling
constant and Fermionic constituents,  certain salient features of
elastic scattering were extracted with limited model dependence.  A
key step in this analysis is the definition of the ``fixed angle"
kinematic limit, which is chosen for simplicity of the theoretical
analysis.  In $2\longrightarrow 2$ elastic
scattering, the fixed angle limit means in terms of
 Mandlestam variables:
\beq
 s \cong const.  \mid t\mid \cong const. \mid u\mid
\eeq

\noindent
 with the ratios reasonably close to unity. It follows that
the fixed angle limit is a ``one--scale" problem.
The resulting dependence on the single scale $s$ (which in terms of the
 lab energy is approximately $2mE_{lab}$) is relatively simple.
This limit is not chosen to make the experimentalist's job easy -- indeed,
 it defines a very difficult regime of measurements -- but it is chosen to make
 the theory comparatively clean.

\medskip

\noindent {\it Quark Counting}
\medskip

        The scaling properties of a perturbative theory in the fixed
angle limit can be done ``counting on one's fingers."  Consider the
diagram shown (Fig. 4.1) for the amplitude of $qq\longrightarrow qq$
scattering.
{}From the Feynman rules, this diagram has an amplitude
\beq
M(2\longrightarrow 2) = g^2 \bar u(p_i) u(p_j)  \bar u(p_k) u(p_l) /t
\longrightarrow g^2 s^{1/2} s^{1/2} / s  = g^2
\eeq

\noindent
where $u(p_i)$ is a spinor of momentum $p_i$. The Dirac traces for $\mid M
\mid ^2$  are readily evaluated.  Even before doing
that, one can anticipate some simple results simply by observing that
in the fixed angle limit, $\bar u u$ scales like a momentum, which must scale
like $\sqrt{s}$.   The amplitude is
a Lorentz scalar function of scalar kinematic variables $s, t,$ and $u$,
which are all proportional.  Counting the $4-u$'s and the vector
propagators, then by elementary  dimensional analysis, the
amplitude is dimensionless and scales like ``1", i.e a constant in the
fixed angle limit.  Note that mass terms, if calculated, would give a
small contribution in the high energy limit:  we can drop $m^2/ s <<1$.
A mass $m$ term is an example of a ``subleading" or higher twist contribution
in a power  series in $m\sqrt{s}$.

The power counting of amplitudes for scattering of more constituents follows
from induction.  (Figure 4.1) For example, add two external Fermion legs,
and find two new propagators: the result by counting dimensions goes like
\beq
M(3\longrightarrow3) = g^4 \bar u u  \bar u u \bar u u /\slash p /t_1 /t_2 =
g^4/ s
\eeq

\noindent
where $t_1$ and $t_2$ are the gluon line transfers. The same pattern
follows for n-constituents, which will give an
amplitude-squared  like $s^{4-n}$.   Allowing for the kinematic flux
factors relating the amplitude squared to a cross section (for
example, in the high energy limit $d \sigma/ dt = \mid M\mid^2/s^2)$, then the
overall power of $s$ measured in a cross section might be used to
``count" the constituents.  The prediction that the $pp\longrightarrow pp$
fixed angle  cross section goes like $s^{-10}$ follows from this counting of
three  quarks in the proton.

        It is sometimes wrongly thought that this methodology
assumes that the proton  state has been modeled to consist of only
three quarks.  No;  the spirit of the approach does not contain this
assumption, or any particularly detailed model of the proton states. When
detailed (and sometimes very thorough\cite{SMZ89}) analysis is done, it might
allow more to be learned about the various contributions.
If higher order Fock State projections are scattered,  then their
contribution will simply be small at higher $s$ (goes the argument).
 The experimentally available values of momentum transfer are not always large
enough to make such contributions small, however. Thus, the argument is an {\it
asymptotic} one, representing an idealized limit which may or may not apply to
a realistic experimental situation.

Is quark counting a rigorous result of QCD? Again, No.
At finite $Q^2$ the entire theory of exclusive processes is controversial
and extraordinarily interesting because it involves the details of how quarks
are really arranged in hadrons.  One reason that the quark counting predictions
are not the same as QCD is that they are at the kinematic level of the old
(sometimes called naive) parton model before scaling violations.
 At large $Q^2$  they are modified in QCD by
renormalization group corrections, typically involving inverse powers of
logarithms \cite{LB80,ER80}{}.  These represent the running coupling and
effects of higher orders
of perturbation theory.
Even more than the case in inclusive processes, the logarithmic corrections are
 quite important because several quarks are involved in a typical exclusive
 process.  As we show later in this section, the effects of the logarithms
are sometimes equivalent to a non-integer power of $1/Q^2$, which can confuse
the interpretation of experimental results.

\medskip
\noindent {\it Independent Scattering}
\medskip

 A  second reason  that quark counting is not the same as QCD is
due to the process known as ``independent scattering"
discovered by Landshoff  \cite{L74}{}.  To see how independent
scattering works, consider a hard quark-quark scattering in its cm system.
Suppose one of the final state quarks goes off in a direction given by $p'$.
Consider a second pair of incoming quarks, unrelated to the first, which happen
to collide independently. There is a chance that the second pair accidentally
sends its final state quark
in the same direction as the first.  The final state quarks can then combine
into a hadron. This is independent scattering.  Independent scattering was
anticipated in Section 2, with Fig. (2.6) showing ``mirrors" tilted to arrange
this.
It is evidently
an unlikely process, but all elastic hard scattering processes are rare.  We
will compare this process to the quark counting process by estimating the phase
space integrals for two hard scatterings to coincide in direction and center of
mass motion.

                The power counting of independent scattering goes as follows.
(Figure 4.2).  The outgoing beams of quarks
must coincide in direction well enough to make hadrons in the final state; any
discrepancy is set by the wave functions, which are defined\footnote{In
perturbation theory it is necessary to separate bound state properties of the
wave function from effects of gluon exchange. To avoid double counting the
gluon exchange which produces large $k_T$, the bound state wave functions
should have large $k_T$ tails subtracted.}  to have small
relative $k_T$.   The allowed $k_T$ values are much smaller than the beam
energies, so we can approximate them as almost zero. Because each independent
on--shell quark-quark scattering amplitude scales like
\beq
g^2 \bar u u  \bar u u /t
\eeq

\noindent
so that the independent scattering matrix element scales like
\beq
[g^2 \bar u u  \bar u u /t ]^{n/4} \sim g^{n/2}
\eeq

\noindent
up to logarithmic corrections.The scaling behaviour
of the desired elastic scattering cross section comes then from the
integration region constraint on 4-momenta set by:  $\delta^4(
k^1+k^2-k^3-k^4)$.  There are three large
momenta
for each scattering, and one out-of plane transverse momentum.  This component
of the transverse momentum is not as big as $\sqrt{s}$ but instead depends
sensitively
on what the hadronic wave function  allows.  It should be of order
$C<k_T^2>^{1/2}$  in
the state's wave function, which for purposes of counting is the same as
$C/<b^2>^{1/2}, b$ being a transverse space separation.

Each delta function of a big momentum counts as $1/\sqrt{s}$, since
\beq
 \delta(p-p') \sim s^{-1/2}\delta(x-x'),
\eeq

\noindent
 where $x$ and $x'$ are dimensionless scaling variables.
The overall probability amplitude  for a pair of quarks to coincide
in final state direction to make a hadron scales like the product of the delta
functions of momentum, namely like $C<b^2>^{1/2} (s)^{-3/2}$.  To calculate the
 cross section, recall that
\beq
d\sigma/dt  = const. \mid M\mid^2/s^2,
\eeq

\noindent
 giving $<b^2> s^{-5}$ for meson-meson scattering.  As $s\rightarrow \infty$,
this beats the quark-counting process, which for meson-meson
scattering goes like $s^{-6}$.

 Consider next proton-proton scattering.   The argument goes the same way, but
requires another quark-quark scattering to coincide with the first ones.
This adds three more delta functions of big momenta, so the amplitude-squared
is smaller by $s^{-3}$.  Independent $pp\longrightarrow pp$ scattering thus has
\beq
d\sigma/dt  = const. (<b^2>)^2 s^{-8}.
\eeq

\noindent
  This again beats the quark-counting process, which (recall)
goes like $s^{-10}$.

 How did this process manage to evade the power counting of the quark counting
process?  It is easy to show that the number of gluons and internal propagators
is fewer than that the assumed in the quark-counting induction; the topologies
of the low order diagrams are not the same.
Because {\it both} quark counting and independent scattering were studied
before QCD was established, early discussion focused on comparisons with data.
At first it seemed as if $pp$ scattering went like $s^{-10}$, creating a
puzzle to explain the absence of the much bigger $s^{-8}$. One argument was
made that quark counting diagrams might be more numerous and would dominate for
that reason. However, when
compared at the same order of perturbation theory, the independent scattering
graphs are myriad and re-emerge inside the quark counting diagrams.
This happens because internal gluons can
 become ``soft": the effects of a soft gluon scales like the same power of $s$
 as if the same gluon was absent.  The upshot is that many quark counting
diagrams contain a region indistinguishable from independent
scattering with a soft gluon. Independent scattering cannot in any sense be
``absent". Similarly, if ``soft" gluons attached to an independent scattering
diagram should receive enough momentum to be counted as ``hard", the diagram
may merge into the quark--counting set. It was finally realized \cite{BS89}{}
 that these physically distinct processes actually boil down to different
integration regions found in the one theory of QCD.

      The independent scattering process had a confused history as these
subtleties were only gradually appreciated.  Closely related (and as much
confused) is the issue of ``Sudakov effects" (Section 4.5), at first thought to
suppress the independent scattering regions, but which were subsequently shown
actually to force the independent scattering to dominate in the limit of
$s\longrightarrow
\infty$. We believe that there is rather convincing evidence that independent
scattering region of QCD has
been observed and plays a major role in color transparency (see Section 3).
However, the subject is unsettled, and the interplay of the independent
scattering regions and the quark counting regions is currently a subject of
active investigation.

\medskip
\subsection{Setting up QCD Calculations: Stage One}
\medskip

        The process of setting up calculations in QCD can be divided
into two stages.  In stage one, generalized blobs representing Green
functions are assembled to describe the process under study.  This
non-specific analysis can reveal integrations regions where
perturbative QCD may or may not be used, the most dangerous being
particles with small momentum.  A primary tool in setting up this
study has been power counting methods  which can often predict in
advance which regions will become dominant at high energy.  Sterman
\cite{S93}{}
has given a thorough review which is barely summarized here. A rule
of thumb is that a process can be ``big" only if it develops a
singularity.  From the Landau rules, singularities in loop integrals
occur for kinematic configurations that are allowed with classical, on-shell
particles. (Off shell particles in diagrams are allowed to
propagate only a short distance; they are ``pinched" short.)
The on-shell momentum
space regions correspond to propagation over long distances and long
times which allow the amplitude to get singular.

\medskip
\noindent {\it Momentum Space Versus Coordinate Space}
\medskip

        Much of the intuition of the high energy limit comes from going
back and forth between momentum space and coordinate space ones
and trying to reconcile both pictures.  Because the transverse
coordinate is invariant under the boost, its interpretation is
straightforward and essentially unchanged from the non-relativistic
one \cite{W66}{}. Thus the transverse
coordinate space valence meson wave function:
\beq
\psi(k,p)=\int dy e^{ik\cdot y}<0|\,T\left(q(y)\,\bar{q}(0)\right)|p>
\eeq

\noindent
where Dirac indices have been suppressed, is related to the momentum space
one by a simple Fourier transform:
\beq
\tilde\psi (\vec b_T)= \int d^2 k_T \psi (\vec k_T,x)
e^{i\vec b_T\cdot\vec k_T}
\eeq

\noindent
The other coordinates of a particle depend on the approach.  In old
fashioned perturbation theory, particles are put on-shell but
energy conservation is not imposed.  In covariant perturbation theory,
energy momentum is conserved, but the particles can go off-shell in
intermediate states.  The space time interpretation changes
depending on the procedure.

\medskip
\noindent {\it Null Plane Coordinates}
\medskip

        Most calculations are done in momentum space.  One sets up a
coordinate system oriented locally along each particle's momentum,
which is large in the ``+z" direction.  The null plane momentum
coordinates are $p^\pm = (p^0 \pm p^3)/\sqrt{2}$. Similarly, the null plane
space
coordinates are $x^\pm = (x^0 \pm x^3)/\sqrt{2}$.  With the particle moving
fast
along the ``+z" direction, the $x^+$coordinate acts like the time
coordinate in the limit of an infinite Lorentz boost.

With these coordinates the invariant dot product is
\beq
A\cdot B = A^+ B^- + A^- B^+- A_T\cdot B_T.
\eeq

\noindent
Thus an ordinary plane wave is given by
\beq
exp(-ip\cdot x) = exp(-i  p^+ x^- -  ip^- x^+ +i \vec x_T\cdot \vec p_T)
\eeq

\noindent
Looking at this expression, one can surmise that $p^-$ is the generator
of null plane time $x^+$, or the ``null plane Hamiltonian." This is correct, as
 can be shown by looking at the algebra among generators.\cite{W66} For an
on-shell
particle, the null plane Hamiltonian is just
\beq
p^-= (p_T^2+ m^2)/2p^+
\eeq

\noindent
It is not an accident that this has a non-relativistic appearance: a
two dimensional Galilean group is one of the subgroups of the null
plane algebra.  The dependence on $1/p^+$ is an elementary result of
relativity: when $p^+ >> m$,  then the scale of the null plane time
generator $p^-$ goes to zero, causing time dilation as a kinematic
effect.

Not much more is needed to understand most arguments
using null plane coordinates in QCD.
There is an active industry in calculating things
directly in null plane perturbation theory -- it turns out to be more
difficult and subtle than previously thought, because of gauge dependent
singularities, and difficult interplay between the ultraviolet and infrared
regions. Covariant calculations in
Feynman gauge seem to have the least number of pathological
singularities and unphysical surprises.

\medskip
\subsection { Setting up QCD Calculations: Stage Two}
\medskip

 In stage two, blobs are related to wave functions and standard
separate gauge invariant matrix elements.  A
convenient way to assure gauge invariance of a correlation function
is to relate it to an on-shell Green function whenever this is
possible.
At this stage covariant integrals are 4 dimensional,
something of a conceptual barrier to interpretation in non-
relativistic quantum mechanical terms.  The impulse approximation,
allows one to finesse some of the ignorance.
\medskip

\noindent {\it The Impulse Approximation }
\medskip

The physical picture of  the impulse approximation is a generic
feature of Hamiltonian dynamics, cutting across both the quantum
and the classical realms. In classical physics a particle's state is
represented by a point in the classical phase space $(q(t), p(t))$.   If at
a time $t = 0$ the system's Hamiltonian changes very suddenly, then
the particle remains at the same point -- namely $(q(0), p(0))$ -- and just
time evolves according to the new Hamiltonian. It does not have time
enough to move across phase space, but after the impulse, simply evolves
according to the new
rules.

         In quantum mechanics the mechanism is much the same.  The
most convenient way to see this is to use the Schroedinger picture.
In this picture the dynamical phase space coordinates of the system
are given by the wave function.  If the old wave function for
time $t < 0$ evolves to $\mid s(0)>$,  and if at $t>0$ we suddenly turn on a
perturbation, then the time evolution of the new state $\mid s'(t)>$ is simply
given by the old state evolved with the new Hamiltonian $H'$:
\beq
\mid s'(t)>=e^{-iH't}\mid s (0)>=e^{-iE'_kt}\mid E_n'><E_n'\mid s(0)>\; .
\eeq

\noindent
where $\mid E_k'>$ are the energy eigenstates of the new Hamiltonian $H'$.
The probability that a particular transition occurs in the impulse
approximation is simply given by wave function overlaps from the
initial to the suddenly perturbed system.

        Note that we do not have to follow the details of time
evolution during the instant the sudden perturbation is switched on,
so long as this change is much faster than the time scales already
present.  Sometimes this causes confusion, and opportunities to
misuse the energy-time uncertainty principle.  It would be a
mistake, for example, to assert that if the perturbation is turned on
in a short time $\Delta t$ then the system after the perturbation will
have energies ranging up to $1/\Delta t$ by the uncertainty principle. Looking
at a generic low--energy wave packet we
would see almost no high energy components, revealing the error. The facts do
not change when the
system is relativistic: in that case, the underlying Hamiltonian
structure still exists, although the form of the Hamiltonian must be
consistent.

\medskip

\noindent {\it The Impulse Approximation in QCD:  Factorization}
\medskip

        Let us now see how the impulse approximation separates the
hard scattering from the rest of the dynamics.  In Feynman diagrams
the configurations of the system to be summed over are represented
by momentum space integrals.  The step where the impulse
approximation is used is in the integration over each particle's
respective $k^-$ when it occurs in loops.  Since the light-cone ``time"
coordinate is $x^+$, then integrating over $k^-$ sets the internal times of
the system to a value, namely zero.

With this in mind, consider the diagrams for a hard exclusive
reaction shown in Fig. (4.3).  Let the scattering amplitude be called
$M$.  Following the Feynman rules, we have a $d^4k = dk^+dk^-d^2k_T$ integral
for every loop.  The momentum $k$ flows through the wave
functions, and into the perturbative part of the diagram (H) where
hard gluons are exchanged.
 Schematically the diagram can be written
\beq
M=\int d^4k_1 d^4k_2 d^4k_3 d^4k_4 \psi(k_1)  \psi(k_2) \psi^*(k_3) \psi^*(k_4)
H(k_i , Q )
\eeq

\noindent
where we omit for simplicity the dependence on external variables. Now,  if the
 impulse approximation can be used we can make a power
series expansion of the hard scattering kernel $H$ in powers of $k^-$ and
integrate each wave function over its $k^-$. This will be a good approximation
if the typical $k^-$ in the wave functions is small compared to momenta
occuring in $H$, which is what was meant by hard scattering. (Recall that $k^-$
scales like $m^2/2k^+$ for an object of invariant mass $m^2$.) The Lorentz
boost
properties make the decoupling in $k^-$ a generically good approximation.

This leaves the transverse momentum integrals.  The independent scattering
 region is tricky and will be discussed separately.
{\it If} the hard scattering kernel allows a power
series expansion to be made in $k_T$,  as assumed in the quark--counting
region, {\it then} we can again de-couple most
of the integrals, obtaining
\beq
 M \sim \int dx_1 dx_2 dx_3 dx_4 \Phi(x_1)  \Phi(x_2) \Phi(x_3) \Phi(x_4)
H(x_i, Q )
\eeq

\noindent
We then have relations for the scattering of massless, on shell
partons, which are gauge invariant, parton--model calculations.

These wonderful tricks are foundations of perturbative QCD.
Ignorance about a hadron's quark wave function is permissible.  Just as in
deeply inelastic scattering, it is
enough to make a list of what the wave functions might be, and
arrange to have the experiment measure the wave functions.  Certain general
features of the wave functions will be discussed below.

\medskip

\noindent {\it Power Behavior of Exclusive Processes}
\medskip

The $H$ part of the quark counting diagram  is now evaluated at the on-shell
point $k^- =  k_T = 0$.  This is gauge invariant in perturbation theory.
If we count the dimensions of the diagram and the number of far off shell
gluons
and fermions, we obtain the quark-counting results for the scaling
of the diagrams with $Q^2$.  The power counting was designed to be
simple and crude and anticipated the QCD--based analysis.  The
overall normalization is not determined;  it is an important, but
extremely complicated quantity. It depends not only on the sum over diagrams,
but also on the distribution amplitudes, defined next.

\medskip

\noindent {\it The Distribution Amplitude}

\medskip
\noindent
The object
\beq
\phi(Q^2,x)=\int^{Q^2}_0d^2k_T \psi (k_T,x)
\eeq

\noindent
is called the distribution amplitude, and was invented by Brodsky
and Lepage \cite{LB80,ER80}{}.  As $Q^2\rightarrow\infty$, the distribution
amplitude tends to evaluate the wave function at the point of zero transverse
space separation, by Fourier analysis. One can see this explicitly by
transforming the integrals to obtain
\beq
\phi(Q^2,x)=2\pi\int^\infty_0 db QJ_1(Qb)\tilde\psi (b_T,x)
\eeq

\noindent
One sees that the $Q^2$ dependence of $\phi(Q^2,x)$ ``maps out" the
$b$-dependence of the wave function in the region $b^2\geq 1/Q^2$.
The wave function in question has had its hard
perturbative ``tail" at large $k_T^2$ removed, and put into the hard
scattering. The working assumption is that the wave function is
unknown but smooth, characterized by hadronic physics scales of a
Fermi or so.  There is also information available from the
renormalization group, which can organize the limit of extremely
short distances in powers of $\ell n(b^2)$.  For laboratory values of $Q^2$ it
is unlikely that this information is enough and it is necessary to model
the wave
function.  Finally one is left with the $x$-integrals, which flow
through the diagrams and cannot be eliminated.

        We have re-obtained a main result: the quarks participating
have been shown to be separated by a transverse distance $b^2\geq 1/Q^2$.
The only essential change from the non--realistic mirror analysis of
Section 2 is
the trading of two Lorentz invarient ``b" variables for three non--realistic
``r's".
As in the non-relativistic case short--distance is a statement about a dominant
integration region, not about any compresive preparation of the initial state.
In free space
scattering,  for example, the cartoon picture is shown in Figure (4.3.5).

\medskip

\noindent {\it Wave Functions}
\medskip

A priori, one knows little about quark wave functions except for symmetry
properties. It seems unjustified to assume on
the basis of non-relativistic quark models that the pion would be
dominated by the ``s-wave" $(m=0)$ component, since non-relativistic
quarks are some kind of quasi-particles and not the same as light-cone quarks.
Let us make a list of the quark wave functions for the
simplest case, a pion.  The wave function to find a quark-antiquark
pair in a pion with momentum $p$ is actually a Green function, the
Bethe Salpeter two point amplitude \cite{GPR95}{}:
\beq
<0\mid T\biggl(\psi_\alpha (0)\bar\psi_\beta (x,b_T)\biggr)\mid
ps>=\bigg\lbrace
A \slash
p\gamma_5+B[\slash b,\slash p]+ C\slash b\gamma_5
+D\gamma_5\bigg\rbrace_{\alpha\beta}
\label{wavef}
\eeq

\noindent
The coefficients $A-D$ depend on position while the Dirac matrices represent
the spin projections. (The expressions are not gauge--invariant, because they
are building blocks to be used in a particular procedure.)
Now assume that the pion is moving fast, namely its
momentum $p >> 1$ GeV.  The first two wave functions $(A, B)$  scale linearly
with $p$ and are the big
ones by ``power counting" - namely, they come from the parts of the
quark Dirac operators which get big in a boost. The other wave functions are
kinematically sub-leading.
The list is also organized by powers of $b$;  it makes it
easier to pick out the short distance ($A,D)$ components.
There is a similar list for
the  Lorentz covariant vector meson and baryon wave functions.

\medskip

\noindent {\it Angular Momentum}
\medskip

One can also make an expansion in $L_z$, the $SO(2)$ orbital angular momentum
of the quarks about the particle's momentum direction, according to
\beq
\tilde\psi(\vec b,x)=\sum\limits_m e^{im\varphi}\psi^m (\mid b\mid, x)
\eeq

\noindent
By continuity and differentiability a wave function with
orbital angular momentum $m$ has to go like $b^m$ as $b\rightarrow 0$;  this
will be  important below.
The normalization of each component of the wave
function is a dynamical question.       One thing from experiment that is
known is the normalization of the pion wave function from pion
decay:
\beq
\tilde\psi(0)=\int {d^2k_Tdx\over (2\pi)^3} \psi (k_T,x)=f_\pi\cdot const.
\eeq

\noindent
This is a solid number because to a very good approximation the $W$-boson only
couples to a pair of quarks.

\medskip
\noindent {\it Logarithmic Corrections}
\medskip

As mentioned before, the overall power behavior of the QCD process is
 kinematically the same as the QED models used  before QCD was invented.
  There is a substantial difference due to the logarithmic corrections.
The typical result can be understood rather easily.
Postponing the discussion of Sudakov double-logarithms to Section 4.5, let us
  see how leading logarithms are resummed into the distribution
amplitude, in a way which is much reminiscent of the now standard
 DGLAP evolution equations \cite{D73}{} for structure functions in deep
inelastic reactions.
For simplicity, we restrict to the simple meson case. In axial gauge, one may
isolate big logarithmic factors in ladder-type diagrams, leading to an
expansion:
\beqn
\Phi_{\rm LL}(x,Q)=
\Phi_0(x)
+\kappa\int_0^1du\,V_{q\bar{q}\rightarrow q\bar{q}}(u,x)\,\varphi_0(u) \\
+{\kappa^2\over2!}\int_0^1duV_{q\bar{q}\rightarrow q\bar{q}}(u,x)
\int_0^1du'V_{q\bar{q}\rightarrow q\bar{q}}(u',u)\,\varphi_0(u')+\ldots
\nonumber
\eeqn

\noindent
 where
\[
\kappa={2\over \beta}\ln{\alpha_S(\mu^2)\over\alpha_S(Q^2)},
\]
\noindent
with $\beta=(11-{{2n_{f}}\over{3}})/4\), \(n_{f}$ being
the number of quark flavors and $V_{q\bar{q}\rightarrow q\bar{q}}$ is the
following kernel
\begin{equation}
V_{q\bar{q}\rightarrow q\bar{q}}(u,x)={2\over3}\left\{
{\bar{u}\over\bar{x}}\left(1+{1\over u-x}\right)_+\theta(u-x)+
{u\over x}\left(1+{1\over x-u}\right)_+\theta(x-u)\right\},
\end{equation}
\noindent
where the $()_+$  distribution comes from the colour neutrality of the
meson.

The equation on  $\Phi$ may be rewritten as
\begin{equation}
\left({\partial\Phi\over\partial\kappa}\right)_x=
\int_0^1duV(u,x)\,\Phi(x,Q),
\end{equation}
\noindent
the solution of which has been known for a long time as:
\begin{equation}
\Phi(x,Q)=x(1-x)\sum_n\Phi_n(Q)C_n^{(3/2)}(2x-1);
\end{equation}
\noindent
where the $C_n^{(m)}$ are Gegenbauer polynomials which satisify:
\begin{equation}
\int_0^1du~u(1-u)\,V(u,x)\,
C_n^{(3/2)}(2u-1)=A_nx(1-x)C_n^{(3/2)}(2x-1),
\end{equation}
\noindent
with eigenvalues $A_n$. One thus gets
\begin{equation}
\Phi_n(Q)=\phi(\mu)e^{A_n\kappa}=\Phi(\mu)\left(
{\alpha_S(\mu^2)\over\alpha_S(Q^2)}\right)^{2A_n/\beta},
\end{equation}

\noindent
where exponents monotonically decrease, beginning with:
\beq
{2A_0\over\beta}=0,\ {2A_2\over\beta}=-0.62,\ \ldots
\eeq

\noindent
Using the previously derived normalization:
\begin{equation}
\int_0^1dx\Phi(x,Q)=\Phi_0(Q)\int_0^1dxx(1-x)={\Phi_0\over6}=f_{\pi},
\end{equation}

\noindent
the expansion may be written as:
\begin{equation}
\Phi(x,Q)=6f_{\pi}x(1-x)+\Phi_2(\ln Q^2)^{-0.62}x(1-x)(2x-1)^2+\ldots
\end{equation}

\noindent
We thus know the asymptotic pion distribution amplitude:
\begin{equation}
\Phi(x,Q)\sim_{Q\rightarrow\infty}6f_{\pi}x(1-x).
\end{equation}
We do not know the realistic distribution amplitude at accessible
energies, since the constants $\Phi_2,\ldots\Phi_n$ cannot be derived from
perturbative QCD.
This is where the QCD sum rule approach enters \cite{CZ84}{} to
 yield values for moments of the distribution

\begin{equation}
\int_0^1dx(2x-1)^{2n}\Phi(x,\mu),\ \ldots
\end{equation}
and then reconstruct the constants $\Phi_2,\ldots\Phi_n$. In practice, the
reconstruction of $\Phi(x,\mu)$ is fairly model
dependent, as research in this area continues.

 A similar result is obtained in the baryon case. The proton valence
 wave-function can be written as a series derived from the leading logarithmic
analysis, similar to Eq. (4.35) for the pion case, but here in terms of
 Appel polynomials \cite{CZ84}{}:

\begin{equation}
\Phi(x_i,Q)=120 x_1 x_2 x_3 \left\{1 + {{21}\over{2}}
\left({\alpha_S(Q^2)\over\alpha_S(Q_0^2)}\right)^{\lambda_1} A_1
 P_1(x_i) + {{7}\over{2}} \left({\alpha_S(Q^2)
\over\alpha_S(Q_0^2)}\right)^{\lambda_2} A_2 P_2(x_i)
+...\right\} .
\end{equation}

\noindent
Here the slow $Q^2$ evolution entirely comes from renormalization group
factors $ \alpha_S(Q^2)^\lambda$, the $ \lambda_i$  being calculated to be
increasing
 numbers:
\[
\lambda_1 = {{20}\over{9\beta}}   ~~~~,~~~~\lambda_2 = {{24}\over
{9\beta}}~~~~ {\rm etc.},
\]

\noindent
and $P_i(x_j) $ are tabulated Appell polynomials
\[
P_1(x_i)=x_1-x_3 ~~~,~~~ P_2(x_i)=1-3x_2 ~~~,...
 \]
 $A_i$ are unknown constants which measure the projection of the
wave-function onto the Appell polynomials:
\begin{equation}
A_i=\int _{0}^{1}dx_1 dx_2 dx_3 ~\delta(x_1+x_2+x_3-1)~\phi(x_i)~P_i(x_i)
\end{equation}

\medskip

\noindent {\it Independent Scattering, Again}

\medskip

Recall that the quark counting analysis contains an assumption which is not
true in general: the assertion that the transverse momentum integrals can be
decoupled from the hard scattering. Because of this problem, the resulting
power counting, etc., are doubtful. Instead, quark--counting should be
understood as a particular model for QCD, a model defined by the region it
emphasizes. The complementary region where the transverse momentum integrals do
not obey the quark--counting assummptions is just the independent scattering
region.

In this region a new singularity occurs near $k_T=0$ as one might anticipate,
because the process goes with all particles on--shell. This ``Landshoff--pinch"
invalidates the assumption that $H$ can be usefully expanded in a power series
in $k_T$, and consequently invalidates the factorization {\it \`a la} Brodsky -
LePage. For some time it seemed that no
factorization existed and that the problem might be
beyond the ``divide and conquer" strategy of standard pQCD. This was finally
solved by Botts and Sterman \cite{BS89}{}, who realized that a different, new
form of factorization had to be created to accomodate the different region.

The basic trick is quite simple, and intuitively appealing. Given that the
amplitude contains unavoidable convolutions over transverse momentum, it is a
good idea to express this in terms of the transverse
spatial coordinate, where one will obtain products. Thus, instead of decoupling
the $k_T$ integrals, one just Fourier transforms from $k_{Ti}$ to $b_i$ and
gets
the fixed angle amplitude:
\begin{equation}
\label{Af}
\hbox{\bf M}(s,t) \sim
\int_0^1dx_i\,db_i\,H(\{x_i\})
\prod_{i=1}^4{\tilde\psi}_i(x_i,b_i),
\end{equation}

A cartoon going with this expression shows that one has a simple overlap
between hadrons with different impact parameters. (Figure 4.3). The terms
inside
the expression just have to be evaluated, much the same way as the
$x$--integrals simply have to be calculated.

The essential feature of the impulse approximation remains. Although the wave
function integrals are more complicated, their slow bound state time--evolution
is successfully separated from the fast time evolution of the hard scattering.
There is no reason why pQCD should not apply to the resulting amplitude, albeit
with more detailed dependence on the transverse overlap integrals. As it turns
out these are extremely interesting, because they contain information we want
to know about where the quarks might be found. We will return to discuss this
in more detail below.
\bigskip
\subsection {Hadron Helicity Conservation as a Test of Models}
\medskip

The almost chiral nature of QCD without heavy quarks leads to interesting
simple properties at the perturbative level, namely the fact that a quark
or antiquark conserves its helicity when emitting or absorbing a vector
particle,
a photon or a gluon. Long time confinement physics allow however orbital
angular momentum to mix with quark spins to construct the hadron polarized
states, so that polarization observables at the hadron level remain an
interesting physics question. The fact that hard exclusive reactions probe the
short distance structure of hadrons leads to interesting consequences.
   We will first present this in the context of the assumption
of the quark counting model, and then examine the differences caused
by independent scattering.

\medskip
\noindent {\it Hadron Helicity Conservation in the Quark Counting Model}
\medskip

Within the quark counting model, note that a particular orbital
angular momentum component of the wave function is selected by Eq.4.21.
Expanding the wave function according to Eq. 4.22, then only the $m = 0$
term contributes in Eq.4.23.

The wave functions used in this approach therefore are selected by
the hard scatttering to give the hadron helicities $\lambda_n$ as the sum
of the
quark  helicities $\lambda_{q,i}$
\beq
\lambda_n=\sum\limits_i\lambda_{q,i}
\eeq

\noindent
In the high energy limit the quark helicities do not flip when gluons
are exchanged: this is because pQCD is almost perfectly chirally
symmetric.  It is quite easy to see this.  The rule for Dirac quark
spinors is
\beq
\gamma^5 u(p, \lambda) = \lambda u(p, \lambda)\; \; ; \; \; \bar
u(p\lambda)\gamma_5=-\lambda\bar u(p,\lambda)\; .
\eeq

\noindent
One can insert $1 =\lambda\gamma_5$ on an initial state spinor, then permute
$\gamma_5$ through all the Dirac matrices until it reaches the final state. One
immediately finds that the ``$\slash p$" term in propagator preserves helicity,
while the mass term flips it. Terms of order $m/p$ can be dropped
when $m$ is only a few MeV.
Putting these results together gives the famous hadron helicity
conservation rule \cite{BL81}{}. In a reaction between hadrons
$A+B\rightarrow C+D$, then the
sum  of the  helicities going in to the reaction is the same as the
sum going out:
$$\lambda_A+ \lambda_B = \lambda_C+ \lambda_D
$$

This rule is evidently as general as the factorization; it is an exact
dynamical symmetry of the quark-counting model.

        The hadron helicity conservation rule is a remarkable
achievement: it allows us to test the general model independent of
detailed assumptions about the unknown wave functions.
Unfortunately, its success when confronted by experiments is very
uneven.  There currently seems to be no evidence of strong violations
of the rule when applied to electromagnetic form factors.   Yet in
experiments involving scattering of hadrons, it is violated  in
almost every case tested!  Despite this checkered record, we will
later discuss reasons to believe that the hadron helicity conservation
 rule will be important in  the study of color transparency.

\medskip
\noindent {\it Hard Scattering Hadronic Helicity Flip}
\medskip

        Recall that the hadron helicity conservation rule represents an
exact symmetry of the quark - counting factorization, because the
hard scattering is ``small" and ``round".   But perhaps it is not true in
free space, if independent scattering is really contributing there.
The crucial question whether (or not) the symmetry of the model is a
property of the entire perturbative theory. For this we follow the
treatment of Ref. 70.

        First, note that the non-perturbative Hamiltonian of QCD does
not conserve spin and orbital angular momentum separately, but
instead generates mixing between them.  This Hamiltonian operates
over the long times in which wave functions are formed.  Its effects
are largely unknown; is why a list of all the orbital and spin projections in a
meson wave function was given earlier.  Thus if a non-zero orbital
angular momentum component somehow enters the hard scattering -
and this is a crucial point - then the long-time evolution before or
after the scattering can convert this angular momentum into the
observed hadron spin.  It is not necessary to flip a quark spin in the
hard interaction, because the asymptotic hadron spin fails to equal
the sum of the quark spins.  Such a mechanism is totally consistent
with the impulse approximation.

        It turns out that  the independent scattering processes are not
``round" but instead are ``flat" (Fig (4.2)).
The origin of the asymmetry is kinematic.  As before let $x$ be the
direction perpendicular to the scattering plane and $y$ be a vector in
the scattering plane. In the independent scattering mechanism, the
two (or more) uncorrelated scattering planes are separated at the
collision point by a transverse out-of-plane distance $ b_x$.  The in-plane
transverse position separations $\Delta b_y$ that contribute are as small as
possible, namely of order the uncertainty principle estimate $1/Q$,
because this is the direction of large momentum flow carried by the
gluons.  The kinematic ``short-distance" in the problem is, however,
only the in-plane distance.  The relevant out-of plane transverse
distance is set by the hadronic wave functions.

Returning to the kinematic discussion , the ``out-of plane" transverse
momentum is the main focus in studying independent scattering.  To
control this variable, it is convenient to make a Fourier transform to
conjugate space variables $b$.   We let the $x$-direction be out of plane;
this is the same for all the hadrons.  We let the $y_i$ variables be
chosen relative to the hadron momenta directions, so $b_y$ is in the
(vector) $b_x \hat x \times \vec p_i$ direction for each momentum $\vec p_i$.
Botts and Sterman\cite{BS89} then found a  useful formula for the independent
scattering amplitude,
which we write in the meson case for simplicity:
\begin{equation}
\label{A0}
M(s,t)={\sqrt{2}Q\over 2\pi |\sin\theta|}
\int_0^1dx\,(2\pi)^4H(\{ xQv\})\,H'(\{\bar{x}Qv\})\Big|_{\{\alpha\beta\}}
\int_{-\infty}^{+\infty}db\prod_{i=1}^4
{{\cal P}_{\alpha_i\beta_i}(x,b;Qv_i)\over Q},
\end{equation}
where

$${\cal P}_{\alpha\beta}(x,b;Qv)=\int {dy^- \over 2\pi}\,e^{ixQy^-}
<0|\,T\left(q_{\alpha}(y)\,\bar{q}_{\beta}(0)\right)|\pi(Qv)>
\Big|_{y=y^{-}v'+b\eta},$$

\noindent
with Dirac indices $\alpha\beta$.

For the discussion here, note\cite{GPR95} that
the scattering plane breaks  rotational symmetry with
the out-of-plane direction $x$.  Immediately one is struck by the
absence of any selection rule favoring $m =0$; instead, all orbital
angular in the wave functions are allowed.  It is as if hadrons
``flatten" under impact in the in-plane direction $y$, forming a cigar-shaped
hard scattering region.

        We therefore have a new rule of ``hadron helicity
nonconservation" when independent scattering occurs:
\beq
\lambda_A+ \lambda_B \not= \lambda_C+ \lambda_D\
\eeq

\noindent
There are a number of tests of this idea.  Basically, we want to have
spin information on  every possible reaction in free space, and every
similar reaction in the nuclear target.  When one observes hadronic
helicity violation one is observing the orbital angular momentum makeup of
the hadron!  This will be discussed further below.

\bigskip

\subsection {  Sudakov Effects}

\smallskip

{\it QED History}

\medskip

        In 1956, V. Sudakov  \cite{S56}{} studied the large $Q^2$ dependence of
the electromagnetic form factor of an electron in perturbative quantum
electrodynamics.  At  one-loop order, he observed that the form
factor received a ``double-log" correction of order  $\alpha
\ell n^2(Q^2/\Lambda^2)$.  Here $\Lambda << Q$ is an infrared cutoff of
some kind.  These terms come from the integration region where the
internal photon loop momenta are ``soft", with momenta much less
than $Q$.  At two loop order the same contribution was squared, and
multiplied by 1/2!.  Because $\ell n^2(Q^2)$ is a large factor, standard
perturbation theory in $\alpha$ is inadequate when $\alpha
\ell n^2(Q^2)$ becomes large.

         For this reason the soft photon region is calculated to all
orders in perturbation theory. Conversely, it can be shown that this
region is the only one giving terms of order $[\alpha
\ell n^2(Q^2/\Lambda^2)]^J$.  Sudakov showed that the vertex function
from summing terms of this kind to all orders is the exponential of
the one-loop graph, times the bare vertex. The ``Sudakov form factor"
goes like exp($-\alpha\ const\  \ell n^2(Q^2/\Lambda^2)$.  Note the
minus sign in the exponential. For large $Q^2$ the form factor goes
to zero faster than any power of $Q^2$!

        Sceptics have pointed out that the procedure is awkward,
which is true enough.  The value of the sum to all orders is smaller
than some of the low-order terms neglected, causing worries about
self-consistency.  Although asymptotic trends can be identified,
quantitative values of leading-log expressions must be interpreted
with care.   A helpful rule is that at a given order of approximation,
$\ell n( zQ^2) = \ell n(Q^2) $ for any fixed number $z$ as $Q^2$ goes to
infinity, up to subleading corrections.  These technical issues have
dominated much work, but the physical picture that emerged is
beautiful enough to justify the procedure.

\medskip
{\it Physical Picture   }
\medskip

Sudakov correctly interpreted the rapid decrease of the form factor with $Q^2$
as follows.  At large $Q^2$, a struck electron has a chance to radiate
an enormous number of photons into a very large phase space.
Because the phase space is so large, these inelastic events dominate
the probability.  There remains very little probability for an elastic
event.  This explains qualitatively why the elastic form factor must
be small when $Q^2$ is large.  (As a check on the calculation, the
calculation of probability ``one" minus the inelastic events does
indeed recover the double-log exponentiated.\cite{BD65}{}.

        The physical picture indicates that the phenomenon of a
strongly damped elastic process is very general.  It is important to
realize that each (on-shell) charged leg emerging from any Green
function can radiate vector particles copiously,  so that Sudakov
effects are not generally described by simple form factors.
Sudakov's form factor was simply the first example of far-reaching
phenomena, now called ``Sudakov effects" that must occur in many
processes in a gauge theory with massless particles.

        The result seems to be deeper than its modest perturbative
origin..  The exponentiation is not an accidental pattern, but closely
related to the gauge transformation properties of the matter fields.
This follows from  the observation of Grammer and Yennie  \cite{GY73}{} that
the polarizations of soft gauge photons are primarily longitudinal,
and can systematically be summed using Ward identities. These
become phases on the matter fields, because this is how the fields
transform.  In low orders of perturbation theory QCD is much like
QED and the classic procedures of infrared QED can be used almost
without change \cite{CT76,KU77}{}.  Higher orders in QCD are tricky, but
theory can
organize the calculations into ordered sums, which add up all terms
of the form $[\alpha^J\ell n^K(Q^2/\Lambda^2)]^J$, where $J$ and $K$ are
related by some pattern. The case of $J <K$ is called subleading
logarithms. Implementing this systematically in QCD took quite
some work to accomplish, but the underlying gauge symmetry makes
it possible.  The techniques were eventually brought to a high level
of perfection by Collins and Soper \cite{CS81}{}.  Further progress has been
made by Sterman \cite {S93}{}.

\medskip

{\it Running coupling and the chromo--Coulomb phase shift}
\medskip

        Here we will present a fast, heuristic way to understand the
exponential functional form physically.  Consider an $e^+e^-$ pair
approaching each other before they produce a time-like photon.  The
electron and positron are not free particles but are accompanied by
clouds of soft photons.  These can be exchanged before the event, and
the low-momentum region will produce a phase-shift on the
particles' wave function.  That phase should be given by an eikonal
form, in a gauge theory the line integral being $ie\int A\cdot dx$
where $A$ is the vector potential.  For $A$ one might semi-classically insert
the the Coulomb potential, but more generally by
dimensional analysis there will be an integral of the form $ie^2\int
dx/x$ between two endpoints. The closest approach in space between
the particles is of order $1/ Q$, and the furthest relevant point set
by the infrared cut-off, of order $1/\Lambda$.  Thus there is a phase
proportional to a $\ell n(Q^2/\Lambda^2)$.  This can be shown to be the
``Coulomb phase shift" \cite{YFS61}{}. In perturbation theory it is recovered
from summing diagrams to all orders.

        It is convenient to employ analyticity to reconstruct the
function from its analytic properties in the complex plane. The
analytic function of $Q^2$ having a branch cut on the real axis and
going like $exp(i \pi \ell n (Q^2/\Lambda^2)$ for $Q^2>0$  is
 $\exp(- \ell n^2 (- [Q^2-i 0] /\Lambda^2)$.
 This is the Sudakov form factor already
quoted.  This motivates the intimate relation between the eikonal
phase, which is closely tied to gauge symmetry, and the exponential
form with double-logs. Note that in reconstructing a contour
integral knowledge of  the region of $Q^2<\Lambda^2$ is needed.  This
is no problem for QED, but indicates infrared sensitivity of quantity  in QCD.

        In QCD much the same eikonalization occurs for quarks and
gluons \cite{CS81,T85}{}.
The quark's Coulomb field is modified by the running coupling, which
in coordinate space can be written $\alpha_s(x) =
const/\ell n(1/x^2\Lambda_{QCD}^2)$.  Repeating the calculation above,
the QCD ``chromo-Coulomb" phase effects are
\beq
\exp \biggl(-ic\int\limits^{1/\Lambda}_{1/Q^2} {dx\over x}
{1\over \ell n
(x^2\Lambda^2_{QCD})}\biggr) \sim \exp -i{c\over 2}\ell n\ell n
(Q^2/\Lambda^2_{QCD}).
\eeq

\noindent
 While this way of calculating requires justification, the result
suggests that sums in perturbation theory should recover a phase
shift going like
$ \ell n\ell n (Q^2/\Lambda^2_{QCD})$, which is
indeed true.   A bonus is that bothersome infrared cutoff effects
produce a simple constant phase that might be ignored. The
appearance of $\Lambda_{QCD}$ as opposed to the infrared cutoff is
an indication of a perturbatively calculable result.
The region of momentum from the infrared cutoff to a
renormalization point can be held fixed, while the region from the
renormalization point to the big scale evolves with $Q^2$ by
the renormalization group.\cite{PR82}   Thus, the QCD
 ``Chromo-Coulomb" phases can be calculated as purely imaginary
anomalous dimensions \cite{PR82}{}, on the same perturbative footing as other
anomalous dimensions.  These physical arguments have been confirmed by
careful and sophisticated work  \cite{S83}{}, and now there is no doubt of
the calculability of these phases.

\medskip
{\it Experimental Evidence for Sudakov Effects}
\medskip

        Many years ago it was realized that inclusive processes
involving large ratios of scales would probe Sudakov effects as a
novel form of radiative correction in QCD.  The classic experimental
application was Drell Yan lepton pair production at measured
transverse momentum.   Subsequent study has confirmed the
qualitative features of the theory, creating a convincing example
supporting the experimental relevance of the effect.  Several other
processes involving small transverse momentum seem to be equally
well treated by now.

         Even earlier, Polkinghorne \cite{P74}{} had pointed out that the
independent scattering process occured via nearly on-shell
scattering, and its description might require Sudakov resummation
in a theory with vector exchange. This statement lead to claims that the rapid
 decrease with $Q^2$ of Sudakov form factors would eliminate the independent
scattering contributions altogether.  Mueller \cite{M81}{} showed that this
argument was incorrect, and that independent scattering actually dominates at
asymptotic energies, as we review below.

        Another approach is to test experimentally for the presence of
Sudakov effects by looking for the Coulomb phases in processes with
sufficient coherence, such as elastic scattering.   Evidence for the
QCD coulomb phase in $pp$ fixed angle scattering was presented in Ref.
 137.  It was proposed that interference between
independent scattering regions and quark counting regions would
explain the energy-dependent oscillations, of about 50\% magnitude of
$d\sigma/dt$, as
experimentally observed.   Similar oscillations have also been
observed in the energy dependence of $\pi-p$ hard elastic scattering.
The disappearance of these oscillations observed in the BNL
experiment is consistent with filtering of the process to short
distance.  This has been discussed in Section 3.

\medskip

{\it QCD Analytic Reconstruction, Bin-by-Bin  in Transverse Space}
\medskip

          Again one can find the analytic function which has the
running coupling-modifed phase reported above.  To leading log order
the function in the exponent has to be
\beq
S( Q^2,  \Lambda_{qcd}, \Lambda^2) ={const\over \pi}\;\ell n (\frac{Q^2}
{\Lambda^2})\ell n\biggl(
\frac{\ell n
(Q^2/\Lambda^2_{QCD})} {\ell n(\Lambda^2/\Lambda^2_{QCD})}\biggr)
\eeq

\noindent
meaning that the complete Sudakov factor is
\beq
 \exp\biggl(-S( Q^2,  \Lambda_{QCD}, \Lambda^2) \biggr) =
\exp\biggl( - {const\over \pi}\ell n (\frac{Q^2}{\Lambda^2})\ell n\biggl(
\frac{\ell n
(Q^2/\Lambda^2_{QCD})} {\ell n(\Lambda^2/\Lambda^2_{QCD})}\biggr)\biggr)
\eeq

\noindent
        This expression contains both the infrared cutoff and
$\Lambda_{QCD}$, showing that its calculation is sensitive both to
unphysically soft gluons and the calculable
perturbative region.  This is the QCD ``Sudakov form factor", which
shows the effects of the running coupling. If the dependence on an
infrared cutoff were taken literally, the expression would be
doubtful, because infrared-divergent gluons do not physically exist.
Curiously, the rate of change of phase of the expression with $Q^2$
does not depend
on any infared cutoff, so at most the $scale$ of infrared cutoff can
be in doubt, reflecting the leading-log character of the result.

        The role of the infrared cutoff in QCD is actually provided by
bound state and confinement effects.  Given a color neutral ``atom"
of quarks and antiquarks, then there is a tendency toward
destructive interference of long wavelength radiation from the
various color charges inside, just as in QED.  This effect can be
incorporated by considering the Sudakov sums from the whole bound
state coherently. By translational symmetry, the radiation of a gluon
of momentum $q$ from an emitter at the origin and an opposite
emitter at a transverse position $b$ are added like $1 - e^{-ibq}$.
Expanding around $q =0$, the  ``monopole" term vanishes, leaving
perturbative dipole emission into the momentum region $q>1/b$. The
upshot is that coherence gives a cutoff on radiation exactly like an
infrared cutoff.

        Up to subleading proportionality constants, for color singlet
hadrons we can trade the infrared cutoff for the transverse size
with  $\Lambda = 1/b$, giving a Sudakov factor of the form  \cite{BS89}{}
\beq
S( Q^2,  \Lambda_{QCD}, b^2) = c\ell n (Qb)\ell n
\biggl({\ell n(Q/\Lambda_{QCD})\over \ell n(1/(b \Lambda_{QCD})}\biggr)
\eeq

\noindent
  The calculable coefficient $c$ depends on the color algebra,
and the number of particles entering and leaving the reaction.  The
chromo-coulomb phase is not shown, but takes the form of the
exponential of some calculable color matrices.  The form of the
expression above can be considered a transverse ``bin-by-bin" model
for the results of bound state coherence, acting much like a wave
function.   Presumably it applies well to the short distance
perturbative region of $b \rightarrow 0$, while application to large
distance regions of order the hadron size is problematic.
Figure (4.4) shows the damping effects of this function on large $b$
values of the wave function.

        In an axial gauge, such leading logs are contained in diagram
topologies which are wave function-like, allowing one to count their
properties without engaging the details of the hard scattering.  In
any gauge the next-to leading logarithms are more difficult and will
not be found to ``factorize" so simply. We refer the reader to
systematic procedures in the literature which re-organize the
pertubation theory without overcounting diagrams and momentum
integration regions.

\medskip

{\it Sudakov-Damped Regions Versus Sudakov Suppression }
\medskip

{\it a. Asymptotics}

        There is no doubt that the Sudakov effects are real, but there is
controversy about their numerical importance.  Early arguments
 claimed that the $Q^2$ dependence would rapidly damp
out processes, such as independent scattering, associated with
Sudakov type corrections.  These arguments have not been supported
by subsequent studies.  The $Q^2$ dependence is simply not
sufficiently rapid to make strong suppression for experimentally
accessible values of $Q^2$.   The term
``Sudakov suppression" is obsolete when applied in this way.

        A different approach is to find which regions of $b$ are
contributing at a given value of $Q^2$.  The interplay of these two
scales causes one to depend on the other. First, let us note that the
regions of large logarithms of $b$ are the regions of $1/Q< b <
1/\Lambda_{QCD}$; these regions are of interest and experimentally
accessible.  Second, inspecting the Sudakov formula, one observes that
 the region of  $b$ scaling like $Q^{-A}$, where $A$ is a positive (but not
necessarily integer) power, gives a constant, unsuppressed exponent
as $Q^2$ is increased. This is a candidate for a dominant region.
With this observation and a saddle-point approximation,  Mueller
 \cite{M81}{} came to the conclusion that independent scattering would
become a short-distance process in the limit of asymptotically
large $Q^2$, namely $Q^2$ going to infinity in a mathematical limit.
In this limit Mueller estimated an effective power behavior for the
amplitude, showing that the contributiuon was larger than the
quark-counting one.

        In the same region and with a similar saddlepoint
approximation, Botts and Sterman \cite{BS89}{}  showed that the Chromo-coulomb
phase effects disappear.  This can be anticipated, because a power-behaved
amplitude does not have an energy dependent phase.

         However, in the absence of truly enormous $Q^2$, are the
saddle point arguments quantitatively reliable?  In numerical
studies, Botts \cite{B91}{} subsequently showed that the asymptotic region
argument sets in at about $Q^2 = 1$ TeV$^2$, which is far too high to
be relevant to experiments.

\medskip
{\it b. Realistic Values of $Q^2$}
\medskip

        There is considerable current interest on the regions of $b$
relevant to current energies.  At accessible energies and momenta
many studies\cite{B91,JaK93} have shown that comparatively large values of $b$
ranging up to $1/\Lambda_{QCD}$ make substantial contributions.  This
can be seen by consulting Fig. (4.4).  In practice, many authors
investigate the effects of the Sudaokov factor multiplying models of
the ``soft" hadronic wave functions which extend out to about a
Fermi. Li and Sterman  \cite{LS92}{} were able to use the Sudakov ``wave
function"
in a calculation of the electromagnetic form factor, illustrating this
effect \cite{JK93}{}.    While the regions of
very large $b$ are completely eliminated by the Sudakov effect, some
worrisome regions which cannot easily be called ``short distance"
remain to give sizable contributions.

        If finite values of $b$ do give substantial contributions, then
the Chromo--coulomb phase should continue to produce oscillations.
One can learn much from spin, because spin dependence is another
piece of evidence that very short distance is simply not achieved in
real reactions.  There is considerable consistency in the picture
between electromagnetic form factor studies and hadron hadron
studies.  Recently, the effects of all wave functions in
$\pi \pi \rightarrow \rho \rho$
 scattering have been explored  \cite{GPR95}{} showing
that hadronic helicity violation might also be attributed to non-short distance
effects even in the presence of Sudakov corrections.

\medskip

\noindent {\it Color Transparency  and Nuclear Filtering as Sudakov
Effects}

\medskip

        Recall that in color transparency, big $b$-zones of the wave
function interact with the nuclear medium, while small $b$-zones
survive.  In Sudakov effects, big $b$-zones of the wave function
radiate inelastically into free space, while small $b$ zones survive to
make exclusive channels. The mechanisms of color transparency and
nuclear filtering have the same basic feature of ``survival of the
smallest".   In comparing the two, the ``filtering" aspects depend on
the medium, which can be vacuum or nuclear matter.    At any
particular value of $Q^2$, the nuclear medium sould be more
effective in selecting short distance than the vacuum because of its
stronger interactions. \cite{RP89,PR912}{}.

        This suggests that repeating the calculation of Sudakov
effects in a nucleus, and in particular summing pattern in
perturbation theory, might lead to a quantitative, perturbative
theory of color transparency. Unfortunately there is no theoretical
agreement on how to model nuclear matter in QCD.

        Let us consider this idea in a medium with the average
properties of nuclear matter, but lacking some of the detailed
correllations. A perturbative gluon can go into free space, or be
absorbed and re-emitted by spectator nucleons.  The result is
renormalization of the gluon propagator in the nuclear medium. There are
some constraints to be respected in modeling this.  The effects of the nucleus
will break Lorentz
symmetry while remaining gauge invariant.  In a derivative
expansion, one could find terms in an effective action ${\cal L}_{eff}$ of the
form
$${\cal L}_{eff} =\epsilon  E^2/2 + B^2/ 2 \mu  + \bar\psi (i\slash \partial -
\slash A)\psi$$
where $E$ and $B$ are the gauge--covariant electric and magnetic fields,
and $\epsilon$ and $\mu$ are contants representing the medium.  In principle
$\epsilon$  and $\mu$ can depend on position.  From causality the theory
would be non-local, a further complication.  Let us investigate the
local form above anyway.  In the temporal $A_0 = 0$ gauge defined by the
vector $n_\mu =(1,0,0,0) $, the renormalized  gluon propagator takes the form
\beq
d_{\mu \nu} = (1/\epsilon )  1 /(w^2 -{\bf k}^2+ (1/(\epsilon \mu) -1 )
{\bf k}^2
) R_{\mu \nu}
\eeq

\noindent
where $R_{\mu \nu}$ is the usual tensor
$$R_{\mu \nu} = -g_{\mu \nu}+ \frac1 {n \cdot k} (n_\mu k_\nu+n_\nu k_\mu)
-\frac
{n^2} {(n \cdot k)^2}k_\mu k_\nu . $$
\noindent
(It is well known that other tensor terms can also contribute in a
medium.)  The point of this example is the renormalization of the
pole value by $1/\epsilon$.  Screening in QED makes $\epsilon >1$.
Antiscreening in QCD would be expected to make $\epsilon <1$,
increasing the value of the gluon propagator near the pole.

 Neglecting ($1/(\epsilon \mu) -1 ){\bf k}^2$ compared to ${\bf k}^2$ in the
denominator, we can quickly anticipate the results.  They are
equivalent to $g^2 \rightarrow  g^2/\epsilon$, which is coupling constant
renomalization to a stronger coupling.  The sum of Feynman diagrams
in leading-log region can also be found using the new propagator.  It
is equivalent to known Sudakov corrections replacing  $g^2\rightarrow
g^2/\epsilon$ in the final answer: a nuclear Sudakov factor of the form
\beq
 \exp\biggl(-S( Q^2,  \Lambda_{QCD}, \Lambda^2) \biggr) =
\exp\biggl( - {const\over {\pi \epsilon}}\ell n (\frac{Q^2}{\Lambda^2})\ell
n\biggl(
\frac{\ell n
(Q^2/\Lambda^2_{QCD})} {\ell n(\Lambda^2/\Lambda^2_{QCD})}\biggr)\biggr)
\eeq

\noindent
  In Figure (4.4) we show comparisons of the relevant $b$-space
profiles of this expression as $\epsilon$ is varied.  For any given value
of $Q^2$, as $\epsilon$ is varied to smaller values, the effects of
filtering become more pronounced, exactly as anticipated in the
coherence arguments.

        Although theory cannot yet demonstrate such a formula
rigorously, we believe that future progress in defining the nuclear
medium may lead to results of this kind.  It would be extremely
interesting to know whether color transparency and nuclear filtering
might be described within such a ``leading log" framework, because
then the theory would be tightly constrained.  Leading log sums are
probably more reliable than the present situation embedded in
higher twist concepts.  One can hope that the perturbative
calculability of aspects of color transparency and nuclear filtering
might be put on a secure footing by more work along such lines.
\newpage

\subsection{Figure Captions}

\begin{itemize}

\item[4.1]  Fermion -  fermion scattering with vector exchange is dimensionless
and
scale invariant in the fixed-angle limit (left).   Adding a pair of Fermions
connected by a hard scattering, the amplitude is smaller by a factor of $1/Q$.
\item[4.2] Coordinate space picture of meson-meson independent scattering.
Scattering planes (dashed) and directions of quarks (arrows) are independent,
but to make a final state hadron the important region is the case where they
are
parallel.  The intersection of the Lorentz contracted hadrons (initial state
pancakes are shown) and the region of integration over the transverse
separation
between quarks $(b)$ inside the hadrons is a cigar shaped overlap region.
Pancake wave functions of outgoing hadrons have been omitted.
\item[4.3] Factorization in QCD.  Small internal momenta ($k$) circulating in
wave
functions (blobs) are separated from the hard scattering kernel H by
integrating
over them.
\item[4.3.5] (a) Hard scattering in the impluse approximation selects a region
where the quarks are close together in a full--sized hadron. (b) Compressed
state hypothesis assumes that the state is quantum--mechancially prepared to
exist in a small spatal region.
\item[4.4]  (a)  The Sudakov factor for meson-meson scattering plotted as a
function of the transverse separation b, in units of Fermi, using
$\lambda_{QCD}
=100$ MeV. The region which survives after coherent emissions of gluons are
taken
into account is restricted to $ b \le 1/\lambda$;  large regions of  $ b$
associated with
large color dipole moments are strongly damped (b) Same as (a) but using $g^2
\rightarrow g^2/\epsilon$ as a renormalized coupling, with $\epsilon = 1,
1/2, 1/3, 1/4$
(top to bottom). Smaller $ \epsilon $ leads to stronger suppression of
large $ b$ - regions.
Both (a) and (b) have been ``flattened" in the short distance region $ b<1/Q$
where
standard Sudakov expressions should be patched into short distance expansions.

\end{itemize}

\newpage

\section{Models of Propagation}
\setcounter{equation}{0}
\medskip

    In the preceeding section we have explained and
developed the concept of color transparency from the pQCD point of view.
These ideas are certainly valid once we go to high enough energies. However
it is not clear exactly at what energies this formalism will be applicable.
It is therefore useful to approach the problem also from lower energy
points of view which, however, necessarily require modeling.
 This problem has been addressed extensively in the literature. We review
in this section some of the ideas and models that have emerged.

\subsection{The Glauber Theory Framework}

    Color transparency effects represent the deviation of the experimentally
measured nuclear cross sections in comparison to Glauber theory \cite{G59},
which provides a framework for calculating the attenuation of fast moving
hadrons
through a nuclear medium using standard hadron--nucleon interactions. It
is based on the assumption that the nuclear potential changes very slowly
in comparison to the center of mass motion of the hadron. We do not cover the
Glauber formalism here since it is reviewed extensively in the literature, see
for
example Ref. \citenum{PE74}.
Let us, however, make a few comments. Glauber theory has been used as a
``bottom line" against
which color transparency effects might be measured.
There exist many different calculations\cite{BFF92,FF92,FMS95,ILS94,KYS92,LM92}
of this type for
both hadro-production and electro-production experiments.
These calculations typically use pQCD predictions or free space data
for the single hard
scattering, and Glauber formulation or some variation
for the soft scatterings with the
nucleus. Although minor differences in assumptions might be expected, it is
somewhat disturbing that
the results of these calculations vary considerably, of the order of
10-30\%, from one another. Nevertheless almost all of these authors agree that
they do not find much $Q^2$ dependence of the transparency ratio purely by
inclusion of nuclear medium effects, and
the results of these ``traditional" calculations disagree with the
BNL hadro-production experimental data. Furthermore, most of these calculations
remain below the BNL results for the transparency ratio.
 The only exception is the work of  Frankel and Frati \cite{FF92} who
surprisingly
reproduce the BNL data purely in
 terms of Glauber calculation (claiming that an important  effect comes when
the finite size of the proton is subtracted from the distance $z$ that the
nucleon traverses in the nucleon, which drastically modifies their calculation
of the suppression
 factor $P(r)={\rm exp}(-\sigma\rho z_i)$,in the survival probability). In
contrast to the hadro-production experiments, the
SLAC electro-production data is reasonably close to Glauber predictions.

  Color transparency effects can be included in
 the Glauber formalism, if it is modified
in several ways. At the simplest level it is sufficient to use a reduced
hadron-hadron interaction strength.
 The reduction in hadron-hadron cross-section is prescribed by QCD and
goes like $1/Q^2$ where $Q$ is the momentum transfer.
 However many authors \cite{FLF88,JM93,KZ912} have argued that because of the
expansion of a proposed
small wave packet,
the interaction strength will also depend on the position of the propagating
hadron. At the laboratory energies of the order of about 10 GeV, this is
claimed to be a large effect not calculable from fundamental theory,
thereby making model calculations necessary. Whether this is true or not is
 controversial. Many such model
calculations will be reviewed. Another important effect,
which is crucial for hadro-production experiments, is the possibility of
sizeable contributions due to long distance components in free space
scattering \cite{RP88}. These contributions are filtered in the nucleus and
introduce an
additional parameter, which for large nuclei might be incorporated as
a shift in the hard interaction strength.

\subsection{A Simple Model Calculation}

  The basic idea of color transparency  is that the process of hard
scattering involves a region of the wave function of very small spatial size
($r\approx 1/Q$)
which then evolves through the nucleus. Most authors have interpreted this to
mean that the entire wave packet is small, a ``compressed state."
Since this wavepacket is not an
eigenstate of the Hamiltonian, it expands and evolves as it travels through
the nucleus. Modeling the wavepacket and its subsequent evolution can be
 treated by introducing a complete set of states. The various matrix
 elements necessary for this calculation can be obtained by using some models
like the non-relativistic quark model or the Skyrme model. Partial information
 about these matrix elements can be also obtained phenomenologically.

 Let us first discuss this approach using a toy model of Blaizot, Venugopalan
and Prakash \cite{BVP92} which is quite illustrative and highly
pedagogical. The main simplifying idea is to use a nonrelativistic harmonic
oscillator model to discuss the evolution of a $c\bar c$ pair produced
in a diffractive photoproduction or electroproduction process, assuming
 factorization of the nonperturbative soft scattering process
from the  hard scattering part. In this case the straightforward impulse
approximation (Section 2) does not apply to diffractive kinematics.
However, Feynman diagrams contain momentum transfers of order $m_c$ or greater,
indicating short distance.
It is natural to describe the $c\bar c$ pair
as a very small wave packet at the time of the hard scattering. The interaction
of the
system with the nuclear medium is then assumed to be of the form
\beq
H_1 = e{\cal E}z
\eeq
where $z$ is the relative coordinate for the charm pair and ${\cal E}$ is a
model color ``electric field."
The wave packet $\phi$ at the hard scattering is parametrized by  a gaussian
with size
$r \approx 1/Q $, $Q$ being the momentum transfer in the hard scattering,
which is much smaller than the size of the ground state wave function.

    The evolution of this wavepacket after the hard scattering is then
determined by the total hamiltonian, that is the sum of the harmonic oscillator
 Hamiltonian plus the perturbation $H_1$ given in equation (5.1).
 One then defines the absorption factor $S$ such that,
$$
{P(T)\over P(0)} = e^{-S}
$$
where $P(T)$ is the probability of finding the system in the ground state
of the unperturbed harmonic oscillator hamiltonian at time $T$, where $T$ is
larger than the time that the charm pair spends inside the nucleus. In the case
when the radius of the small wave packet is much smaller than the radius of
the ground state wave function, $S$ is found to be
\beq
S \approx {2p_0^2\over <p_z^2>_0 + <p_z^2>}\ ,
\eeq
where $p_0= e{\cal E}\tau$, $\tau$ is the time over which the charm pair
interacts with the nucleus and $<p_z^2>\equiv <\phi|p_z^2|\phi>$. In the limit
when the size of the initial wave packet goes to zero, $<p_z^2>$ goes to
infinity and therefore $S\rightarrow 0$. The model therefore leads to color
transparency and gives reduced attenuation as the size of the initial
wave packet becomes smaller. The above result for $S$ (Eqn. (5.2))
is valid
in the case when the charm pair undergoes no free expansion prior to its
interaction with the nucleus.

   The calculation with the compressed state initial conditions can be compared
with one consistent with the impulse approximation, a factor of $\exp (-iQ\cdot
r)$ [Eq. (2.5)].
Markovoz and Miller \cite{MM94} have analyzed such a model to test
the assumptions behind color transparency. They model the ``nucleon'' as being
the ground state of an electrically neutral system of two quarks interacting
via the Coulomb potential. They find that the basic three assumptions namely,
i) selection of a small region in a high momentum transfer reaction,
 ii) reduced interactions between a small ``color'' neutral wave packet and the
nucleus and iii) long expansion time of the wave packet compared with the
time of propagation in nucleus, are satisfied in their model. They find
a rate of expansion of the wave packet that is inversely proportional to the
momentum transfer $Q$ and is therefore extremely small.

\subsection{Classical Geometrical Expansion Model}

   Farrar {\it et al} \cite{FLF88}
studied both hadroproduction and
electroproduction color transparency experiments within the
framework of a classical physics geometric model for the
expansion of the wavepacket. The authors argue that a small size
configuration is formed at the hard scattering and estimate its
transverse expansion during its travel through the nucleus: they propose
two models, the first one, called naive, being driven by the equation:

$$b \sim t \sim (m/E)z\; \; ; \sigma_{abs}\sim z^2$$

\noindent
where $E/m$ is a time dilation factor and $z$ the distance traveled during
the time $t$. The second model, said to be inspired by
perturbative QCD, is given by
$$b \sim  z^{1/2}\; \; \; \; \; \sigma_{abs}\sim z\; .$$

\noindent
Their ansatz for an effective cross-section of the expanding particle is
\beq
\sigma_{eff} = \sigma_{tot}\bigg[ [(z/l)^k + \frac {\langle n^2k_t^2 \rangle}
t
(1-(z/l)^k)] \theta(l-z) +\theta(z-l)\bigg]
\eeq
\noindent
where $k$ is $1$ or $2$ depending on the  model,  $l$ is an
estimated effective length of the Lorentz-contracted nucleus,$n$ is
 the number of valence quarks in the expanding object and $k_t$ is
an average internal transverse momentum. The authors advocate different
expressions for $l$ in their two models. While they favor $k=1$, one notes that
a $\sqrt{t}$ rise of $b(t)$ corresponds to a transverse velocity which is
superluminal and violates causality \cite{JSW92},
casting into doubt its usage as a
classical cross section.

    The merit of this approach is its simplicity and physical motivations. It
does not, however, pretend to be more than an educated guess, a quasi-classical
picture. The resulting predictions of this paper are given in Fig (3.8).  As
discussed
 in Section III, it is no surprise that the BNL data cannot be understood
 within such a model. As for (e,e',p) SLAC data, the $Q^2$ region accessed
is too narrow and the data's precision is too limited to allow definite
conclusions to be drawn.  One can infer that the most
optimistic models in this framework are disfavoured.

\subsection{Hadronic Expansion Models}

   The completeness of the set of hadronic states  remains an assumption of
strong
interaction-physics dating from $S$--matrix days. It is not clear how this is
to work in a confining theory. In most works in the hadronic basis, it is
simply
assumed a priori that one may
expand a small configuration present at the hard scattering as
a series of states selected from a discrete and a continuum spectrum. One very
interesting goal is to understand how coherence among the {\it internal}
coordinates (destructive interference of gluon exchange) might be understood in
terms of coherence between the {\it external} (hadronic) coordinates. To
realistically implement this idea requires a judicious choice of approximations
which may be difficult to justify. This lead various groups to propose diverse
descriptions of the expansion of the mini-states.

   In the specific example of $eA\rightarrow e'p(A-1)$, Jennings and Miller
write
the amplitude ${\cal M}_\alpha$ to knock a hadron out of the nuclear shell
 model orbit $\alpha$ as a sum of two terms,
$$
{\cal M}_\alpha = B_\alpha  + ST_\alpha
$$
where $B_\alpha$ refers to the Born term and $ST_\alpha$ refers to the
scattering terms shown diagramatically in fig. (5.1). The Born term
corresponds to no interaction with the nucleus and is given by,
$$
B_\alpha = <p|T_H(Q)|\alpha> = F(Q^2) <p-q|\alpha>\ ,
$$
where $T_H$ refers to the photon-proton hard scattering, $q$ is the momentum
carried by the photon, $p$ is the momentum of the outgoing proton and $F(Q^2)$
is a generic proton form factor (taken as the free space magnetic form factor
 $G_M$).

    The scattering term is written as
$$
ST_\alpha = <p|UGT_H(Q)|\alpha> \ ,
$$
where $G$ is the Green's operator for the wave packet emerging from the
hard scattering and $U$ represents the nuclear interactions of the ejected
hadron.
This matrix element may be evaluated by inserting a complete set of hadronic
states. The resulting expression for the scattering amplitude is then written
as:
\beq
ST_\alpha = \sum_{m}\int d^2B dZ dZ' {e^{iqZ}\over (2\pi)^{3/2}}
U_{Nm}(B,Z)G_m(Z,Z')
e^{iqZ'}<B,Z'|\alpha> F_{mN}(Q^2)\ ,
\eeq
where the sum over $m$ extends to all the hadronic states.
The matrix elements of the operator $U$ that appear in this expression are
 written as,
$$
U_{N,m}(B,Z) = \int d^2b<N|b>U(B,Z;b)<b|m>\ ,
$$
where $|m>$ are the eigenstates of the internal Hamiltonian and $|N>$ is
the nucleon state. The variables $B$ and $Z$ are the transverse and
longitudinal
distances between the ejected hadron and the nuclear center and the variable
$b$ refers to the internal coordinates of the hadron, the
transverse component of the interquark separation. The inelastic form factor
is given by
$$F_{mN} (Q^2) = <m|T_H(Q)|N>$$
Finally the eikonal Green's function $G$ is taken to be equal to
$$G_m(Z,Z') = \theta(Z-Z') {e^{ip_m(Z-Z')}\over 2ip_m}\ ,$$
where $p_m^2 = p^2+M^2-M_m^2$ and $M$ is the nucleon mass.

         To compute the scattering amplitude $ST_\alpha$, the interaction
matrix
$U_{N,m}$
is modeled as
$$
U(B,Z;b) = -i2E{\sigma\over 2}\rho(B,Z){b^2\over b_H^2}\ .
$$
where $\sigma$ is the nucleon-nucleon cross section, $\rho$ is the nuclear
density, $b_H$ is the size of the hadron and $b$ is the position operator
measuring the transverse separation of the quarks inside the hadron.
The nuclear interaction amplitude, given in equation (5.4) can then be computed
within some specific model for the hadron wave functions.

       Jennings and Miller \cite{JM91} (JM)
illustrate their ideas by using a simple harmonic
oscillator model in two dimensions and restricting the sums to only
the ground state $N$ and the first excited state $N^*$ contributions to the
scattering amplitude. The contribution of both these states turns out to be
important to calculate the scattering amplitude. Within this model and using
 different values for the excited state mass the authors
concluded that the results of the calculation do not agree with the BNL data
on $pp$
scattering. In particular they do not obtain the decrease in the transparency
ratio experimentally observed above
10 GeV.

       In Ref. \citenum{JM921},
JM repeated their calculation by including in
the effects due to the independent scattering contribution \cite{RP88}
 and also the charm threshold proposal \cite{BT88}.
They concluded that both of these effects
go in the right direction for explaining the data, but that
their calculation still did not yield a sufficiently large  oscillation.

          The authors also performed their calculation
using a more realistic treatment of baryonic components of the wave packet
created at the hard scattering. This treatment \cite{JM922} involves using as
much information as possible directly from the experimental data to compute
the hadronic matrix elements thereby reducing the model dependence.
Assuming first that at the hard scattering a point like
configuration  is formed which has no interaction with the nucleus,
$$
UT_H(Q^2) |N> = 0 $$
 and inserting a complete set of hadronic states, they obtained
\beq
\sigma_p + \sum_\alpha\int_{(M+m_\pi)^2} dM_X^2 <N({\vec
q})|\hat\sigma_p|\alpha,M_X^2>
{<\alpha,M_X^2|T_H(Q^2)|N>\over F(Q^2)} = 0 \eeq
 where $\sigma_p$ is the proton-nucleon total cross section. As the wave
packet propagates through the nucleus its interaction with the nucleus  no
longer
vanishes. Defining an effective {\it mini hadron}--nucleon cross section,
$\sigma_{\rm eff}(l)$,
\beq
\sigma_{\rm eff}(l) \equiv \sigma_p + \sum_\alpha\int_{(M+m_\pi)^2} dM_X^2
<N({\vec q})|\hat\sigma_p|\alpha,M_X^2>e^{i(p_X-p)l} {<\alpha,M_X^2|T_H(Q^2)
|N>\over F(Q^2)} \eeq
 where each of the components has picked up a phase factor $e^{iP_Xl}$ with
$$p_X^2 = p^2 + M_N^2 - M_X^2\ ,$$
 one is then lead to compute $\sigma_{\rm eff}(l)$.
 One thus needs the
two matrix elements, for which only  partial experimental information is
available. In particular the relative phases of these matrix elements are
unknown. These matrix elements are then written in terms of
cross sections,
$$
<N({\vec q})|\hat\sigma_p|\alpha,M_X^2>_{DD}  =
\left[{d^2\sigma^{DD}(\alpha)\over
dt dM_X^2}\right]^{1/2}$$

$$
_{DIS}<\alpha,M_X^2|T_H(Q^2)|N> = \left[{1\over \sigma_M}
{d^2\sigma^{DIS}(\alpha)
\over
d\Omega dE}\right]^{1/2}
$$
\noindent where $DD$ and $DIS$ refer to diffractive dissociation and deeply
inelastic
scattering. (A subtlety in Eq. (5.6) is that the experimentally measured final
states in diffractive dissociation and deeply inelastic scattering are not the
same, either theoretically or experimentally, so going from Eq. (5.5) to Eq.
(5.6) requires dynamical assumptions.)
Further modelling is still needed: one may parametrize the sum
of matrix elements as a
simple function $g(M_X^2) = (M/M_X)^\beta$.
The resulting calculated results, however, still do not agree well with the
data.

        Finally in Ref. \citenum{JM93}, JM further improve their calculation
by including several kinematic and dynamical effects which had previously
been ignored. The new effects included were a proper treatment of the
longitudinal (parallel to the virtual photon or incident proton momentum)
component of the momentum  of the detected protons
and careful inclusion of Fermi motion.
  The authors then find that their results do agree with the experimental
results on $pA$ quasielastic scattering as published by the BNL group.

      Anisovich et al. \cite{ADN92} have also performed a dynamical calculation
of
 color transparency in $pA\rightarrow p'p''(A-1)$ process with the inclusion
of the mechanism  \cite{RP88}  for interference of soft and hard hadronic
components. The authors use the
 diquark model of the proton, which allows for a considerable simplification
of the calculation. This paper can be recommended as a reasonably complete
calculation including proper kinematics and attention to detail.
Their results, with the inclusion of color transparency effects, agree with BNL
data \cite{CEA88}.
Similar results have also been obtained by Kohama et al. \cite{KYS95}.

\subsection{Nonrelativistic Quark Model Calculations}

    Kopeliovich, Nikolaev and collaborators
\cite{KZ911,KZ912,KNN93,NSS941,NNZ94}
have applied the nonrelativistic
quark model
to color transparency calculations. The model is expected to be
closer to reality in the case of heavy quark systems like charmonium than for
light quark hadrons. Let us
take the concrete example of diffractive charmonium production
\cite{KZ911,BKM92} in the process
$\gamma N\rightarrow c\bar c N$ to describe this type of calculation.
Using a harmonic oscillator model to describe the interaction
between quarks, the Lagrangian of the quark antiquark system in interaction
with the nucleus may be written as
\beq
L_{\rm eff}(\tau,\dot{\tau},t) = L(\tau,\dot{\tau}) + {iv\gamma\over 2}
\sigma(\tau_T)
\rho_A({\bf r}(t)) \eeq
where $\tau$ is the interquark separation, $\sigma$ is the total $c\bar c$
nucleon cross section, $\rho_A$ is the nuclear density, $\gamma$ and $v$
are the Lorentz factor and the velocity of the $c\bar c$ pair in the laboratory
frame and $L(\tau,\dot{\tau})$ is the Lagrangian of the $c\bar c$ system in
free
space.
$$
L(\tau,\dot{\tau}) = {\mu\dot{\tau}^2 \over 2} - {\mu\omega^2\tau^2\over 2}
$$
Here $\mu$ is the reduced mass $\mu = m_c/2$ of the $c\bar c$ system and
the frequency $\omega= (M_{\psi'} - M_\psi)/2$
 is adjusted to fit the mass difference of the low lying
charmonium states. The evolution operator
representing the  propagation of the
$c\bar c$ pair through the nucleus is then given by
$$
U = \int D^3\tau {\rm exp}\left[i\int dt L_{\rm eff}(\tau,\dot{\tau},t)\right]
$$
This path integral is estimated  by breaking it up into
several smaller steps such that for each step the nuclear density can
be taken to be constant. This facilitates the calculation since then an
analytic expression is available for each step of constant density for the
case of harmonic oscillator. The full evolution can be calculated by
convoluting the different steps.

The remaining unknown is the initial wave function for the $c\bar c$ pair.
Neglecting the interquark interaction, the authors assume the
transverse part of the wave function as
$$
\Psi_\gamma(\alpha, k_T) \propto (m_c^2 + k_T^2)^{-1}
$$
where $\alpha$ is the light cone momentum fraction carried by one of the
quarks.
In position space  $b$ this model wave function is
$$
\Psi^T_{\rm in}(b) \approx {\rm const} \times [{\rm exp}(-b^2/A^2) -
{\rm exp}(-b^2/B^2)]
$$
where $A=0.536$ fm and $B=0.11$ fm.
In the low energy region, the authors model the wave function differently as,
$$
\Psi^T_{\rm in}(\rho) \propto {\rm exp}(-b^2/a^2)
$$
where the parameter $a$ is allowed to vary in a reasonable range,
 $a=1/m_c,\ 2/m_c,\ 3/m_c$.

The transparency is then calculated by taking the ratio of nuclear
to free space cross sections,
$$
T_N(E) = {\int d^3r \rho_A(r) |<\Psi_f|\hat U|\Psi_{\rm in}>|^2\over A|<\Psi_f|
\Psi_{\rm in}>|^2}
$$
The final results of Kopeliovich et al's calculation for charmonium are shown
in fig. (5.2). The authors have also used their formalism to calculate
transparency ratio for
photo-production of $\rho$ \cite{KNN93} and $\phi$ mesons \cite{BZN95} and
their radial excitations. The results for $\rho$ and $\rho'$ are shown in fig.
(5.3).
In this case the use
of a nonrelativistic oscillator model is, of course, not very credible.
Nevertheless
their results for $\rho$ meson were found to be in good agreement with
the E665 experimental results. However they also predicted substantial
difference between the transparency ratio for $\rho$ and $\rho'$, whereas
experimentally not much difference was found between these two particles
\cite{F94}.
Their predictions for $\phi$ and $\phi'$ mesons \cite{BZN95}
 for CEBAF energies remain
to be tested. Finally
this procedure has also been applied \cite{KZ912}
 to the calculation of transparency
in $pA\rightarrow p'p''(A-1)$ in quasielastic collision.
Within their model, the authors show that their theory does not reproduce the
BNL experimental results. This result is obtained both with and without the
inclusion of independent scattering contributions.

Gardner \cite{G93} has also done a detailed study of color transparency in
photoproduced charmonium using a non-relativistic harmonic oscillator
model. The initial state is assumed to be a small wave packet formed out
of superposition of many eigenstates. The nucleons in the nucleus are modelled
as strings which carry uniform U(1) color field.
The calculation shows that the quantum mechanical coherence effects
in the final state interactions are very important.

In most of these calculations the true initial state is guessed in a resonable
way, but actually unknown. Because of the sensitivity of the calculation to
coherence of the wave packet, it might be interesting to try different
variations of the initial state. One idea which has not been explored is to
tune the initial state in various ways while observing which choices propagate
best. In this way, one might explore Nature's quantum mechanically selected
states that actually will survive.

The Skyrme model may be used \cite{JSW92,EK92} to test for the existence of
color transparency.
Within this model the baryons emerge as solitons within an effective
theory of mesons. The model is known to describe quite well the low
energy properties of strong interactions. Although color transparency is a high
energy phenomenon such model calculations may qualitatively describe the
onset of color transparency at low energies. Both the studies find that
Skyrme model supports the appearance of color transparency.

\subsection{Fermi Motion Effects}

Jennings and Kopeliovich \cite{JK93} (JK)
have argued that color transparency of the
nominal elastic point is possible
only due to Fermi motion. The authors proceed within the hadronic basis
by taking the specific example of electroproduction. They argue that in the
limit Bjorken $x_{Bj}=1$ only the proton can be produced on shell, if the
Fermi momentum is neglected. Therefore no excited states are produced, and
within their model it is not possible to produce an ejectile with small size
which
requires interference of several amplitudes.

This argument appears to be wrong
for elementary kinematic reasons \cite{NSS93}.
The excited states needed in the model are immediately created
diffractively when the nucleus makes the ``measurement" by being probed by the
fast ejectile system.  However the authors make interesting experimental
sugggestions, which focus attention on the observability of elastic events at
$x_{Bj}\not= 1$ due to Fermi motion.
The authors then conclude that the transparency effects become considerably
smaller
with increase in $x_{Bj}$.
By using a two component model, JK conclude that
color transparency effects will be negligible in electroproduction for
momentum $Q^2<7$ GeV$^2$, whereas considerable effects are predicted at
$Q^2>10$ GeV$^2$. JK further argue that the unexpected behavior of the
BNL (p, 2p) scattering experiment is understandable once the effects of
Fermi motion (as measured by a momentum fraction $x$) are carefully included.
Their  argument is that for a fixed
beam energy, higher center of mass energy $\sqrt{s}$ corresponds to higher
Fermi
momenta in the direction opposite to the incoming proton. This means that
for fixed
beam energy, $x$ is larger for larger $\sqrt{s}$. This explains the decrease in
transparency observed at 12 GeV since according to their arguments transparency
decreases with increasing $x$. The authors argue that this does not happen at
the lower energy of 6 GeV due to ``strong mixing of eigenstates." The authors
also concede that the explanation of the
10 GeV points is problematic within this framework.

Bianconi, Boffi and Kharzeev \cite{BBK931,BBK932,BBK94} have also studied the
phenomenon of color
transparency within the hadronic basis. They  agree on a crucial role of
Fermi motion in exciting higher mass states which can interfere to form
a small sized wave packet that can exhibit color transparency.

Others do not  agree with these conclusions.
Anisovich, Dakhno and Giannini \cite{ADG94} calculate transparency in
electroproduction
experiment using the deuteron as the target nucleus. This is one of the
simplest
systems that can be studied which is expected to show some color transparency
effects. The calculation involves approximating the struck proton and the
spectator neutron interaction by means of one-pomeron exchange. Anisovich
{\it et al} also allow for the possibility that the photon produces resonances
which
 subsequently emerge as protons after diffractive scattering. No energy
dependence
 of transparency due to Fermi motion effects emerge from their calculation.
They further find that the onset of color transparency is very slow in all
cases considered, and suggest that for the investigation of color
transparency in electroproduction experiments both intermediate-energy as
well as next generation higher energy accelerators will be necessary.

Nikolaev et al \cite{NSS93} have also
developed a detailed theory of Fermi motion and have studied the
possibility of the strong $x_{Bj}$ dependence proposed in Ref. \citenum{JK93}.
 Working within the framework of the diffraction
operator technique, which treats the evolution and the final state
interactions of the wave packet consistently, the authors find that the
$x_{Bj}$ dependence is much weaker than what was claimed in Ref.
\citenum{JK93}.

We note that the conclusion of Refs. \citenum{JK93,BBK932}, namely that color
transparency
is not possible without Fermi motion, appears to be result of the picture
that a small sized wave packet is formed because of the large momentum transfer
process. The phenomenon of color transparency, however, does not require the
formation of such a wave packet. It only requires (Section 2) that short
distance
components of hadronic wave functions dominate the exclusive process under
consideration. This is discussed in more detail below. Although
it is very important to properly account for Fermi motion effects in order
to extract signals of color transparency, there is no agreement in the
literature with the conclusions
of Refs. \citenum{JK93,BBK932} concerning any crucial link between Fermi motion
and color
transparency.

\subsection{Quantum Chromotransparency}

We have so far outlined a variety of approaches
which aim to describe color transparency experiments
using hadronic basis or non-relativistic quark models for hadron
structure. In this section we show that color transparency can be obtained
directly from QCD in the high energy limit. We further show how the
factorization of hard scattering and the nonperturbative wave function
is achieved in this limit. A considerable advantage of the pQCD approach is
that it represents a unified
theory which can be held accountable.  The effects of coherence are what the
theory predicts, and not imposed from the outside by a model. The initial state
is also not modeled as an outside input, but is obtained from a definite
calculation.

We follow Ref. \citenum{RP90} by assuming factorization for the ampitude $M$
between  the hard and soft
components. The
 amplitude $M$ for a hard scattering with 2 legs (Fig. 5.4) can be written
\beq
M(Q^2, A) = \bigg\lbrace\prod_{i,f}\int [dx_i d^2k_{T,i}]
\bigg[\psi_A^{(f)*}(x_f,k_{Tf})
H(x_i,x_f,k_{Ti},k_{Tf},Q^2)\psi_A^{(i)}(x_i,k_{Ti})\bigg]\bigg\rbrace \eeq
where $H(x,k_T,Q^2)$ represents the hard scattering,
$\psi_A(x,k_{T})$ represents the hadronic wave function inside the
nuclear medium,
and $Q^2$ is the characteristic momentum scale
in the collision.  In this equation we are emphasising the internal quark
coordinates
and suppressing integration over center of mass hadron variables, indicated
by curly brackets.
The wave function represents the overlap
of the short distance hadronic wave function with its propagation
through the nuclear medium and can be expressed as
\beq
\psi_A(x',{k'}_T^2) = \bigg\lbrace\int dx d^2k_T\bigg(
\delta(x-x')\delta^2(k_T-k_T')-
F_A\bigg[s,\bigg({x\over x'}(k_T-k_T')\bigg)^2
\bigg]\bigg)\psi_0(x,k_T^2)\bigg\rbrace \eeq
where $\psi_0(x,k_T^2)$ is the free space hadronic wave function and
$F_A$ is the nuclear
scattering amplitude. This is nothing more than the light--cone region of
$1-G_A$ used in Section 2. For
simplicity here we suppress the $i,f$ indices.
The nuclear filtering of soft components of the wave function will imply that
only the short distance part of the wave function survives. Therefore
the function $\psi_{0A}(x,k_T^2)$ may be approximated very well
by its short distance form, in contrast to the analogous quantity in
free space. Going to the quark transverse separation $b$ space we get:
\beq
\tilde \psi_A(x,b) = \tilde f_A(s,x,b)\tilde \psi_0(x,b) ,\eeq
where the tilde's denote the Fourier transforms and $\tilde f_A=1-\tilde
F_A$ is the nuclear survival amplitude. In Eq. (5.9) we have suppressed a
kernel
for transitions from $x\rightarrow x'$ and left the $x$--dependence as
diagonal.
The distribution amplitude
can then be written as
\beqn
\phi_A(x,Q^2)& =& \int^{Q}_0d^2k_T\int d^2b_T e^{i\bf b_T\cdot\bf k_T}
\tilde f_A(s,x,b^2)\tilde\psi_0(x,b)\nonumber \\
& = &(2\pi)^2 Q\int^{\infty}_0 db J_1(Qb)
\tilde f_A(s,x,b^2)\tilde\psi_0(x,b). \\
\nonumber
\eeqn
The above integral will get its dominant contribution from the $b\lsim 1/Q$
region of the wave function. This is exactly analogous to the selection of
short distance by $\exp (-iQ\cdot r)$ in the non--relativistic case (Section
2).

We assume that the wave function $\psi$ is a slowly varying
function of $b$ for small $b$. This assumption is supported by bound state and
renormalization group studies. We can
then approximate the above integral by replacing the wave function
by its value at $b\approx 1/Q$ and cutting off the integral at
the same value of $b$. By substituting this expression into Eq. (5.8)
we see that we have a factorized form for the amplitude $M$:
\beqn
M(Q^2,A) & = &\int [dx] \bigg\lbrace\tilde f_A(b\approx 1/Q)
 \bigg[\psi_{A0}^{(1)}(x_1,1/Q)
H(x,0,Q^2)
\psi_{A0}^{(2)}(x_2,1/Q)\bigg]\bigg\rbrace\nonumber \\
&\equiv& <\bigg\lbrace\tilde f_A(b^i\approx 1/Q)\ldots\tilde
f_A(b^f\approx 1/Q)\bigg\rbrace H(x,0,Q^2)>, \\
\nonumber
\eeqn
where the pointed and curly brackets indicate restoration of the integrals
over $x$ and
hadron center of mass coordinates, respectively, which code information
on nuclear size and density. Note that $H(x,0,Q^2)$ is independent of $A$ and
hadron center of mass coordinates for $A>>1$. That allows us to factor the
hard scattering out of the integral over center of mass coordinates.
Generally one cannot factor the $x$
dependence of the cross section. One might hope
that the nuclear medium effects do not much change the $x$ integrals over the
wave
functions.

The transparency ratio T can then be written as
\beqn
T(Q^2,A)& =& {d\sigma/dt|_A\over A\ d\sigma/dt|_{free\ space}}\nonumber \\
& \cong& {|<\lbrace\tilde f_A(b^i\approx 1/Q)\ldots \tilde f_A(b^f\approx
1/Q)\rbrace H(x,0,Q^2)>|^2\over
Ad\sigma/dt_{free\ space}}\nonumber \\
& \to& P_N(Q^2,A) R(Q^2),\\
\nonumber
\eeqn

where $N$ is the number of participating particles that cross the nucleus. The
first line is an approximation based on the expectation that the
$x$--convolutions more or less factor.
If
so, then
the transparency ratio can be factored into two pieces:
the survival probability $P(Q^2,A)$ and the ratios of the hard scatterings
$R(Q^2)$. Furthermore,
the leading $Q^2$ dependence of the nuclear hard scattering
can be obtained by considering the leading order perturbation theory
diagrams convoluted with the short distance hadronic wave function.

Let us now demonstrate
that perturbative QCD predicts color transparency at asymptotic
$Q^2$ \cite{RP90}. For this purpose, let us consider equation (5.11) at large
$Q^2$, and show that the nuclear-filtered wave function reduces to the short
distance
free space wave function as $Q^2\rightarrow \infty$.  This may be done
systematically
by using a Mellin transform, defining
$$
\phi_{A,N}(x) = \int_0^\infty{dQ\over Q}Q^{-N}\phi_A(x,Q)$$

$$\ \ \ \ \ = (2\pi)^2 2^{-N}{\Gamma(1-N/2) \over \Gamma(1+N/2)}
\tilde\psi_{A,N}(x)$$
using the Mellin transform of the Bessel function. Here $\tilde\psi_{A,N}(x)$
is the impact parameter moment
$$\psi_{A,N}(x) = \int_0^\infty{db\over b}b^{N}\tilde\psi_A(x,b).$$
Note that the $b$ moments are defined with the negative $-N$ of the $Q$
moments;
this is because powers of $b$ will turn out to map into powers of $1/Q$
(modulo logarithms).

Inverting $\phi_N$, complex $N$-plane singularities at $N=0$ are leading
twist; those at $N=-2$ give $1/Q^2$ higher twist, and so on. The wave function
$\tilde\psi_A(b)$ determines the $N=0$ singularity:
$$\tilde\psi_{A,N}(x) = \int_0^\infty {db\over b} b^N\tilde\psi_0 (b)
[1 - b^2 A^{1/3}n \sigma'_{\rm eff} + ....]$$
Note that the $N\rightarrow 0$ behavior is fixed by the short distance free
space wave function $\tilde \psi_0(b)$. Here the effect of interaction with
the nucleus is a factor ``1''. The $A^{1/3}\sigma'_{\rm eff}$ dependence
affects the $N\rightarrow -2$ behavior, and is thus suppressed by $1/Q^2$.
This is transparency. It is remarkably general in the asymptotic
limit $Q^2\rightarrow \infty$.

To illustrate the idea that at large
$Q^2$ nuclear wave functions reduce to the short distance free space
wave functions, let us assume a typical example of the hadronic wave function
$\tilde \psi_0(b) = {\rm exp}(-mb)$, and model the nuclear filtering
amplitude $\tilde f_A $ by $\tilde f_A(b) \sim {\rm
exp}[-(b^2A^{1/3}\sigma'_{\rm eff})^{1/2}]$.
Inserting this into equation (5.11), we get
$$
\phi_A(Q^2) = \phi_0(Q^2) \left(\left[1-{m+(A^{1/3}\sigma'_{\rm eff})^{1/2}
\over\sqrt{[m+(A^{1/3}\sigma'_{\rm eff})^{1/2}]^2 + Q^2}}\right]\bigg/
\left[1-{m\over \sqrt{(m^2+Q^2)}}\right]\right)
$$
which goes to the free space distribution $\phi_0(Q^2)$ as $Q^2\rightarrow
\infty$. By adjusting parameters $m$ and $\sigma'_{rm eff}$ many models can be
made. 	More sophisticated models are of course possible.  The advantage (and
purpose)
of the factorization/distribution amplitude formalism is that certain universal
functions are identified.  These are definite matrix elements, acccessible to
short distance renormalization group techniques.  If one also accepts that
Pomeron exchange is calculable for short distance dominated processes, (and
certainly there is a great deal of work in the literature claiming this is so),
then the entire effects of interactions, and all of color transparency, are
perturbative problems. Part of the underlying protection from infrared effects
is due to large $Q^2$, the original motivation\cite{B82,M82} for color
transparency.  The other part is large A, which also acts like an infrared
cut-off in the nuclear filter.\cite{RP88,RP90} A totally perturbative treatment
is evidently
realistic, and a very exciting prospect.  However, its exploration would take
us
into new research and remains for future work.

\subsection{Resum\'e}

We have established that pQCD predicts color transparency in the limit
$Q^2\rightarrow
\infty$. In order to compare with experiments, however,
one needs numbers for the magnitude of this effect at finite energy. The
usual approach requires modeling the nuclear medium which introduces
considerable
uncertainty into the calculation.

As emphasized in Ref. \citenum{PR911},
model uncertainty can be considerably reduced by noting that the nuclear
effects should follow a scaling law. The scaling law arises because
 the survival probability,
being dimensionless, can only depend on a dimensionless variable.
In the high energy limit the only dimensionless variable that is
relevant is the effective number of nucleons encountered by the hadron,
which is proportional to $n\sigma_{eff}A^{1/3}$, where $n$ is the nuclear
density,  and $\sigma_{eff}$ is the effective attenuation cross section.
 In the high energy limit $\sigma_{eff}$ is set by the dominant $b$ region
proportional
to $1/Q^2$, and therefore the scaling law follows. We have discussed
applications of this in Section 3.

In closing this section on Models of Propagation, a comment might be suitable.
If the various models can be understood well enough to extract other broad
regularities, such as modified scaling laws or results that are independent of
adjustable parameters, then clearly their scientific discriminating value
might be enhanced. Science would be well served by first establishing definite
trends, and afterwards fitting models.

\medskip
\noindent {\bf Acknowledgements:} This research was supported in part by DOE
Grant Number FG0285ER40215 and by the {\it Kansas Institute for Theoretical and
Computational Science}. We thank numerous colleagues for helpful discussions
and Don Geesaman for comments.
PJ thanks Ecole Polytechnique, the University of Tubingen, and M. Virasoro
along with ICTP, Trieste
for hospitality. CPT is unit\'e propre de CNRS.
\vfill
\eject

\subsection{Figure Captions }

\begin{itemize}
\item[5.1] The Born $B_\alpha$ and first scattering $ST_\alpha$ terms for
$(\gamma*p,p)$ scattering at high $Q^2$, Ref. \citenum{JM91}
\item[5.2] The predictions of Ref. \citenum{KNN93} for photoproduction
of charmonium.
\item[5.3] The predictions of Ref. \citenum{KNN93} for photoproduction
of $\rho$ and $\rho'$. The experimental results \cite{AEA95,F93,F94}
for these mesons, shown in Figs. (3.12-14), agree nicely with the predictions
 of Ref. \citenum{KNN93} for the case of the $\rho$ meson but fail to agree
for the $\rho'$.
\item[5.4] Factorization of the hard scattering $H(k,k',...Q)$ from the
soft evolution of the wave functions $\psi$ and $\psi^*$.
\end{itemize}


\newpage




\begin{thebibliography}{9999999}

\bibitem{AEA95}
M. R. Adams et al., Phys. Rev. Lett. {\bf 74}, 1525 (1995).




\bibitem{ADN92}
V. V. Anisovich and M. M. Giannini, Z. Phys. C {\bf 50}, 441 (1991).
V.V. Anisovich, L.G. Dakhno, V.A. Nikonov, and M.G. Ryskin,
Phys. Lett. {\bf B292}, 169 (1992).

\bibitem{ADG94}
V.V. Anisovich, L.G. Dakhno, and M.M. Giannini, Phys. Rev. {\bf C49}, 3275
(1994).

\bibitem{AP95} {\it Physics with a 15-30 GeV high intensity continuous beam
electron
accelerator: The ELFE project,} J. Arvieux and B. Pire, Progress in Particle
and Nuclear Physics, {\bf 30}, 299  (1995).

\bibitem{AS93} {\it ``The ELFE Project: an Electron Laboratory for
Europe"}, in {\it  Proceedings of the Italian
Physical Society},Vol. 44 (Mainz, 1992) ed. by J. Arvieux and E. De Sanctis
(Bologna, Italy, 1993).

\bibitem{B92}
R.G. Badalian, Phys. Lett. B {\bf 296}, 440, (1992).

\bibitem{BFF92} O. Benhar, A. Fabrocini, S. Fantoni, V. R. Pandharipande, and
I. Sick, Phys. Rev. Lett. {\bf 69}, 881 (1992); Nucl. Phys. A {\bf 532},
277 (1991).

\bibitem{BKM92} O. Benhar, B. Z. Kopeliovich, C. Mariotti, N. N. Nikolaev,
and B. G. Zakharov, Phys. Rev. Lett. {\bf 69}, 1156 (1992).


\bibitem{BZN95} O. Benhar, B. G. Zakharov, N. N. Nikolaev, and
S. Fantoni,  Phys. Rev. Lett. {\bf 74}, 3565 (1995).

\bibitem{BEA81} G. Bertsch, S. J. Brodsky, A. S. Goldhaber, and J.C.
Gunion, Phys. Rev.
Lett. {\bf 47}, 297 (1981).

\bibitem{BBK931}
A. Bianconi, S. Boffi, and D.E. Kharzeev,
Phys. Lett. B {\bf 305}, 1 (1993).

\bibitem{BBK932}
   By A. Bianconi, S. Boffi, and D.E. Kharzeev, Nucl. Phys. A {\bf 565},
767 (1993).

\bibitem{BBK94}
A Bianconi, S. Boffi, and D.E. Kharzeev, Phys. Lett. B {\bf 325}, 294 (1994).

\bibitem{BD65} {\it Relativistic Quantum Fields}, J. D. Bjorken and S. D. Drell
(McGraw-Hill 1965).

\bibitem{BVP92}
J. P. Blaizot  , R. Venugopalan, and M. Prakash,  Phys. Rev. D {\bf 45},
814 (1992).


\bibitem{B91} J. Botts, Nucl. Phys. B {\bf 353}, 20 (1991); Phys. Rev. D
{\bf 44}, 2768, (1991).


\bibitem{BS89} J. Botts and G. Sterman, Phys. Lett. B {\bf 224}, 201 (1989);
Nucl. Phys. B {\bf 325}, 62 (1989).

\bibitem{BF75} S. J. Brodsky and G. R. Farrar, Phys. Rev. D {\bf 11},
1309 (1975).

\bibitem{BL81} S. J. Brodsky and G. P. Lepage, Phys. Rev. D {\bf 24}, 2848
(1981).

\bibitem{B82} S. J. Brodsky, in {\it Proceedings of the Thirteenth
International Symposium on Multiparticle Dynamic}, (Vollendam, 1982) ed. by
W. Kittel {\it et al} (World Scientific, 1982).

\bibitem{BM88} S. J. Brodsky and A. H. Mueller, Phys. Lett. B 206, 685 (1988).


\bibitem{BH91}
S. J. Brodsky, P. Hoyer, Phys. Rev. Lett. {\bf 63}, 1566 (1989).
S. J. Brodsky and Paul Hoyer, Nucl. Phys. A {\bf 532}, 79 (1991);
S. J. Brodsky, in {\it Electronucler Physics with Internal Targets} (SLAC,
1989)
 ed. by R. Arnold (World Scientific, 1990).



\bibitem{BT88}
S. J. Brodsky and G.F. de Teramond, Phys. Rev. Lett. {\bf 60}, 1924 (1988).




\bibitem{BNL850} {\it BNL--Experiment 850}; spokesmen A. Carroll and S.
Hepplemann.



\bibitem{BFG94}
S. J. Brodsky, L. Frankfurt, J.F. Gunion, A.H. Mueller, and M. Strikman,
Phys. Rev. D {\bf 50}, 3134 (1994).

\bibitem{CCM92} C. Carlson, V. Chashkuhnashvilli and F. Myhrer, Phys. Rev. D
{\bf 46}, 2891 (1992).

\bibitem{CM91}
C.E. Carlson and Joseph Milana, Phys. Rev. D {\bf 44}, 1377 (1991).

\bibitem{CEA88} A. S. Carroll et al., Phys. Rev. Lett. {\bf 61}, 1698 (1988).

\bibitem{CZ84} V. L. Chernyak and A. R. Zhinitsky, Phys. Rep.
{\bf 112}, 173 (1984).

\bibitem{CS81} J. C. Collins and D. E. Soper, Nucl. Phys. B {\bf 193}, 381
(1981).

\bibitem{CT76} J. Cornwall and G. Tiktopolous, Phys Rev D {\bf 13}, 3370
(1976).

\bibitem{CEA81} E. A. Crosbie et al., Phys. Rev. D {\bf 23}, 600 (1981).


\bibitem{D73} Yu. L. Dokshitzer, Soviet JETP {\bf 73}, 1216 (1971);
V. N. Gribov and L. Lipatov, Sov. J. Nucl. Phys. {\bf 15}, 78 (1972);
G. Altarelli and G. Parisi, Nucl. Phys. {\bf 126}, 298 (1977).

\bibitem{DL79} A. Donnachie and P. V. Landshoff, Z. Phys. C {\bf 2}, 55 (1979).

\bibitem{ER80} A. V. Efremov and A. V. Radyushkin, Phys. Lett. B {\bf 94},
245 (1980).

\bibitem{EFG94} K. Egiian, L. Frankfurt, W. R. Greenberg, G. A. Miller, M.
Sargsian and M. Strikman, Nucl. Phys. A {\bf 580}, 365 (1994).

 \bibitem{EK92} J.M. Eisenberg and G. Kalbermann,  Phys. Lett. B {\bf 286},
24, (1992).

\bibitem{E92} R. Ent, presented at the {\it Conference on High Energy Probes
of QCD and Nuclei}, (University Park, PA, 1992 (unpublished)).

\bibitem{F93}
G. Y. Fang et al., E665 Collaboration, in
{\it Particle and Nuclei 13th International Conference},
(Perugia, Italy, 1993) ed. by A. Pascolini (World Scientific, Singapore
(1994)).

\bibitem{F94} G. Fang, in {\it Proceeding of the 2nd Workshop on Small-x
and Diffractive Physics at
the Tevatron} (Fermilab, 1994), M. Albrow and A.R. White, co-chairs
(Fermilab,1994).

\bibitem{FLF88} G. R. Farrar, H. Liu, L. L. Frankfurt,
and M. I. Strikman, Phys. Rev. Lett. {\bf 61}, 686 (1988).
G. R. Farrar, L. L. Frankfurt, M. I. Strikman and H. Liu,
Phys. Rev. Lett. {\bf 64}, 2996 (1990).

\bibitem{FFS91}
G. R. Farrar, L. L. Frankfurt, M. I.
Strikman, Phys. Rev. Lett. {\bf 67}, 2111 (1991).


\bibitem{FP56} E. L. Feinburg and I. J. Pomeranchuk, Suppl. Nu. Cim {\bf
3}, 652 (1956).

\bibitem{F72}  {\it Photon Hadron Interactions} by R. P. Feynman,
(Benjamin) 1972.


\bibitem{FF92} S. Frankel and W. Frati, Phys. Lett. B {\bf 291}, 368 (1992).

\bibitem{FFW94}
  S. Frankel, W. Frati and N. Walet,
Nucl. Phys. A {\bf 580}, 595 (1994).

\bibitem{FMS92} L. L. Frankfurt, G. A. Miller, and M. Strikman, Comm.
Nucl. Part. Phys.
{\bf 21}, 1 (1992).

\bibitem{FMS93} L. L. Frankfurt, G. A. Miller, and M. Strikman, Phys. Lett. B
{\bf 304}, 1 (1993).

\bibitem{FMS94} L. L. Frankfurt, G. A. Miller, and M.
Strikman, Annu. Rev. Nucl. Part. Sci. {\bf 45}, 501 (1994).

\bibitem{FMS95} L. L. Frankfurt, E. J. Moniz,
   M. M. Sargsian, and M. I. Strikman, 1995 preprint, nucl--th 9501019.




\bibitem{G93} S. Gardner,   Phys. Rev. C {\bf 48}, 3011 (1993).

\bibitem{GEA92} G. Garino, {\it et al.} Phys. Rev. {\bf C45}, 780 (1992).

\bibitem{GRB94} R. Gilman, P. M. Rutt, E. J. Brash, S. Nanda {\it et al},
CEBAF proposal 94-120.

\bibitem{G59} R. J. Glauber in {\it Lectures in Theoretical Physics}, W. E.
Brittin and L. G. Dunham, Eds. (Interscience, New York, 1959).

\bibitem{GW60} M. L. Good and W. D. Walker, Phys Rev. {\bf 120}, 1857 (1960).

\bibitem{GPR95} T. Gousset,  B. Pire, and J. P. Ralston,
preprint DAPNIA/SPhN 95 19, to be published in Phys. Rev. D.

\bibitem{GY73} G. R. Grammer, Jr. and D. Yennie, Phys. Rev. D {\bf 8},
4332 (1973).

\bibitem{GM94}
 W. R. Greenberg and G. A. Miller,  Phys. Rev. D {\bf 47}, 1865, (1993).
 W. R. Greenberg and G. A. Miller, Phys. Rev. C {\bf 49}, 2747 (1994).

\bibitem{GS77} J. F. Gunion and D. E. Soper,
Phys. Rev. D {\bf 15}, 2617 (1977).

\bibitem{HEA77} CLMN collaboration: J. L. Hartmann {\it et al}: Phys. Rev.
Lett. {\bf 39}, 975 (1977); S. Conetti {\it et al}, Phys. Rev. Lett. {\bf 41},
924 (1978).

\bibitem{HBB91}
H. Heiselberg, G. Baym, B. Blaettel, L. L. Frankfurt, and M. Strikman,  Phys.
Rev. Lett. {\bf 67}, 2946  (1991).

\bibitem{H90}
S. Heppelmann, Nucl. Phys. B, Proc. Suppl. {\bf 12}, 159 (1990).
S. Heppelmann {\it et al.}, Nucl. Phys. A {\bf 527}, 581c (1991).

\bibitem{HERMES} {\it Hermes proposal}, Report DESY--PRC 90/01.

\bibitem{ILS94} D.G. Ireland, L. Lapikas, and G. van der Steenhoven,
   Phys. Rev. C {\bf 50}, 1626 (1994).

\bibitem{IL84} N. Isgur and C. Llewelyn-Smith, Phys. Rev. Lett. {\bf 52},
1080 (1984); Phys. Lett. B {\bf 217}, 535 (1989).

\bibitem{JPR92} P. Jain, B. Pire, and J. P. Ralston, in {\it DPF92: the
Fermilab Meeting}, ed. by
C. H. Albright et al, (World Scientific, 1993).

\bibitem{JR92}
P. Jain and J. P. Ralston,  Phys. Rev. D {\bf 46}, 3807 (1992).

\bibitem{JR931}
P. Jain and J. P. Ralston, Phys. Rev. D {\bf 48}, 1104 (1993).

\bibitem{JR932}  P. Jain and J. P. Ralston, in {\it QCD and High Energy
Interactions} (XXVIII International
Rencontre de Moriond, Les Arcs, France 1993), ed. by J. Tr\^an Thanh V\^an
(Editions
Fronti\`eres, 1994);
P. Jain and J. P. Ralston, in proceedings of {\it International Conference on
Elastic and
Diffractive Scattering} (5th Blois Workshop) (Providence, RI, 1993)
ed. by H. Fried, K. Kang and C.--I. Tan (World Scientific, 1994).

\bibitem{JSW92} P. Jain, J. Schechter, and H. Weigel, Phys. Rev. D {\bf 45},
1470 (1992).

\bibitem{JaK93} See R. Jakob and P. Kroll, Phys. Lett. B {\bf 315}, 463 (1993),
and references therein.

\bibitem{JK93} B. K. Jennings and B. Z. Kopeliovich,
Phys. Rev. Lett. {\bf 70}, 3384 (1993).

\bibitem{JM90} B. K. Jennings and G. A. Miller,
  Phys. Lett. B {\bf 236}, 209 (1990).

\bibitem{JM91}
B. K. Jennings and G. A. Miller,
 Phys. Rev. D {\bf 44}, 692 (1991).

\bibitem{JM921} B. K. Jennings and G. A. Miller,
  Phys. Lett. B {\bf 274}, 442, (1992).

\bibitem{JM922}  B. K. Jennings and G. A. Miller,
  Phys. Rev. Lett. {\bf 69}, 3619 (1992).

\bibitem{JM93}
   B. K. Jennings and G. A. Miller,
   Phys. Lett. B {\bf 318}, 7 (1993).

\bibitem{JM94}
B. K. Jennings and G. A. Miller,   Phys. Rev. C {\bf 49}, 2637 (1994).

\bibitem{KEA77} CHHAV collaboration: H. de Kerret {\it et al}, Phys. Lett.
B {\bf 68}, 374 (1977); E. Nagy {\it et al}, Nucl. Phys. B {\bf 150}, 221
(1979).

\bibitem{K50} D.T. King, private communication (1950) cited by Perkins (Ref.
135).

\bibitem{KU77} T. Kinoshita and A. Ukawa, Phys. Rev. D {\bf 16}, 332 (1977).



\bibitem{KYS92}A. Kohama, K. Yazaki,  and R. Seki,   Nucl. Phys. A {\bf
536}, 716
(1992); A. Kohama, K. Yazaki, and R. Seki, Nucl. Phys. A {\bf 551}, 687 (1993);
   Nucl. Phys. A {\bf 575}, 645 (1994).

\bibitem{KYS95} A. Kohama, K. Yazaki, and R. Seki, Phys. Lett. B {\bf 344}, 61
(1995).

\bibitem{KZ911} B. Z. Kopeliovich and B. G. Zakharov, Phys. Rev. D {\bf 44},
3466 (1991).


\bibitem{KZ912} B. Z. Kopeliovich and B. G. Zakharov,
Phys. Lett. B {\bf 264}, 434 (1991).

\bibitem{K92} B. Z. Kopeliovich,
Sov. J. Nucl. Phys. {\bf 55}, 752 (1992).

\bibitem{KNN93}
B. Z. Kopeliovich, J. Nemchick, N. N. Nikolaev, and B. G. Zakharov,   Phys.
Lett. B {\bf 309}, 179, (1993); Phys. Lett. B {\bf 324} 469 (1994).


\bibitem{L74} P. V. Landshoff, Phys. Rev.  D{\bf 10}, 1024 (1974).

\bibitem{LISS95} {\it LISS: a Light Ion Spin Synchrotron} (IUCF Report, 1995).

\bibitem{LM92}
T. S. H. Lee and G. A. Miller, Phys. Rev. C {\bf 45}, 1863 (1992).

\bibitem{LB80} G. P. Lepage and S. J. Brodsky, Phys. Rev. D {\bf 22}, 2157
(1980);
in {\it Perturbative QCD}, edited by A. H. Mueller (World Scientific,
Singapore, 1990).

\bibitem{LS92}
H.-n. Li and G. Sterman, Nucl. Phys. B {\bf 381}, 129 (1992);
H.-n. Li, Phys. Rev. D {\bf 48}, 4243 (1993).



\bibitem{L75} F. E. Low, Phys. Rev. D {\bf 12}, 163 (1975).


\bibitem{MEA94} N. Makins {\it et al.}, Phys. Rev. Lett. {\bf 72}, 1986
(1994).

\bibitem{MEET} See, e.g., Ref.\citenum{AS93}; also {\it Hadronic Physics with
Electrons Beyond 10GeV} (Dourdan, France 1990) Ed. by B. Frois and J.F. Mathiot
(Nucl. Phys. A {\bf 532}, 1c (1991); {\it Proceedings of the Workshop on Future
Directions in Particle and Nuclear Physics at Multi--GeV Hadron Facilities}
(Brookhaven 1993) BNL--52389, Ed. by D. Geesaman.



\bibitem{MM94}
D. Makovoz and G. A. Miller,  Phys. Rev. C {\bf 51}, 2716 (1995).


\bibitem{MM95} D. Makovoz and G. A. Miller, 1994 preprint, nucl-th/9411004.


\bibitem{MMT73} V. A. Matveev, R. M. Muradyan and V. A. Tavkhelidze, Lett.
Nuovo
Cimento {\bf 7}, 719 (1973).


\bibitem{MEL91} R. Milner et al., CEBAF proposal PR-91-007 (1991).

\bibitem{M82} A. H. Mueller, in {\it Proceedings of the Seventh Rencontres
de Moriond}
(Les Arcs, France (1982)) ed. by J. Tr\^an Thanh V\^an (Editions Fronti\`eres,
Gif-sur Yvette 1982).

\bibitem{M81} A. H. Mueller, Phy. Rep. {\bf 73}, 237 (1981).

\bibitem{M89} {\it Perturbative QCD}, by A. H. Mueller, World Scientific,
Singapore (1989).

\bibitem{NZ92}
N. N. Nikolaev and B. G. Zakharov,
   Z. Phys. C {\bf 49}, 607, (1991); C {\bf 53}, 331 (1992).


\bibitem{NSS93} N. N. Nikolaev, A. Szczurek, J. Speth, J. Wambach, B. G.
Zakharov, and V. R. Zoller, Phys. Lett. B {\bf 317}, 287, (1993).


\bibitem{N94}
N. N. Nikolaev, Int. J. Mod. Phys. {\bf E3}, 1 (1994); ERRATUM-ibid. {\bf E3},
995 (1994);  Comm. Nucl. Part. Phys. {\bf 21}, 41 (1992); JETP Lett. 57
(1993) 88. (Pisma Zh. Eksp. Teor. Fiz. {\bf 57}
82 (1993).

\bibitem{NSS941}
 N. N. Nikolaev, A. Szczurek, J. Speth, J. Wambach, B. G.
Zakharov, and V. R. Zoller, Nucl. Phys. A {\bf 567}, 781 (1994);
 Phys. Rev. C {\bf 50}, 1296 (1994); ERRATUM-ibid.C {\bf 51}, 1041 (1995).

\bibitem{NNZ94}
J. Nemchik, N. N. Nikolaev, and B. G. Zakharov, {\it Workshop on CEBAF at
Higher Energies},
(Newport News, VA, 1994) preprint nucl--th 9406005;
N. N. Nilolaev  and J. Speth in ``The ELFE Project" ed. by J. Arvieux and E.
De Sanctis, Ref. \citenum{AS93}.

\bibitem{N75} S.  Nussinov, Phys. Rev. Lett. {\bf 34}, 1286 (1975).

\bibitem{OEA94} T. G. O'Neill et al., Phys. Lett. B {\bf 351}, 87 (1995).

\bibitem{P55} D. Perkins, Phil. Mag. {\bf 46}, 1146 (1955).

\bibitem{Pe74} {\it High Energy Hadron Physics}, by M. Perl, (Wiley, 1974).

\bibitem{PR82} B. Pire and J. P. Ralston, Phys. Lett. B {\bf 117}, 233
(1982); J. P. Ralston and B. Pire, Phys. Rev. Lett. {\bf 49}, 1605 (1982).

\bibitem{PR89} B. Pire and J. P. Ralston, in {\it Electronucler Physics
with Internal Targets} (SLAC)
 ed. by R. Arnold (World Scientific, 1990).



\bibitem{PR911} B. Pire and J. P. Ralston, Phys. Lett. B {\bf 256}, 523
(1991).

\bibitem{PR912} B. Pire and J. P. Ralston,
 Nucl. Phys. A {\bf 525}, 419, (1991);  Nucl. Phys. News {\bf 1}, 23 (1991);
 in {\it ``Nuclear Physics at Intermediate Energies''}, (Trieste, Italy, 1991)
 ed. by S. Boffi et al.(World Scientific 1992);  B. Pire, in {\it ``Proceedings
 of Workshop on Physics at SuperLEAR''}, (Zurich, Switzerland, 1991) ed. by
 C. Amsler and D. Urner (Institute of Physics, Bristol 1992);
{\it ``Winter Workshop on
Hadronic Physics with Multi-GeV Electrons''}, (Les Houches, France, 1990)
ed. by B. Desplanques and D. Goutte (Nova Science Publishers, New York 1991).

\bibitem{P74} J.C. Polkinghorne  Phys. Lett. {\bf 49}B, 277 (1974).

\bibitem{PC65} J.L. Powell and B. Craseman, {\it Quantum Mechanics} (Addison
Wesley 1965) Ch. 8.

\bibitem{R84} A. V. Radyushkin, Acta Phys. Polonica, B {\bf 15},
403 (1984);
Nucl. Phys. A {\bf 532}, 141 (1991);
A. P. Bakulev and A. P. Radyushkin, Phys. Lett. B {\bf 271}, 223
(1991).

\bibitem{RP88} J. P. Ralston and B. Pire, Phys. Rev. Lett. {\bf 61},
1823 (1988).

\bibitem{RP89} J. P. Ralston and B. Pire, in {\it QCD and High Energy
Interactions} (XXVIII International
Rencontre de Moriond, Les Arcs, France 1993), ed. by J. Tr\^an Thanh V\^an
(Editions
Fronti\`eres, 1994); B. Pire and J. P. Ralston, in {\it Electronucler Physics
 with Internal Targets} (SLAC) ed. by R. Arnold (World Scientific, 1990);
{\it Proceedings of the European Physical Society High Energy Physics}
Madrid 1989) Nucl.Phys. B (Proc. Supp.)  {\bf 16}, 264 (1990).


\bibitem{RS79} For general properties of the quark distributions, see J. P.
Ralston and
D.E. Soper, Nucl. Phys. B {\bf 152}, 109 (1979) and references therein.

\bibitem{R94} J. P. Ralston, in {\it Perspectives in the Structure of Hadronic
Systems} (NATO
Advanced Study Institute, Dronten, the Netherlands 1993) edited by M. Harakeh,
Th. Bauer, J. H. Koch and E.Scholten (Plenum, NY 1994).


\bibitem{R91}  J. P. Ralston,  Phys. Lett. B {\bf 269}, 439  (1991).

\bibitem{RP90}  J. P. Ralston and B. Pire, Phys. Rev. Lett. {\bf 65}, 2343
(1990);
 Nucl. Phys. A {\bf 532}, 155, (1991).

\bibitem{RP912} J. P. Ralston and B. Pire,
Phys. Rev. Lett. {\bf 67}, 2112, (1991).

\bibitem{RS92} G. P. Ramsay and D. Sivers, Phys. Rev. D {\bf 45}, 79 (1992).

\bibitem{RJ94} A. S. Rinat and B. K. Jennings, Nucl. Phys. A {\bf 568},
873, 1994.

\bibitem{SEA91} A. Saha et al., CEBAF proposal PR-91-006 (1991).

\bibitem{S83} A. Sen, Phys. Rev. D {\bf 28}, 860 (1983).

\bibitem{S93} {\it An Introduction to Quantum Field Theory}, by  G. Sterman
(Cambridge University Press, 1993).

\bibitem{S56} V. Sudakov, Soviet JETP {\bf 3}, 65 (1956).

\bibitem{SMZ89} A. Schafer, L. Mankiewicz, and Z.
Dziembowski,  Phys. Lett. B {\bf 233}, 217 (1989).

\bibitem{T85} R. Tucci, Phys Rev D {\bf 32}, 945 (1985).

\bibitem{W66}   S. Weinberg, Phys Rev {\bf 150},1313 (1966);  L. Susskind, Phys
Rev
{\bf 165}, 1535 (1968); see also D. E. Soper, SLAC-REPORT 137 (1971)
(unpublished).

\bibitem{YFS61} D. C. Yennie, S. C. Frautschi, and H. Suura, Ann Phys (NY) {\bf
13}, 379 (1961).

\bibitem{ZK93} {\it QCD 20 years later} ed. by P. M. Zerwas and H. A.
Kastrup (World Scientific, Singapore, 1993).


\end{thebibliography}
\end{document}